\documentclass[preprint,sort&compress,12pt]{elsarticle}
\newcounter{bla}

\journal{Computer Physics Communications}
\usepackage{amsmath}
\usepackage{amsfonts,amsbsy}
\usepackage{amssymb}
\usepackage{graphicx}
\usepackage{xcolor}
\usepackage{amsmath}
\usepackage{hyperref} 
\usepackage{subfigure}
\usepackage{bm}
\usepackage{sidecap, caption}
\usepackage{relsize}
\usepackage{listings}
\usepackage{tabularx}
\usepackage{inconsolata}
\DeclareMathOperator{\arcsinh}{arcsinh}

\def\empile#1\over#2{\mathrel{\mathop{\kern 0pt#1}\limits_{#2}}}

\def\beq{\begin{equation}}
\def\eeq{\end{equation}}
\def\bea{\begin{eqnarray}}
\def\eea{\end{eqnarray}}

\newcommand{\xbj}{$x_{{\!}_{Bj}}$}

\newcommand{\Tr}{{\rm Tr}}

\newcommand{\Lb}{\left(}
\newcommand{\Rb}{\right)}
\def\p{{\boldsymbol p}}

\def\x{{\boldsymbol x}}

\def\r{{\boldsymbol r}}

\def\v{{\boldsymbol v}}

\def\d3p{\frac{d^3\p}{(2\pi)^3}E_\p}
\definecolor{darkgreen}{rgb}{0.0, 0.45, 0.0}
\usepackage[normalem]{ulem}

\def\d{{\rm d}}

\begin{document}

\begin{frontmatter}



\title{\texttt{SIDIS-RC EvGen}: a Monte-Carlo event generator \\
of semi-inclusive deep inelastic scattering with \\
the lowest-order QED radiative corrections} 


\author[label1,label2]{Duane Byer}
\author[label1,label2]{Vladimir Khachatryan}
\author[label1,label2]{Haiyan Gao}
\author[label1]{Igor Akushevich}
\author[label3,label4]{Alexander Ilyichev}
\author[label5]{Chao Peng}
\author[label6,label7]{Alexei Prokudin}
\author[label1,label2]{Stan Srednyak}
\author[label1,label2]{Zhiwen Zhao}

\cortext[author] {Corresponding author.\\\textit{E-mail address:} vladimir.khachatryan@duke.edu (V. Khachatryan)}
\address[label1]{Department of Physics, Duke University, Durham, NC 27708, USA} 
\address[label2]{Triangle Universities Nuclear Laboratory, Durham, NC 27708, USA}
\address[label3]{Belarusian State University, Minsk, 220030, Belarus}
\address[label4]{Institute for Nuclear Problems, Belarusian State University, Minsk, 220006, Belarus}
\address[label5]{Physics Division, Argonne National Laboratory, Lemont, IL 60439, USA}
\address[label6]{Division of Science, Penn State Berks, Reading, PA 19610, USA}
\address[label7]{Theory Center, Thomas Jefferson National Accelerator Facility, Newport News, VA 23606, USA}

\begin{abstract}
\texttt{SIDIS-RC EvGen} is a C\texttt{++} standalone Monte-Carlo event generator for studies of semi-inclusive deep inelastic 
scattering (SIDIS) processes at medium to high lepton beam energies. In particular, the generator contains binary and library 
components for generating SIDIS events and calculating cross sections for unpolarized or longitudinally polarized beam and unpolarized, 
longitudinally or transversely polarized target. The structure of the generator incorporates transverse momentum-dependent parton
distribution and fragmentation functions, whereby we obtain multi-dimensional binned simulation results, which will facilitate
the extraction of important information about the three-dimensional nucleon structure from SIDIS measurements. 
In order to build this software, we have used recent elaborate QED calculations of the lowest-order radiative effects, 
applied to the leading order Born cross section in SIDIS. In this paper, we provide details on the theoretical formalism as well as the
construction and operation of \texttt{SIDIS-RC EvGen}, e.g., how we handle the event generation process and perform multi-dimensional 
integration. We also provide example programs, flowcharts, and numerical results on azimuthal transverse single-spin asymmetries.
\end{abstract}


\end{frontmatter}


{\bf PROGRAM SUMMARY}

\begin{small}
\noindent
{\em Program Title: {\rm SIDIS-RC EvGen}}  \\
{\em Licensing provisions: {\rm GNU General Public License Version 3 (GPLv3)}}  \\
{\em Developer's repository link: {\rm \url{https://github.com/duanebyer/sidis}}}  \\
{\em Maintainers:} Duane Byer; Department of Physics, Duke University  \\
{\em Distribution format: {\rm GitHub repository}}  \\
{\em Programming language: {\rm C\texttt{++}, Python}}  \\
{\em Operating System: {\rm Linux, macOS and their distributions}}  \\
{\em External packages: {\rm FOAM, ROOT, GSL, VEGAS, Cubature, Cog, WW-SIDIS, MSTWPDF}}  \\
{\em Nature of problem: 
{\rm The task is to first create a code for calculations of the leading order Born cross section as well as radiative corrections 
(RCs) at the next-to-leading order (NLO) of the cross section of lepton-hadron semi-inclusive deep inelastic scattering (SIDIS) at
medium to high beam energies with incident unpolarized or longitudinally polarized lepton beam and unpolarized, longitudinally or 
transversely polarized target, enabling to compute azimuthal single-target and double-beam-target spin asymmetries. 
Afterwards, a Monte-Carlo event generator based upon this code is developed, where in the coding and simulation processes 
multi-dimensional integrals need to be calculated precisely to obtain the exact NLO RCs to the SIDIS cross section with high 
precision beyond ultra-relativistic limit, which means that the lepton mass is taken into account.}
} \\
{\em Solution method: 
{\rm 
The project for building the \texttt{SIDIS-RC EvGen} software is divided into a library component called \texttt{libsidis}, for the 
purpose of calculating the inelastic tail to the SIDIS total cross section, and a binary component called \texttt{sidisgen}, for generating 
events. The Monte-Carlo event generation is performed  through the usage of the FOAM library, by applying a spatial partitioning method 
with hyper-cubical ``foam of cells". The multi-dimensional numerical integration is carried out by the GSL, VEGAS, and Cubature packages. 
The SIDIS structure functions that describe the SIDIS scattering process, and which are given in terms of transverse momentum-dependent 
parton distribution and fragmentation functions, are calculated in Gaussian and Wandzura–Wilczek-type approximations and implemented 
in the generator accordingly.} 
} \\
{\em Additional comments including restrictions and unusual features: 
{\rm The restrictions depend on the complexity of problems, limited by CPU time. As a consequence, the program running time is 
of order of several minutes to hours. \texttt{SIDIS-RC EvGen} does not comprise yet the contribution coming from the radiative 
tail of exclusive lepton-hadron reactions, which is a separate contribution involving exclusive structure functions that are 
currently not well known.}
} \\
   \\


\end{small}




\tableofcontents

\section{\label{sec:intro} Introduction and physics motivation}
A considerable amount of our knowledge on the nucleon's quark-gluon composition has been obtained via experimental and theoretical 
studies of the collinear parton distribution functions (PDFs) \cite{Ethier:2020way} and fragmentation functions (FFs) \cite{Metz:2016swz}. 
The precise knowledge of PDFs is essential for understanding of Quantum Chromodynamics (QCD), and for making quantitative predictions for 
the observables related to the strong interaction sector of the Standard Model. In turn, the search of beyond Standard Model physics
also requires precise understanding of QCD.  Within the so-called collinear factorization of deep inelastic scattering (DIS),
leading-twist (one-dimensional) PDFs can be interpreted at the leading order as probability densities for finding an
unpolarized/longitudinally polarized parton in a fast moving unpolarized/longitudinally polarized nucleon\footnote{Collinear FFs in 
turn describe how quarks and gluons transform into color-neutral particles such as hadrons.}. They depend on a fraction \xbj of the
nucleon momentum carried by the parton and on the ``resolution scale" that is associated with the hard scale $Q^{2}$ of a lepton-nucleon
scattering process. These PDFs are well-known, and over the course of more than two decades the frontier of exploration has been extended
to the three-dimensional (3D) nucleon structure \cite{Belitsky:2003nz}.

The 3D nucleon structure in the momentum space is encoded in the so-called transverse momentum dependent distribution and fragmentation
functions, TMD PDFs and TMD FFs 
\cite{Boer:1997nt,Goeke:2005hb,Bastami:2018xqd,Bacchetta:2006tn,Barone:2015ksa,Aybat:2011zv,Angeles-Martinez:2015sea,Metz:2016swz}. 
These partonic functions are generalizations of the aforementioned collinear distribution and fragmentation functions appearing in the 
standard collinear factorization \cite{Collins:1989gx}. Both of them depend on two independent variables: \xbj and $k_{\perp}$ in the 
case of TMD PDFs, as well as  $z_{h}$ and $p_{\perp}$ in the case of FFs. In the latter case, $z_{h}$ is the fraction of the quark
momentum transferred to the produced (final-state) hadron, and $p_{\perp}$ is the transverse momentum of the same final-state hadron 
but with respect to the direction of the fragmenting quark. QCD factorization theorems have been proven for processes with two distinct 
measured scales, $Q_1 \ll Q_2$, such as the semi-inclusive deep inelastic scattering (SIDIS) \cite{Bacchetta:2006tn,Bastami:2018xqd}, 
Drell-Yan process \cite{Arnold:2008kf}, and production of two hadrons in $e^+e^-$ annihilation \cite{Metz:2016swz}. In case of SIDIS, 
the observed transverse momentum $P_{hT}$ of the produced charged hadron with respect to the virtual photon momentum 
plays the role of the small scale $Q_1$, while the virtual photon virtuality $Q$ plays the role of the large scale $Q_2$. In the TMD 
description, the observed transverse momentum $P_{hT}$ is related to the active parton (quark) intrinsic transverse momentum $k_{\perp}$ 
and fragmenting quark $p_{\perp}$ as well as to the effects of the gluon radiation encoded in the evolution equations \cite{Barone:2015ksa}. 
By utilizing juxtapositions of the lepton beam and nucleon target polarizations in SIDIS experiments, we will be capable of extracting 
pivotal information on various TMDs, whereby one can quantify the quark transverse motion inside the nucleon and observe spin-orbit 
correlations, as well as obtain quantitative insight on the quark orbital angular momentum (OAM) contribution to the proton spin. Note 
that the results coming from advances in target polarization techniques, like those employed by HERMES, COMPASS, LHCb and JLab experiments 
\cite{HERMES:2015,Andrieux:2022zmk,Aidala:2019pit,Goertz:2002vv,Keller:2020wan}, are crucial in studies of TMDs together with the additional 
information that can be accessed from the Drell-Yan process \cite{Arnold:2008kf} and $e^{+}e^{-}$ annihilation \cite{Metz:2016swz}.

However, in SIDIS experiments devoted to data extractions on TMDs, such as 
HERMES \cite{Airapetian:2004tw,HERMES:2009lmz,HERMES:2010mmo,HERMES:2012kpt,HERMES:2012uyd}, 
COMPASS \cite{Alexakhin:2005iw,COMPASS:2012dmt,COMPASS:2014bze,COMPASS:2014kcy,COMPASS:2017mvk},
and JLab \cite{Qian:2011py,CLAS:2008nzy}, 
it is indispensable to have a very good control over systematic uncertainties. In this regard, one of the dominant sources of the systematic 
uncertainties in these experiments, with and without polarization of the lepton beam and the nucleon target, are the radiative corrections 
(RC) due to the radiation of photons off leptons. Radiative corrections are calculated based on Quantum Electrodynamics (QED). Exact 
analytical formulas for the lowest-order model-independent part of QED RCs to the SIDIS cross section, with a longitudinally polarized 
lepton beam and arbitrarily polarized nucleon target, beyond ultra-relativistic approximation\footnote{It means that the incident beam 
lepton mass is taken into account in all calculations.}, are calculated in Ref.~\cite{Akushevich:2019mbz}. The methodology developed in that 
paper is anchored upon the covariant approach for RC calculations, so that the obtained formulas can be directly applied to any coordinate 
system. Besides, the Bardin-Shumeiko approach \cite{Bardin:1976qa,Shumeiko:1978cn} is applied for covariant extraction and cancellation 
of the infrared divergence that stem from the real and virtual photon emissions in the SIDIS process. Within this approach, the RC 
cross-section results do not depend on an artificial parameter, which is usually introduced to separate the photon emission on hard and soft 
parts \cite{Mo:1968cg}. The calculations of Ref.~\cite{Akushevich:2019mbz} have been performed in a model-independent manner, involving the 
SIDIS cross section that contains eighteen SIDIS structure functions \cite{Bastami:2018xqd,Bacchetta:2006tn}\footnote{The results of 
Ref.~\cite{Akushevich:2019mbz} also include a contribution of the exclusive radiative tail to the lowest-order RC to the SIDIS process, 
however, currently the corresponding exclusive structure functions are not known yet. More theoretical \cite{Akushevich:2022} and 
\cite{Goloskokov:2009ia} experimental efforts are needed for their determination, as well as for the complete derivations of all SIDIS 
structure functions.}. Note that an alternative method of accounting for QED radiative corrections in SIDIS based on factorization approach was introduced in Ref.~\cite{Liu:2021jfp}. It will be interesting to compare results of the two methods \cite{Karki:2022}.

There exist other Monte-Carlo (MC) frameworks for evaluation of RC corrections in SIDIS. One can successfully evaluate RC effects by having 
information obtained during event generation, and such a generator can be constructed by utilizing a code for RC in SIDIS, in a similar way 
as the program RADGEN 1.0 \cite{Akushevich:1998ft,RadCorr} was constructed based on the program POLRAD 2.0 \cite{Akushevich:1997di,RadCorr,Codes}. 
RADGEN 1.0 is a MC generator of polarized/unpolarized DIS radiative events, which can be applied for RC generation in inclusive, semi-inclusive 
and exclusive DIS processes. POLRAD 2.0 is a FORTRAN code for treating experimental data with an implemented RC procedure, in polarized inclusive 
and semi-inclusive DIS. Note that there is also HAPRAD 2.0, a FORTRAN code for RC studies in the semi-inclusive hadron electroproduction 
processes, which has available versions with MC and numerical integration \cite{Akushevich:2007jc,Codes}. HAPRAD 2.0 in turn stems from the 
original HAPRAD program \cite{Akushevich:1999hz,HAPRAD,RadCorr,Codes}, which performs RC calculations to the five-fold differential cross 
section of unpolarized particles, $\d\sigma_{\rm SIDIS}/\d x_{{\!}_{Bj}} \d y \d z_{h} \d P_{hT}^{2} \d \phi_{h}$. In addition to the already
specified variables $x_{{\!}_{Bj}}$, $z_{h}$, $P_{hT}$, here $y$ is the fraction of the lepton beam energy carried by the virtual photon, and 
$\phi_{h}$ is the hadron azimuthal angle measured with respect to the lepton scattering plane in the SIDIS process. We will discuss the SIDIS 
process in details in Sec.~\ref{sec:SIDIS_kinematics}.

However, we now have exact and more general results on the six-fold differential cross section, 
$\d \sigma_{\rm SIDIS}/\d x_{{\!}_{Bj}} \d y \d z_{h} \d P_{hT}^{2} \d \phi_{h} \d \phi_{{\!}_{S}}$ (obtained in \cite{Akushevich:2019mbz}), where $\phi_{{\!}_{S}}$ 
is another azimuthal angle describing the target-spin direction (spin-vector), if transversely polarized targets are applied. With all the past developments 
our task was to create a MC event generator that for the SIDIS process would be the following:
\begin{itemize}
\item[(i)]  generate radiative and non-radiative channels of scattering: namely, the generated events could be 
selected to be either radiative or non-radiative, with a probability of being proportional to the radiative/non-radiative
cross section; 
\item[(ii)]  generate scattered lepton kinematics (i.e., $Q^{2}$ and \xbj) and final-state hadron kinematics 
(i.e., $z_{h}$, $P_{hT}$, and $\phi_{h}$); 
\item[(iii)]  generate radiation real photon kinematics;
\item[(iv)]  calculate the full SIDIS cross section  in any generated phase-space point with RCs included. 
\end{itemize}
In this paper, we introduce the event generator \texttt{SIDIS-RC EvGen}, which includes implementation of the items (i) -- (iv), based on the SIDIS RC 
studies in \cite{Akushevich:2019mbz}, and on using TMD PDFs and TMD FFs discussed extensively in \cite{Bastami:2018xqd}. The full package of 
\texttt{SIDIS-RC EvGen} can be found in \cite{SIDIS-RC_EvGen:2020}, along with the installation and running instructions.

The generated final-state hadrons can be fed into detector simulations, allowing precise predictions and verification for some key aspects of entire experimental 
setups, such as SoLID, CLAS12 at JLab \cite{Chen:2014psa,Burkert:2008rj,Burkert:2018nvj}, EIC at BNL \cite{Accardi:2012qut,AbdulKhalek:2021gbh}, etc..
\texttt{SIDIS-RC EvGen} can be considered as one of the tools similar to HAPRAD 2.0, mLEPTO, LEPTO-PHI, the NJL-jet model-based generator 
(see \cite{Avakian:2015vha,Avakian:2015faa} for more details), which are aimed to help in multifaceted efforts study the internal spin structure, TMD evolution 
effects, spin-orbit and quark-gluon correlations, and the 3D momentum structure of the nucleon in general.

In this regard and by having all the above discussion in mind, let us present how this paper is structured in what follows. In Sec.~\ref{sec:theory}, we 
discuss the SIDIS process, including the TMD PDFs, TMD FFs, and structure functions that are essential to study the 3D tomography of the nucleon in momentum 
space. We then discuss some key aspects of the lowest-order QED radiative effects in SIDIS, describing each cross-section component that contributes 
to the total inelastic cross section. In Sec.~\ref{sec:MC_gen}, we show how \texttt{SIDIS-RC EvGen} is constructed, how the events are generated and cross 
sections computed. Our focus is in particular on a library component \texttt{libsidis} for calculating the total SIDIS cross section, and on a generator (binary) 
component \texttt{sidisgen} for generating events. We continue this discussion in Sec.~\ref{sec:programs}, where we exhibit some numerical results obtained from 
the event generator. We notably demonstrate several results on the SIDIS final-state charged-hadron azimuthal transverse target single-spin asymmetries (SSAs). 
In the final Sec.~\ref{sec:concl}, we give a summary of our work and also outline some prospects for a further development.

\section{\label{sec:theory} Some basics of the theoretical framework for studying 3D momentum structure of the nucleon with SIDIS}

In this section, we discuss several aspects of a theoretical framework by which nucleon tomography is studied with SIDIS. On the other hand, we show all
necessary formulas and variables/definitions, which underlie the structure of \texttt{SIDIS-RC EvGen}, and whereby the event generation and cross-section 
calculations are carried out. The discussion of Sec.~\ref{sec:nucl_struc} is partially based upon \cite{Bastami:2018xqd}. Its results are represented
in a pertinent \texttt{MATHEMATICA}-implemented library called WW-SIDIS \cite{WW-SIDIS:2018}, which we use in \texttt{SIDIS-RC EvGen}. 
In Sec.~\ref{sec:QED_RC}, we aim to discuss the primary results of \cite{Akushevich:2019mbz} since the main goal for building \texttt{SIDIS-RC EvGen} is 
to produce the next-to-leading order (NLO) RCs to the SIDIS cross section. We also extensively refer to \cite{Bastami:2018xqd} and \cite{Akushevich:2019mbz}, 
wherever it is relevant, such that the reader can obtain complete information regarding any specific point or matter under discussion.

\subsection{\label{sec:nucl_struc} SIDIS process and functions describing the nucleon structure}
In this Section we will outline the basics of the SIDIS process following the formalism and notations described in Refs.~\cite{Bacchetta:2006tn,Bastami:2018xqd}.

\subsubsection{\label{sec:SIDIS_kinematics} Kinematics and cross section of the SIDIS process}

\begin{figure}[hbt]
\centering
\includegraphics[width=11.0cm]{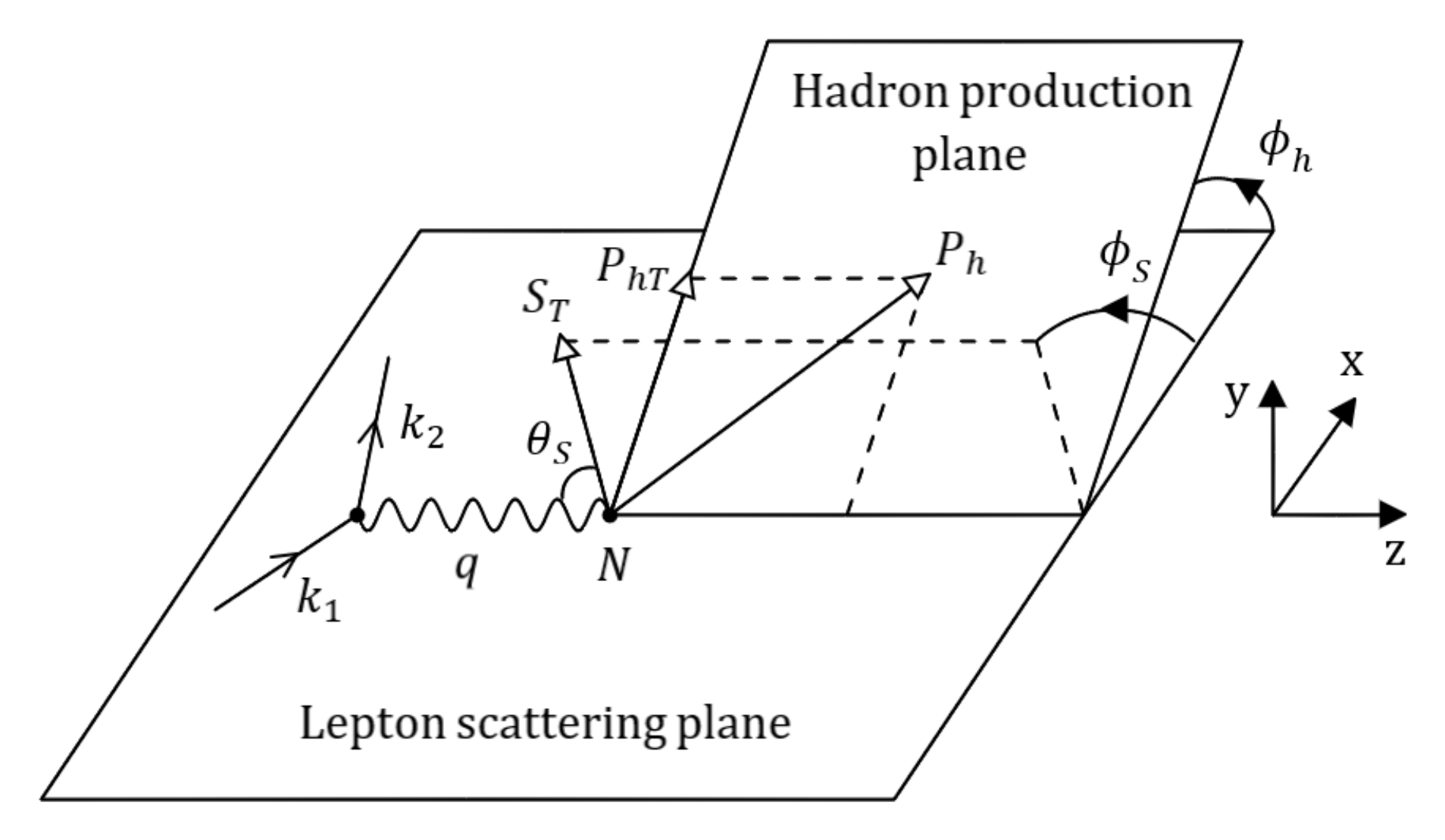}
\caption{Kinematics of the SIDIS process in the one-photon exchange approximation, and the sketch is made by following the Trento conventions in 
\cite{Bacchetta:2004jz}. The overall orientation of the lepton scattering plane around the incident lepton direction is characterized by an azimuthal 
angle $\phi_{l}$ taken with respect to an arbitrary reference frame, which however, in the DIS limit and for a transversely polarized target is 
$\phi_{l} \approx \phi_{{\!}_{S}}$.}
\label{fig:fig_SIDIS}
\end{figure}

We are interested in the SIDIS process, in which a final-state hadron $h$ is detected in coincidence with a lepton $\ell^{\prime}$ scattered off a target $N$. 
That process is given by \cite{Bastami:2018xqd,Bacchetta:2006tn,Barone:2015ksa,Bacchetta:2004jz} 
\beq
\ell(k_{1}) + N(P) \rightarrow \ell^{\prime}(k_{2}) + h(P_{h}) + X(P_{X}) ,
\label{eqn_eq:SIDIS_kin1}
\eeq
where $k_{1}$ and $P$ are the four-momenta of both polarized/unpolarized incident letpon and nucleon target, $k_{2}$ and $P_{h}$ are the four-momenta of the 
scattered lepton $\ell^{\prime}$ and the produced detected hadron $h$, $P_{X}$ is the four-momentum of the unobserved state consisting of all undetected 
hadrons produced in the reaction. In this scattering process, the respective nucleon and hadron masses are given by $M_{N}$ and $M_{h}$. This process is 
depicted in Fig.~\ref{fig:fig_SIDIS}, made under the assumption of the one-photon exchange approximation, where the reference frame is adopted to be the 
$\gamma^{\ast}$-$N$ center-of-mass frame, in which the virtual photon moves in the positive $z$ direction and the azimuthal angles are counted from the 
lepton scattering plane. The cross section of the process is expressed by the following kinematic invariants:
\beq
\mbox{\xbj} = \frac{Q^{2}}{2P \!\cdot\! q} ,~~~~~y = \frac{P \!\cdot\! q}{P \!\cdot\! k_{1}} ,~~~~~ z_{h} = \frac{P \!\cdot\! P_{h}}{P \!\cdot\! q} ,~~~~~ \gamma
= \frac{2M_{N}\mbox{\xbj}}{Q} ,
\label{eqn_eq:SIDIS_kin2}
\eeq
with $q$ defined to be the virtual photon momentum $q \equiv k_{1} - k_{2}$, and $Q^{2}$ is the virtuality (hard scale)  $Q^{2} \equiv -q^{2}$. 

Thus, assuming the one-photon exchange approximation, the SIDIS differential cross section is expressed by a set of eighteen structure functions 
(which can computed within the TMD factorization framework in the region of validity of TMD factorization itself) in a model-independent way \cite{Bacchetta:2006tn,Diehl:2005pc}:
\begin{eqnarray*}
& & \!\!\!\!\!\!\!\!\!\!\!\!\!\!\!\!\!\!\!\!\!\!\!\!\!\!\!\!\!\!\!\!
\frac{d\sigma_{\rm SIDIS}}{dx_{{\!}_{Bj}}\,dy\,dz_{h}\,dP_{hT}^{2}\,d\phi_{h}\,d\phi_{{\!}_{S}}} = \frac{\alpha^{2}}{x_{{\!}_{Bj}} y Q^{2}} 
\Lb 1 + \frac{\gamma^{2}}{2x_{{\!}_{Bj}}} \Rb \times
\nonumber\\
& & \!\!\!\!\!\!\!\!\!\!\!\!\!\!\!\!\!\!\!\!\!\!\!
\times \Bigg\{ \bigg[ c_{1}\,F_{UU,T} + c_{2}\,F_{UU,L} + c_{3} \cos{\!(\phi_{h})}\,F_{UU}^{\cos{\!(\phi_{h})}} +
\nonumber\\
& & \!\!\!\!\!\!\!\!\!\!
+ c_{2} \cos{\!(2\phi_{h})}\,F_{UU}^{\cos{\!(2\phi_{h})}} + \lambda_{e} c_{4} \sin{\!(\phi_{h})}\,F_{LU}^{\sin{\!(\phi_{h})}} \bigg] +
\end{eqnarray*}
\vskip 0.10truecm
\begin{displaymath}
\!\!\!\!\!\!\!\!\!\!\!\!\!\!\!\!\!\!\!\!\!\!\!\!\!\!
+ S_{L} \bigg[ c_{3} \sin{\!(\phi_{h})}\,F_{UL}^{\sin{\!(\phi_{h})}} + c_{2} \sin{\!(2\phi_{h})}\,F_{UL}^{\sin{\!(2\phi_{h})}} \bigg] +
\end{displaymath}
\vskip 0.35truecm
\begin{displaymath}
\!\!\!\!\!\!\!\!\!\!\!\!\!\!\!\!\!\!\!\!\!\!\!\!\!\!\!\!\!\!\!\!\!\!\!\!\!\!\!\!\!\!\!\!\!\!\!\!\!\!\!\!\!\!\!\!\!\!\!\!\!\!\!
+ S_{L}\lambda_{e} \bigg[ c_{5}\,F_{LL} + c_{4}  \cos{\!(\phi_{h})}\,F_{LL}^{\cos{\!(\phi_{h})}} \bigg] +
\end{displaymath}
\vskip -0.05truecm
\begin{eqnarray*}
& &~~~~
+ S_{T} \bigg[ \sin{\!(\phi_{h} - \phi_{{\!}_{S}})} \Lb c_{1}\,F_{UT,T}^{\sin{\!(\phi_{h} - \phi_{{\!}_{S}})}} + c_{2}\,F_{UT,L}^{\sin{\!(\phi_{h} - \phi_{{\!}_{S}})}} \Rb +
\nonumber\\
& &~~~~~~~~~~~~
+ c_{2} \sin{\!(\phi_{h} + \phi_{{\!}_{S}})}\,F_{UT}^{\sin{\!(\phi_{h} + \phi_{{\!}_{S}})}} + 
c_{2} \sin{\!(3\phi_{h} - \phi_{{\!}_{S}})}\,F_{UT}^{\sin{\!(3\phi_{h} - \phi_{{\!}_{S}})}} + 
\nonumber\\
& &~~~~~~~~~~~~
+ c_{3} \sin{\!(\phi_{{\!}_{S}})}\,F_{UT}^{\sin{\!(\phi_{{\!}_{S}})}} + 
c_{3} \sin{\!(2\phi_{h} - \phi_{{\!}_{S}})}\,F_{UT}^{\sin{\!(2\phi_{h} - \phi_{{\!}_{S}})}} \bigg] + 
\end{eqnarray*}
\vskip -0.30truecm
\begin{eqnarray}
& & \!\!\!\!\!\!\!\!\!\!\!\!\!\!\!\!\!\!\!\!\!\!\!\!\!
+ S_{T} \lambda_{e} \bigg[ c_{5} \cos{\!(\phi_{h} - \phi_{{\!}_{S}})}\,F_{LT}^{\cos{\!(\phi_{h} - \phi_{{\!}_{S}})}} +
c_{4} \cos{\!(\phi_{{\!}_{S}})}\,F_{LT}^{\cos{\!(\phi_{{\!}_{S}})}} +
\nonumber\\
& & \!\!\!\!\!
+ c_{4} \cos{\!(2\phi_{h} - \phi_{{\!}_{S}})}\,F_{LT}^{\cos{\!(2\phi_{h} - \phi_{{\!}_{S}})}} \bigg] \Bigg\} ,
\label{eqn_eq:SIDIS_kin3}
\end{eqnarray}
where $\alpha$ is the fine-structure constant, and $\lambda_{e}$ is the helicity of the lepton beam. In the $XY$ subscripts of the most structure functions, 
the first one $X = U/L$ refers to the unpolarized or longitudinally polarized beam (with $\lambda_{e}$). The second one, $Y = U/L~{\rm or}~(= U/T)$, 
is correspondingly ascribed to the unpolarized or longitudinally polarized (or transversely polarized) target with respect to $q$. In the $XY,Z$ subscripts of 
the remaining four structure functions, $Z = T/L$ specifies the virtual photon polarizations. The superscripts in structure functions (e.g., $\cos(\phi_h)$ 
in $F_{UU}^{\cos(\phi_h)}$) denote the corresponding azimuthal dependence, and when the superscript is absent (i.e., in $F_{LL}$), the dependence is flat.

All structure functions in the R.H.S. of Eq.~(\ref{eqn_eq:SIDIS_kin3}) are functions of $x_{{\!}_{Bj}}$, $Q^{2}$, $z_{h}$, and $P_{hT}^{2}$. 
Eq.~(\ref{eqn_eq:SIDIS_kin3}) corresponds to a SIDIS process, where the final hadron polarization is not measured, or a spin-0 hadron is 
produced such as the pion. The factors $c_{1}$, $c_{2}$, $c_{3}$, $c_{4}$, and $c_{5}$ (see eqs.~(2.9)-(2.13) of \cite{Bacchetta:2006tn} for 
more details) are given by
\begin{displaymath}
c_{1} = \frac{y^{2}}{2(1 - \varepsilon)} , \,\,\,\,\,c_{2} = \frac{y^{2}}{2(1 - \varepsilon)}\varepsilon ,\,\,\,\,\,c_{3} = \frac{y^{2}}{2(1 - \varepsilon)}\,
\sqrt{2\varepsilon(1 + \varepsilon)} ,
\end{displaymath}
\beq
c_{4} = \frac{y^{2}}{2(1 - \varepsilon)}\sqrt{2\varepsilon(1 - \varepsilon)} ,\,\,\,\,\,c_{5} = \frac{y^{2}}{2(1 - \varepsilon)}\sqrt{1 - \varepsilon^{2}} ,
\label{eqn_eq:SIDIS_kin4}
\eeq
where $\varepsilon$  is the ratio of longitudinal and transverse photon fluxes:
\beq
\varepsilon = \frac{1 - y - \Lb \gamma^{2}y^{2}/4 \Rb}{1 - y + \Lb y^{2}/2 \Rb + \Lb \gamma^{2}y^{2}/4 \Rb} .
\label{eqn_eq:SIDIS_kin5}
\eeq
The hadron azimuthal angle is defined in \cite{Bacchetta:2006tn,Bacchetta:2004jz}:
\beq
\cos{\!(\phi_{h})} = -\frac{k_{1\mu}P_{h\nu}\,g_{\perp}^{\mu\nu}}{\sqrt{k_{1T}^{2}P_{hT}^{2}}} ,~~~~~~~~~~
\sin{\!(\phi_{h})} = -\frac{k_{1\mu}P_{h\nu}\,\epsilon_{\perp}^{\mu\nu}}{\sqrt{k_{1T}^{2}P_{hT}^{2}}} ,
\label{eqn_eq:SIDIS_kin6}
\eeq
with $k_{1T}^{\mu} = g_{\perp}^{\mu\nu}k_{1\nu}$ and $P_{hT}^{\mu} = g_{\perp}^{\mu\nu}P_{h\nu}$ to be the transverse components of $l$
and $P_{h}$ with respect to $q$. The tensors $g_{\perp}^{\mu\nu}$ and $\epsilon_{\perp}^{\mu\nu}$ are respectively expressed by
\beq
g_{\perp}^{\mu\nu} = g^{\mu\nu} - \frac{q^{\mu}P^{\nu} + P^{\mu}q^{\nu}}{P\!\cdot\!q  \Lb 1 + \gamma^{2} \Rb} + \frac{\gamma^{2}}{1 + \gamma^{2}}
\Lb \frac{q^{\mu}q^{\nu}}{Q^{2}} - \frac{P^{\mu}P^{\nu}}{M_{N}^{2}} \Rb ,
\label{eqn_eq:SIDIS_kin7}
\eeq
\beq
\epsilon_{\perp}^{\mu\nu} = \epsilon^{\mu\nu\rho\sigma}\,\frac{P_{\rho}\,q_{\sigma}}{P\!\cdot\!q\,\sqrt{1 + \gamma^{2}}} ,
\label{eqn_eq:SIDIS_kin8}
\eeq
where the definition of the totally antisymmetric tensor is $\epsilon^{0123} = +1$, and the  non-zero components of these tensors are 
$g_{\perp}^{11} = g_{\perp}^{22} = -1$,
$\epsilon_{\perp}^{12} = -\epsilon_{\perp}^{21} = +1$. Note that the target covariant spin-vector, $Sv$, is decomposed as
\bea
& & Sv^{\mu} = S_{L}\,\frac{P^{\mu} - \left[ q^{\mu}M_{N}^{2}/(P\!\cdot\!q) \right]}{M_{N}\,\sqrt{1 + \gamma^{2}}} + S_{T}^{\mu} ,
\nonumber\\
& &
\mbox{with}~~~S_{L} = \frac{Sv\!\cdot\!q}{P\!\cdot\!q} \frac{M_{N}}{\sqrt{1 + \gamma^{2}}} ,
~~~\mbox{and}~~~S_{T}^{\mu} = g_{\perp}^{\mu\nu}\,Sv_{\nu} ,
\label{eqn_eq:SIDIS_kin9}
\eea
and for the transversely polarized targets, $\phi_{{\!}_{S}}$ is defined as
\beq
\cos{\!(\phi_{{\!}_{S}})} = -\frac{k_{1\mu}S_{\nu}\,g_{\perp}^{\mu\nu}}{\sqrt{k_{1T}^{2}S_{T}^{2}}} ,~~~~~~~~~~
\sin{\!(\phi_{{\!}_{S}})} = -\frac{k_{1\mu}S_{\nu}\,\epsilon_{\perp}^{\mu\nu}}{\sqrt{k_{1T}^{2}S_{T}^{2}}} .
\label{eqn_eq:SIDIS_kin10}
\eeq

In the partonic description of the SIDIS process, the functions $F_{UU,T}$, $F_{UU}^{\cos{\!(2\phi_{h})}}$, $F_{UL}^{\sin{\!(2\phi_{h})}}$, 
$F_{LL} $, $F_{UT,T}^{\sin{\!(\phi_{h} - \phi_{{\!}_{S}})}}$, $F_{UT}^{\sin{\!(\phi_{h} + \phi_{{\!}_{S}})}}$, $F_{UT}^{\sin{\!(3\phi_{h} - \phi_{{\!}_{S}})}}$, 
and $F_{LT}^{\cos{\!(\phi_{h} - \phi_{{\!}_{S}})}}$, are ``twist-2" (leading-twist) structure functions at leading order in the $1/Q$ expansion  in 
Eq.~(\ref{eqn_eq:SIDIS_kin3}). The functions $F_{UU}^{\cos{\!(\phi_{h})}}$, $F_{LU}^{\sin{\!(\phi_{h})}}$, $F_{UL}^{\sin{\!(\phi_{h})}}$, 
$F_{LL}^{\cos{\!(\phi_{h})}}$, $F_{UT}^{\sin{\!(2\phi_{h} - \phi_{{\!}_{S}})}}$, $F_{UT}^{\sin{\!(\phi_{{\!}_{S}})}}$, $F_{LT}^{\cos{\!(\phi_{{\!}_{S}})}}$, 
and $F_{LT}^{\cos{\!(2\phi_{h} - \phi_{{\!}_{S}})}}$ are ``twist-3" (subleading-twist) structure functions at subleading order in the $1/Q$ expansion of the 
same formula. The other remaining two structure functions, namely $F_{UU,L}$ and $F_{UT,L}^{\sin{\!(\phi_{h} - \phi_{{\!}_{S}})}}$ contribute to the process 
at the order of $\mathcal{O}(1/Q^{2})$ due to the longitudinal virtual-photon polarization, and hence can be neglected if one considers $\mathcal{O}(1/Q)$ accuracy.

\subsubsection{\label{sec:SIDIS_struc_func} TMD PDFs, TMD FFs, and SIDIS structure functions}

In the Bjorken limit ($Q^{2} \rightarrow \infty$ and $P \!\cdot\! q \rightarrow \infty$ with $x_{Bj}$ fixed), the SIDIS structure functions in 
Eq.~(\ref{eqn_eq:SIDIS_kin3}) are described at tree level in terms of convolution integrals of the unpolarized TMD PDF, $f^{a}(x, k_{\perp}^{2})$, and TMD FF, 
$D^{a}(z, p_{\perp}^{2})$\footnote{The light-cone momentum fractions $x = k^{+}/P^{+}$ and $z = P_{h}^{-}/\kappa^{-}$ are identified 
with $x_{{\!}_{Bj}}$ and $z_{h}$ up to the order of $k_{\perp}/Q$, where $\kappa$ is the fragmentation quark momentum \cite{Barone:2015ksa}.}, for a 
given quark flavor $a$, Ref.~\cite{Bastami:2018xqd}, as:
\bea
& & \mathcal{C}[\omega\,f\,D] = x \mathlarger{\sum}_{a} e_{a}^{2} \mathlarger{\int} d^{2}\bm{k}_{\perp}\,d^{2}\bm{p}_{\perp}\,\delta^{(2)}\!\Lb z\bm{k}_{\perp}
+ \bm{p}_{\perp} - \bm{P}_{hT} \Rb \times
\nonumber\\
& & ~~~~~~~~~~~~~~~~~~~~~~~~~~~~
\times \omega\!\Lb \bm{k}_{\perp},\, -\frac{\bm{p}_{\perp}}{z} \Rb f^{a}(x, k_{\perp}^{2})\,D^{a}(z, p_{\perp}^{2}) ,
\label{eqn_eq:SIDIS_kin11}
\eea
where $\omega$ is a weight function depending on $\bm{k}_{\perp}$ and $\bm{p}_{\perp}$ generally, along with a given defined unit vector 
$\bm{\hat{h}} = \bm{P}_{hT}/P_{hT}$. The integrals of the eight leading-twist and eight subleading-twist structure functions are all given in \cite{Bastami:2018xqd}
by eqs.~(2.17a)-(2.17h) and eqs.~(2.18a)-(2.18h), respectively, along with the weight functions defined in eq.~(2.19).

We first discuss the Gaussian Ansatz for the TMDs and FFs\footnote{We use shorthand notations: TMDs for TMD PDFs, and FFs for TMD FFs.}, which
is well supported by phenomenological analyses, such as 
\cite{Anselmino:2005nn,Schweitzer:2010tt,Anselmino:2008jk,Anselmino:2008sga,Anselmino:2013lza}. All convolution integrals of the same type as in Eq.~(\ref{eqn_eq:SIDIS_kin11}) 
can be solved analytically with this ansatz. In this case, $f^{a}(x_{{\!}_{Bj}}, k_{\perp}^{2})$ and $D^{a}(z_{h}, p_{\perp}^{2})$ are expressed as
\beq
f^{a}(x_{{\!}_{Bj}}, k_{\perp}^{2}) = f^{a}_{c}(x_{{\!}_{Bj}})\,\frac{ e^{-k_{\perp}^{2}/\langle k_{\perp}^{2} \rangle} }{ \pi \langle k_{\perp}^{2} \rangle } ,
~~~~~D^{a}(z_{h}, p_{\perp}^{2}) =  D^{a}_{c}(z_{h})\,\frac{ e^{-p_{\perp}^{2}/\langle p_{\perp}^{2} \rangle} }{ \pi \langle p_{\perp}^{2} \rangle } ,
\label{eqn_eq:SIDIS_kin12}
\eeq
where $f^{a}_{c}(x_{{\!}_{Bj}})$ is the collinear PDF, and $D^{a}_{c}(z_{h})$ is the collinear FF. For both collinear PDF and FF, we utilize the grid 
files taken from WW-SIDIS \cite{WW-SIDIS:2018} and MSTWPDF \cite{Martin:2009iq,mstwpdf} libraries. Alternatively, one could use the LHAPDF CJ15lo set 
from \cite{LHAPDF} for the PDF, and DSSFFlo set used in \cite{Liu:2019} for the FF. The Gaussian widths $\langle k_{\perp}^{2} \rangle$ and 
$\langle p_{\perp}^{2} \rangle$ may have different forms of kinematic dependence, however, we will assume them to be flavor-, $x_{{\!}_{Bj}}$-, 
and $z_{h}$-independent for simplicity.

All leading- and subleading-twist SIDIS structure functions in Eq.~(\ref{eqn_eq:SIDIS_kin3}) are described in terms of a basis of six TMDs and two FFs under 
the assumption of validity of WW(-type) approximations to be discussed shortly. Two of these basis functions are given by Eq.~(\ref{eqn_eq:SIDIS_kin12}), 
and the rest are $f_{1T}^{\perp a}$, $g_{1}^{a}$, $h_{1}^{a}$, $h_{1}^{\perp a}$, $h_{1T}^{\perp a}$, $H_{1}^{\perp a}$ (see their explicit forms in 
eqs.~(4.5c)-(4.5h) of \cite{Bastami:2018xqd}). Now, as examples, let us write down the general analytical forms of three structure functions in the Gaussian 
approximation, which are used to produce some numerical results from \texttt{SIDIS-RC EvGen}, to be shown in Sec.~\ref{sec:programs}:
\bea
& & \!\!\!\!\!\!\!\!\!\!\!\!\mbox{(i)~Leading-twist}~~F_{UU}(x,z,P_{hT}) =
~~~~~~~~~~~~~~~~~~~~~~~~~~~~~~\left\{ F_{UU} \equiv F_{UU,T} \right\}
\nonumber\\
& & ~~~~~~~~
= x_{{\!}_{Bj}} \mathlarger{\sum}_{a} e_{a}^{2}\,f^{a}_{c}(x_{{\!}_{Bj}})\,D^{a}_{c}(z_{h})\,
\frac{ e^{-P_{hT}^{2}/\langle P_{hT}^{2} \rangle} }{ \pi \langle P_{hT}^{2} \rangle } ,
\label{eqn_eq:SIDIS_kin13}
\eea
where $\langle P_{hT}^{2} \rangle = \langle p_{\perp}^{2} \rangle_{D} + z_{h}^{2}\langle k_{\perp}^{2} \rangle_{f}$. See table~1 of appendix~A in \cite{Bastami:2018xqd} 
for the values of $\langle k_{\perp}^{2} \rangle_{f}$ and $\langle p_{\perp}^{2} \rangle_{D}$.
\bea
& & \!\!\!\!\!\!\!\!\!\!\!\!\!\!\!\!\!\!\!\!\!\!\!\!\!\mbox{(ii)~Leading-twist}~~F_{UT}^{\sin{\!(\phi_{h} + \phi_{{\!}_{S}})}}(x,z,P_{hT}) = 
\nonumber\\
& & ~~~~~
= x_{{\!}_{Bj}} \mathlarger{\sum}_{a} e_{a}^{2}\,h_{1}^{a}(x_{{\!}_{Bj}})\,H_{1}^{\perp(1)a}(z_{h})\,b_{A}^{(1)}
\left[ \frac{z_{h}P_{hT}}{\langle P_{hT}^{2} \rangle} \right]\,\frac{ e^{-P_{hT}^{2}/\langle P_{hT}^{2} \rangle} }{ \pi \langle P_{hT}^{2} \rangle } ,
\label{eqn_eq:SIDIS_kin14}
\eea
where $\langle P_{hT}^{2} \rangle = \langle p_{\perp}^{2} \rangle_{H_{1}^{\perp}} + z_{h}^{2}\langle k_{\perp}^{2} \rangle_{h_{1}}$,
$\langle p_{\perp}^{2} \rangle_{H_{1}^{\perp}} = \left[ \langle p_{\perp}^{2} \rangle_{D}M_{C}^{2} \right] / \left[ \langle p_{\perp}^{2} \rangle_{D} + M_{C}^{2} \right]$,
and $b_{A}^{(1)} = 2M_{h}$. The values of $\langle k_{\perp}^{2} \rangle_{h_{1}}$ and $M_{C}^{2}$ are given in Table~3, and the first moment of the Collins 
fragmentation function, $H_{1}^{\perp(1)a}(z_{h})$, is given in eq.~(A.12), of appendix~A in \cite{Bastami:2018xqd}.
\bea
& & \!\!\!\!\!\!\!\!\!\!\!\mbox{(iii)~Leading-twist}~~F_{UT}^{\sin{\!(\phi_{h} - \phi_{{\!}_{S}})}}(x,z,P_{hT}) = 
~~~~~\left\{ F_{UT}^{\sin{\!(\phi_{h} - \phi_{{\!}_{S}})}} \equiv F_{UT,T}^{\sin{\!(\phi_{h} - \phi_{{\!}_{S}})}} \right\}
\nonumber\\
& & ~~~~~
= -x_{{\!}_{Bj}} \mathlarger{\sum}_{a} e_{a}^{2}\,f_{1T}^{\perp(1)a}(x_{{\!}_{Bj}})\,D^{a}_{c}(z_{h})\,b_{B}^{(1)}
\left[ \frac{z_{h}P_{hT}}{\langle P_{hT}^{2} \rangle} \right]\,\frac{ e^{-P_{hT}^{2}/\langle P_{hT}^{2} \rangle} }{ \pi \langle P_{hT}^{2} \rangle } ,
\label{eqn_eq:SIDIS_kin15}
\eea
where $\langle P_{hT}^{2} \rangle = \langle p_{\perp}^{2} \rangle_{D} + z_{h}^{2}\langle k_{\perp}^{2} \rangle_{f_{1T}^{\perp}}$,
$\langle k_{\perp}^{2} \rangle_{f_{1T}^{\perp}} = \left[ \langle k_{\perp}^{2} \rangle_{f}M_{1}^{2} \right] / \left[ \langle k_{\perp}^{2} \rangle_{f} + M_{1}^{2} \right]$,
and $b_{B}^{(1)} = 2M_{N}$. The value of $M_{1}^{2}$ is given in Table~2, and the first moment of the Sivers function, $f_{1T}^{\perp(1)a}(x_{{\!}_{Bj}})$, 
is given in eq.~(A.4), of appendix~A in \cite{Bastami:2018xqd}.

Twist-2 TMDs, like $g_{1}^{b}(x)$ ($b = q,g,\bar{q}$) and $h_{1}^{c}(x)$  ($c = q,\bar{q}$), have partonic interpretation. Twist-3 TMDs, like $g_{T}^{b}(x)$ 
and $h_{L}^{c}(x)$, can respectively be expressed by $g_{1}^{b}(x)$ and $h_{1}^{c}(x)$ plus additional $\bar{q}gq$ correlations. These correlations may have
new insights on hadronic structure but their contributions to $g_{T}^{b}(x)$ and $h_{L}^{c}(x)$ are small. In the case of TMDs, FFs and collinear PDFs, one can assume 
that the $\bar{q}gq$-correlation contributions are negligible compared with those of $\bar{q}q$ correlations, which makes an approximation named Wandzura-Wilczek (WW)
\cite{Wandzura:1977qf}: $| \langle \bar{q}gq \rangle / \langle \bar{q}q \rangle | \ll 1$. Each correlation comprises matrix elements of different operators, and the nature of 
neglected matrix elements is different. Based upon this difference, one generally considers WW-type approximations in the context of TMDs/FFs. At this point we refer to
section~3.2 of \cite{Bastami:2018xqd} for more detailed discussion of the WW-type approximations for TMDs/FFs. Besides, in sections~4.1~and~4.2 of the same reference, 
it is shown how some of the leading-twist and all the subleading-twist structure functions are treated in these approximations.

Except for $F_{UU}$, $F_{UT}^{\sin{\!(\phi_{h} + \phi_{{\!}_{S}})}}$, and $F_{UT}^{\sin{\!(\phi_{h} - \phi_{{\!}_{S}})}}$ described in the paragraphs (i), (ii), and (iii), 
the other structure functions given in the Gaussian Ansatz with combination of the WW-type approximations wherever relevant, can be found in \cite{Bastami:2018xqd} 
as follows (see also Sec.~\ref{sec:SF}):

For leading-twist:
\begin{itemize}
\item[$\bullet$] $F_{UU}^{\cos{\!(2\phi_{h})}}$ - see section~5.5; eqs.~(A.12),~(A.18); appendix B.5;
\item[$\bullet$] $F_{UL}^{\sin{\!(2\phi_{h})}}$ - see section~6.2; eqs.~(A.12),~(B.9b),~(3.6b); appendix B.5;
\item[$\bullet$] $F_{LL}$ - see section~5.2; appendix A.2; eq.~(4.5c);
\item[$\bullet$] $F_{UT}^{\sin{\!(3\phi_{h} - \phi_{{\!}_{S}})}}$ - see section~5.6; eq.~(A.12); appendix A.6, appendix B.5;
\item[$\bullet$] $F_{LT}^{\cos{\!(\phi_{h} - \phi_{{\!}_{S}})}}$ - see section~6.1; eqs.~(B.9a),~(3.6a); appendix B.5.
\end{itemize}

For subleading-twist:
\begin{itemize}
\item[$\bullet$] $F_{UU}^{\cos{\!(\phi_{h})}}$ - see section~7.8; eq.~(B.9j); appendix B.5;
\item[$\bullet$] $F_{LU}^{\sin{\!(\phi_{h})}}$ - see section~7.1 (also section~III of \cite{Mao:2012dk});
\item[$\bullet$] $F_{UL}^{\sin{\!(\phi_{h})}}$ - see section~7.5; eqs.~(A.12),~(B.9f),~(3.6b); appendix B.5;
\item[$\bullet$] $F_{LL}^{\cos{\!(\phi_{h})}}$ - see section~7.4; eq.~(B.9e); appendix B.5;
\item[$\bullet$] $F_{UT}^{\sin{\!(2\phi_{h} - \phi_{{\!}_{S}})}}$ - see section~7.7; eqs.~(B.9g),~(B.9h),~(B.9i); appendix B.5;
\item[$\bullet$] $F_{UT}^{\sin{\!(\phi_{{\!}_{S}})}}$ - see section~7.6; eqs.~(A.12),~(B.9g),~(B.9h),~(3.3g),~(3.3h);
\item[$\bullet$] $F_{LT}^{\cos{\!(\phi_{{\!}_{S}})}}$ - see section~7.2; eqs.~(B.9c),~(3.2a);
\item[$\bullet$] $F_{LT}^{\cos{\!(2\phi_{h} - \phi_{{\!}_{S}})}}$ - see section~7.3; eq.~(B.9d),~(3.3d),~(3.6a); appendix B.5.
\end{itemize}
All the other details necessary for making our event generator is provided in Sec.~\ref{sec:lib}.

\subsection{\label{sec:QED_RC} Lowest-order QED radiative effects in SIDIS}
In this section, we discuss the main results of \cite{Akushevich:2019mbz} for calculation of the NLO RCs to the SIDIS base cross section.

\subsubsection{\label{sec:QED_RC1} Leading order cross section}

\begin{SCfigure}[][hbt!]
\hspace{3.0cm}
\includegraphics[width=4.5cm]{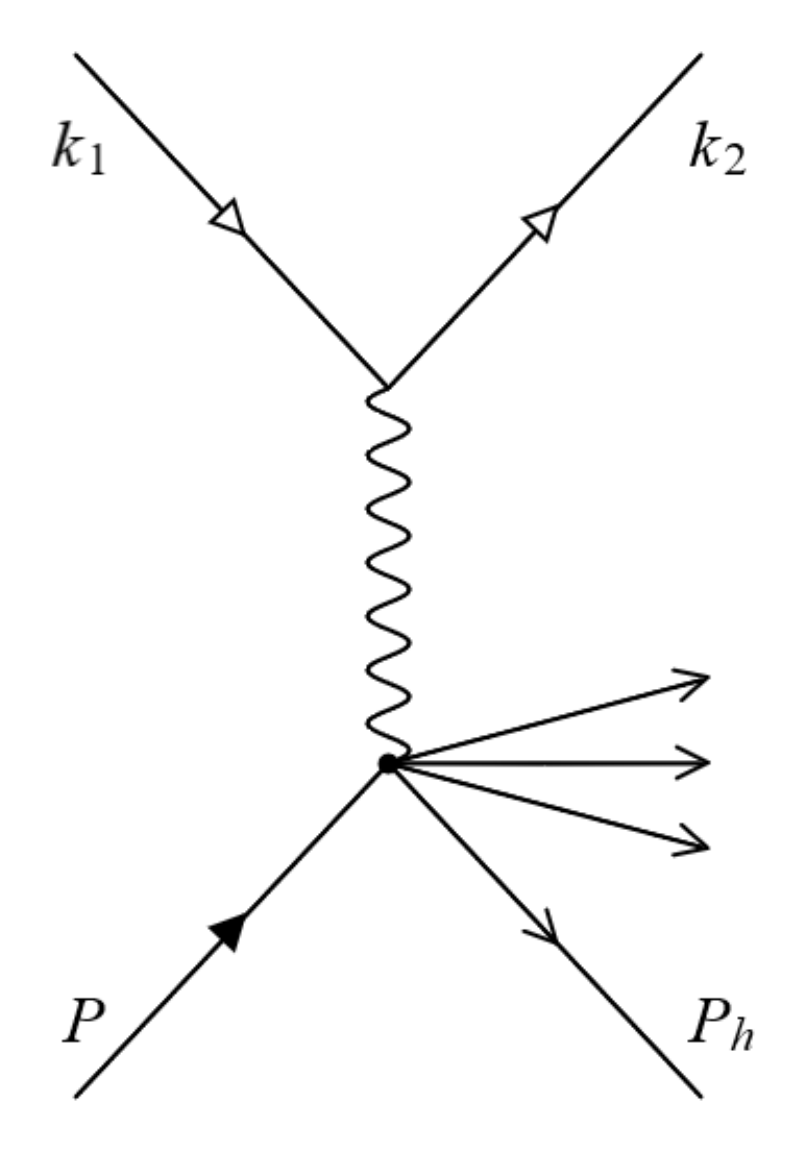}
\caption{A Feynman diagram describing the base (lowest-order QED) SIDIS process.}
\label{fig:fig_Born}
\end{SCfigure}

Let us represent Eq.~(\ref{eqn_eq:SIDIS_kin1}) in a bit modified way:
\beq
\ell(k_{1},\xi) + N(P,\bm{\eta}) \rightarrow \ell^{\prime}(k_{2}) + h(P_{h}) + X(P_{X}) ,
\label{eqn_eq:SIDIS_RC1}
\eeq
where $\xi$ and $\bm{\eta}$ are the incident lepton and target nucleon polarization vectors, respectively, described by $x_{{\!}_{Bj}}$, $y$, 
$z_{h}$, $\phi_{h}$, $\phi_{{\!}_{S}}$, and the variable $t = (q - P_{h})^{2}$. Fig.~\ref{fig:fig_Born} shows the lowest-order QED (Born) 
contribution to SIDIS, and the differential of the cross section of this process is given by a convolution of the hadronic $(W_{\mu\nu})$ 
and leptonic tensors ($L_{B}^{\mu\nu}$):
\beq
d\sigma_{\rm SIDIS}^{B} = \frac{(4\pi\alpha)^{2}}{2\sqrt{\lambda_{S}}Q^{4}}\,W_{\mu\nu}\,L_{B}^{\mu\nu}\,d\Gamma_{B} ,
\label{eqn_eq:SIDIS_RC2}
\eeq
where 
\beq
\lambda_{S} = (2P \!\cdot\! k_{1})^{2} - 4M_{N}^{2}m_{l}^{2} , 
\label{eqn_eq:SIDIS_RC3}
\eeq
with $m_{l}$ being the lepton mass. 
The phase-space is parameterized by
\bea
& & \!\!\!\!\!\!\!\!\!\!\!\!
d\Gamma_{B} = (2\pi)^{4}\,\frac{d^{3}k_{2}}{(2\pi)^{3} 2k_{20}} \frac{d^{3}P_{h}}{(2\pi)^{3} 2P_{h0}} =
\nonumber\\
& &
= \frac{1}{4(2\pi)^{2}} \frac{S S_{x}\,dx_{{\!}_{Bj}}\,dy\,d\phi_{{\!}_{S}}}{2\sqrt{\lambda_{S}}} \frac{S_{x}\,dz_{h}\,dP_{hT}^{2}\,d\phi_{h}}
{4M_{N}P_{hL}} ,
\label{eqn_eq:SIDIS_RC4}
\eea
where $k_{20}$ is the scattered lepton energy, $P_{h0}$ ($P_{hL}$) is the energy (longitudinal momentum) of the charged hadron. Also, 
\bea
& & \!\!\!\!\!\!\!\!\!\!\!\!\!\!\!
S = 2P \!\cdot\! k_{1} ,~~~S_{x} = 2P \!\cdot\! q ,~~~P_{h0} = \frac{z_{h}S_{x}}{2M_{N}} ,~~~P_{hT} = \sqrt{P_{h0}^{2} - P_{hL}^{2} - M_{h}^{2}} ,
\nonumber\\
& & \!\!\!\!\!\!\!\!\!\!
P_{hL} = \frac{z_{h}S_{x}^{2} + 2M_{N}^{2} \Lb t + Q^{2} - M_{h}^{2} \Rb }{2M_{N}\sqrt{\lambda_{Y}}} ,~~~~~\mbox{with}~~\lambda_{Y} =
S_{x}^{2} + 4M_{N}^{2}Q^{2} .
\label{eqn_eq:SIDIS_RC5}
\eea
The incident lepton is adopted to be longitudinally polarized. Its polarization vector reads as
\beq
\xi = \frac{\lambda_{e}S}{m_{l}\sqrt{\lambda_{S}}}\,k_{1} - \frac{2\lambda_{e}m_{l}}{\sqrt{\lambda_{S}}}\,P = \xi_{0} + \xi_{1} ,
\label{eqn_eq:SIDIS_RC6}
\eeq
and the leptonic tensor is represented as
\bea
& & \!\!\!\!\!\!\!\!\!\!\!\!
L_{B}^{\mu\nu} = \frac{1}{2}\,\Tr\!\left[ \Lb \hat{k}_{2} + m_{l} \Rb\!\gamma_{\mu}\!\Lb \hat{k}_{1} + m_{l} \Rb\!\Lb 1 + \gamma_{5}\hat{\xi} \Rb\!\gamma_{\nu} \right] = 
\nonumber\\
& &
= 2\left[ k_{1}^{\mu}k_{2}^{\nu} + k_{2}^{\mu}k_{1}^{\nu} - \frac{Q^{2}}{2}g^{\mu\nu} + \frac{i \lambda_{e}}{\sqrt{\lambda_{S}}}\epsilon^{\mu\nu\rho\sigma} 
\Lb S\,k_{2\rho}k_{1\sigma} + 2m_{l}^{2}q_{\rho}P_{\sigma} \Rb \right] .
\label{eqn_eq:SIDIS_RC7}
\eea
As regards the hadronic tensor, it is partitioned to spin-independent, $H_{ab}^{(0)}$, and spin-dependent, $H_{abi}^{(S)}$, scalar structure functions:
\beq
W_{\mu\nu} = \mathlarger{\sum}_{a,b = 0}^{3} e_{\mu}^{\gamma(a)} e_{\nu}^{\gamma(b)} \Lb H_{ab}^{(0)} +  \mathlarger{\sum}_{\rho,i = 0}^{3} \eta^{\rho} 
e_{\rho}^{h(i)} H_{abi}^{(S)} \Rb ,
\label{eqn_eq:SIDIS_RC8}
\eeq
where $e_{\mu}^{\gamma(a)}$ is the complete set of the basis for polarization four-vectors of the virtual photon, and $e_{\rho}^{h(i)}$ is that of the target 
nucleon\footnote{For the explicit expressions of both basis vectors, see eq.~(A1) and eq.~(A2) of \cite{Akushevich:2019mbz}.}. There are five spin-independent 
$H_{00}^{(0)}$, $H_{11}^{(0)}$, $H_{22}^{(0)}$, Re$H_{01}^{(0)}$, Im$H_{01}^{(0)}$, as well as thirteen spin-dependent $H_{002}^{(S)}$, Re$H_{012}^{(S)}$, 
Im$H_{012}^{(S)}$, Re$H_{021}^{(S)}$, Im$H_{021}^{(S)}$, Re$H_{023}^{(S)}$, Im$H_{023}^{(S)}$, $H_{112}^{(S)}$, Re$H_{121}^{(S)}$, Im$H_{121}^{(S)}$, 
Re$H_{123}^{(S)}$, Im$H_{123}^{(S)}$, $H_{222}^{(S)}$, scalar functions. 

The cross section of the Born contribution to SIDIS is given by
\beq
\frac{d\sigma_{\rm SIDIS}^{B}}{dx_{{\!}_{Bj}}\,dy\,dz_{h}\,dP_{hT}^{2}\,d\phi_{h}\,d\phi_{{\!}_{S}}} = \frac{\alpha^{2} S S_{x}^{2}}{8M_{N} Q^{4} P_{hL}\lambda_{S}}
\,\mathlarger{\sum}_{i = 1}^{9} \theta_{i}^{B} \mathcal{H}_{i} ,
\label{eqn_eq:SIDIS_RC9}
\eeq
where $\theta_{i}^{B}$ is expressed via the combination of various kinematic variables; $\mathcal{H}_{i}$ is the generalized structure function, expressed through 
$H_{ab}^{(0)}$ and $H_{abi}^{(S)}$ that is based on the nucleon polarized three-vector $\boldsymbol{\eta} = (\eta_{1},\eta_{2},\eta_{3})$ decomposition 
over the basis in eq.~(A2). The components of $\boldsymbol{\eta}$ are the following:
\beq
\eta_{1} = \cos{\!(\phi_{{\!}_{S}} - \phi_{h})}S_{T} ,~~~~~\eta_{2} = \sin{\!(\phi_{{\!}_{S}} - \phi_{h})}S_{T} ,~~~~~\eta_{3} = S_{L} .
\label{eqn_eq:SIDIS_RC10}
\eeq
$\mathcal{H}_{i}$ and $\theta_{i}^{B}$ are explicitly shown in eq.~(14) and eq.~(16) of \cite{Akushevich:2019mbz}. 

At this point, it is relevant to show here the relations of the scalar functions with the structure functions discussed in Sec.~\ref{sec:theory}:
\begin{displaymath}
H_{00}^{(0)} = C_{1} F_{UU,L} ,
\end{displaymath}
\begin{displaymath}
H_{01}^{(0)} = -C_{1} \Lb F_{UU}^{\cos{\!(\phi_{h})}} + i F_{LU}^{\sin{\!(\phi_{h})}} \Rb ,
\end{displaymath}
\begin{displaymath}
H_{11}^{(0)} = C_{1} \Lb F_{UU}^{\cos{\!(2\phi_{h})}} + F_{UU,T} \Rb ,
\end{displaymath}
\begin{displaymath}
H_{22}^{(0)} = C_{1} \Lb  F_{UU,T} - F_{UU}^{\cos{\!(2\phi_{h})}} \Rb ,
\end{displaymath}
\begin{displaymath}
H_{002}^{(S)} = C_{1} F_{UT,L}^{\sin{\!(\phi_{h} - \phi_{{\!}_{S}})}} ,
\end{displaymath}
\begin{displaymath}
H_{012}^{(S)} = C_{1} \Lb F_{UT}^{\sin{\!(\phi_{{\!}_{S}})}} - F_{UT}^{\sin{\!(2\phi_{h} - \phi_{{\!}_{S}})}} - i \Lb F_{LT}^{\cos{\!(\phi_{{\!}_{S}})}} -
F_{LT}^{\cos{\!(2\phi_{h} - \phi_{{\!}_{S}})}} \Rb \Rb ,
\end{displaymath}
\begin{displaymath}
H_{021}^{(S)} = C_{1} \Lb F_{UT}^{\sin{\!(2\phi_{h} - \phi_{{\!}_{S}})}} + F_{UT}^{\sin{\!(\phi_{{\!}_{S}})}} - i \Lb F_{LT}^{\cos{\!(2\phi_{h} - \phi_{{\!}_{S}})}} +
F_{LT}^{\cos{\!(\phi_{{\!}_{S}})}} \Rb \Rb ,
\end{displaymath}
\begin{displaymath}
H_{023}^{(S)} = C_{1} \Lb F_{UL}^{\sin{\!(\phi_{h})}} - i F_{LL}^{\cos{\!(\phi_{h})}} \Rb ,
\end{displaymath}
\begin{displaymath}
H_{121}^{(S)} = C_{1} \Lb -F_{UT}^{\sin{\!(3\phi_{h} - \phi_{{\!}_{S}})}} - F_{UT}^{\sin{\!(\phi_{h} + \phi_{{\!}_{S}})}} + 
i F_{LT}^{\cos{\!(\phi_{h} - \phi_{{\!}_{S}})}} \Rb ,
\end{displaymath}
\begin{displaymath}
H_{123}^{(S)} = C_{1} \Lb -F_{UL}^{\sin{\!(2\phi_{h})}} + i F_{LL} \Rb ,
\end{displaymath}
\begin{displaymath}
H_{112}^{(S)} = C_{1} \Lb F_{UT}^{\sin{\!(3\phi_{h} - \phi_{{\!}_{S}})}} + F_{UT,T}^{\sin{\!(\phi_{h} - \phi_{{\!}_{S}})}} - 
F_{UT}^{\sin{\!(\phi_{h} + \phi_{{\!}_{S}})}} \Rb ,
\end{displaymath}
\beq
H_{222}^{(S)} = C_{1} \Lb F_{UT}^{\sin{\!(\phi_{h} + \phi_{{\!}_{S}})}} + F_{UT,T}^{\sin{\!(\phi_{h} - \phi_{{\!}_{S}})}} - 
F_{UT}^{\sin{\!(3\phi_{h} - \phi_{{\!}_{S}})}} \Rb .
\label{eqn_eq:SIDIS_RC11}
\eeq
We  emphasize that if one uses the formulas in eq.~(14) (for $\mathcal{H}_{i}$), and eq.~(16) (for $\theta_{i}^{B}$) of \cite{Akushevich:2019mbz}, 
as well as the formulas in Eq.~(\ref{eqn_eq:SIDIS_RC11}) and inserts them in the R.H.S. of Eq.~(\ref{eqn_eq:SIDIS_RC9}), then the result will 
be equivalent to the R.H.S. of Eq.~(\ref{eqn_eq:SIDIS_kin3}). Therefore our two ways of expressing the SIDIS cross-section are equivalent.

\subsubsection{\label{sec:QED_RC2} Lowest-order radiative corrections to the SIDIS cross section}

\begin{figure}[hbt!]
\centering
\includegraphics[width=4.5cm]{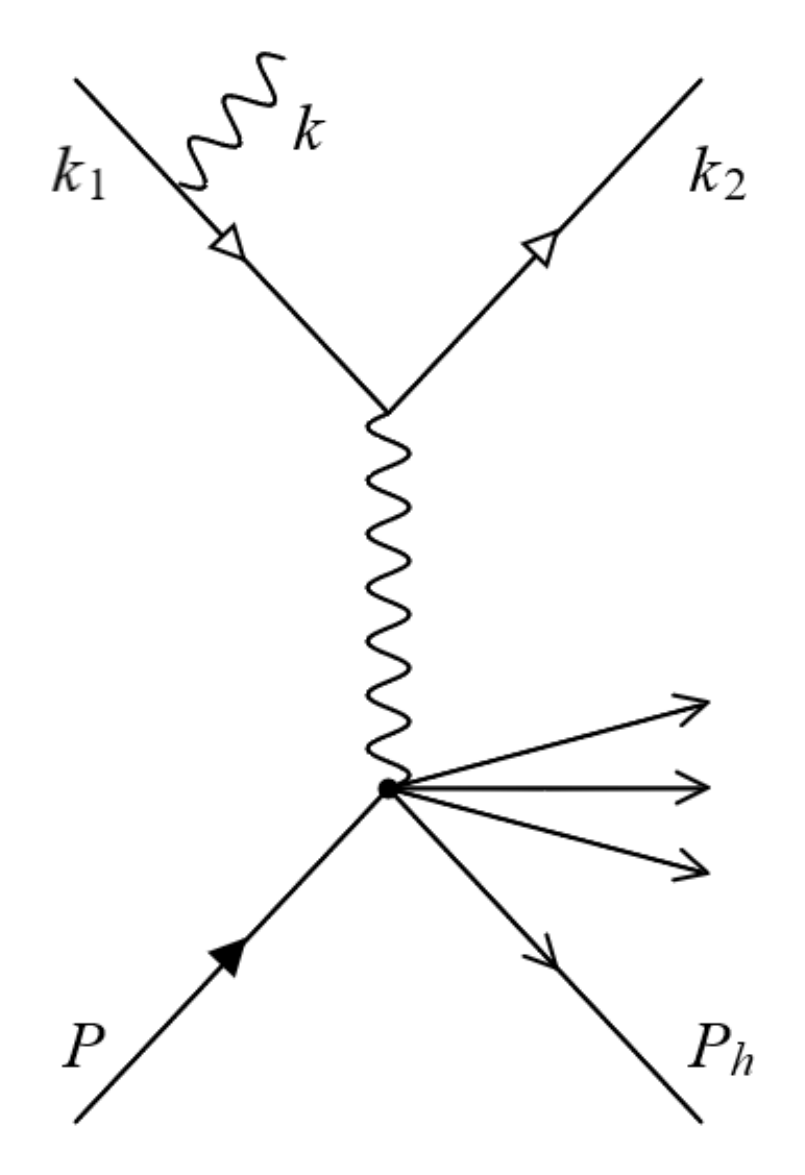}\label{fig:fig_SIDIS_RC1}
\includegraphics[width=4.5cm]{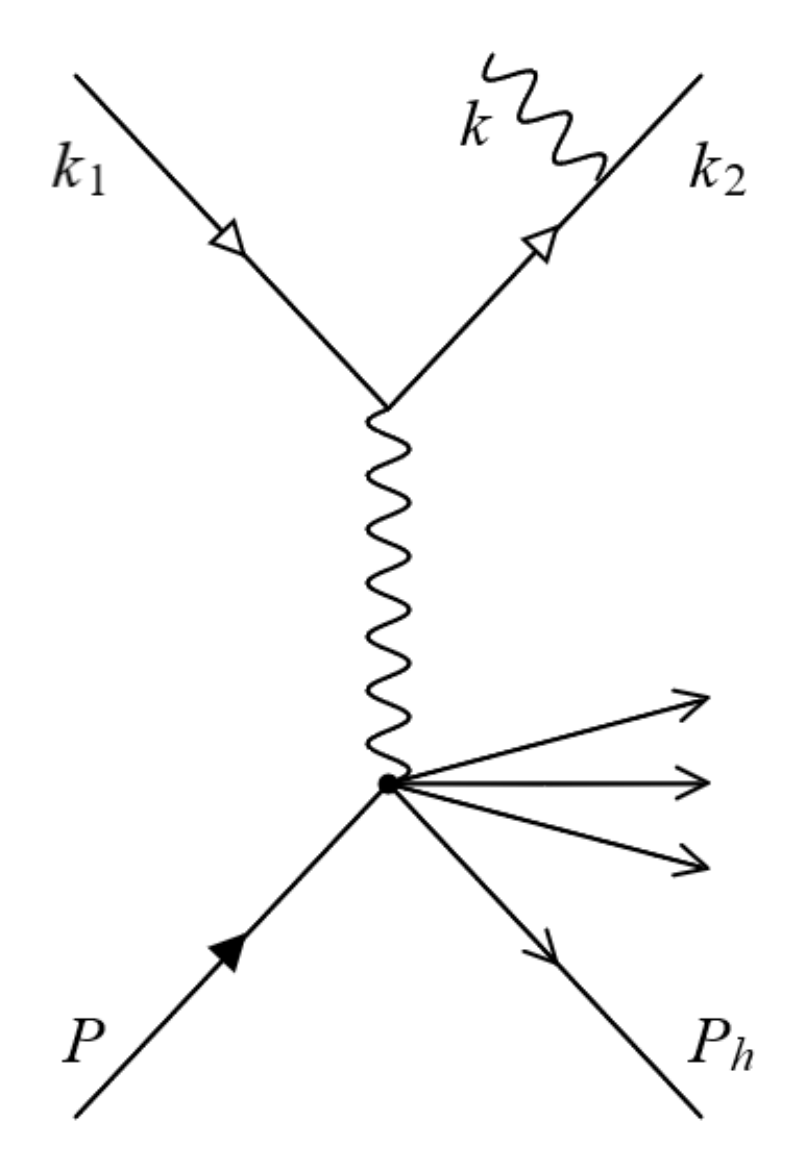}\label{fig:fig_SIDIS_RC2}
\\[-0.1cm]
{\bf (a) \hspace{4.5cm} (b) \hspace{1.5cm}}
\\[0.1cm]
\includegraphics[width=4.5cm]{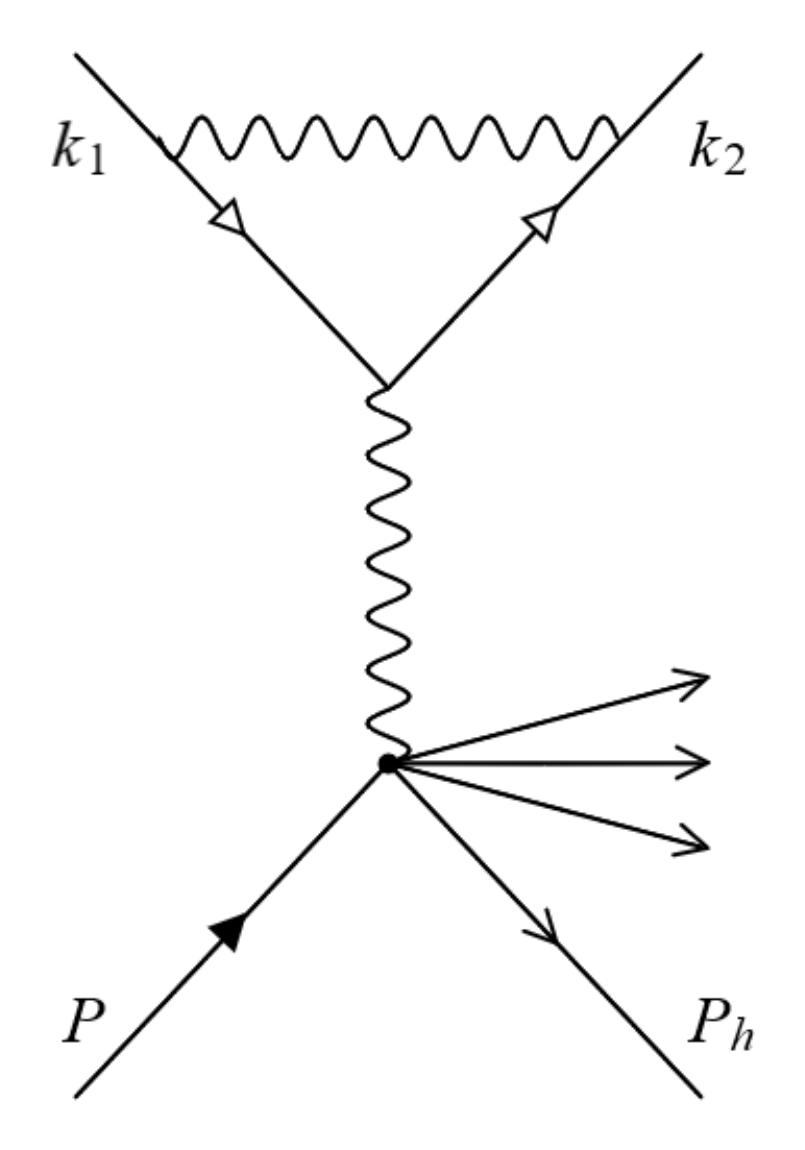}\label{fig:fig_SIDIS_RC3}
\includegraphics[width=4.5cm]{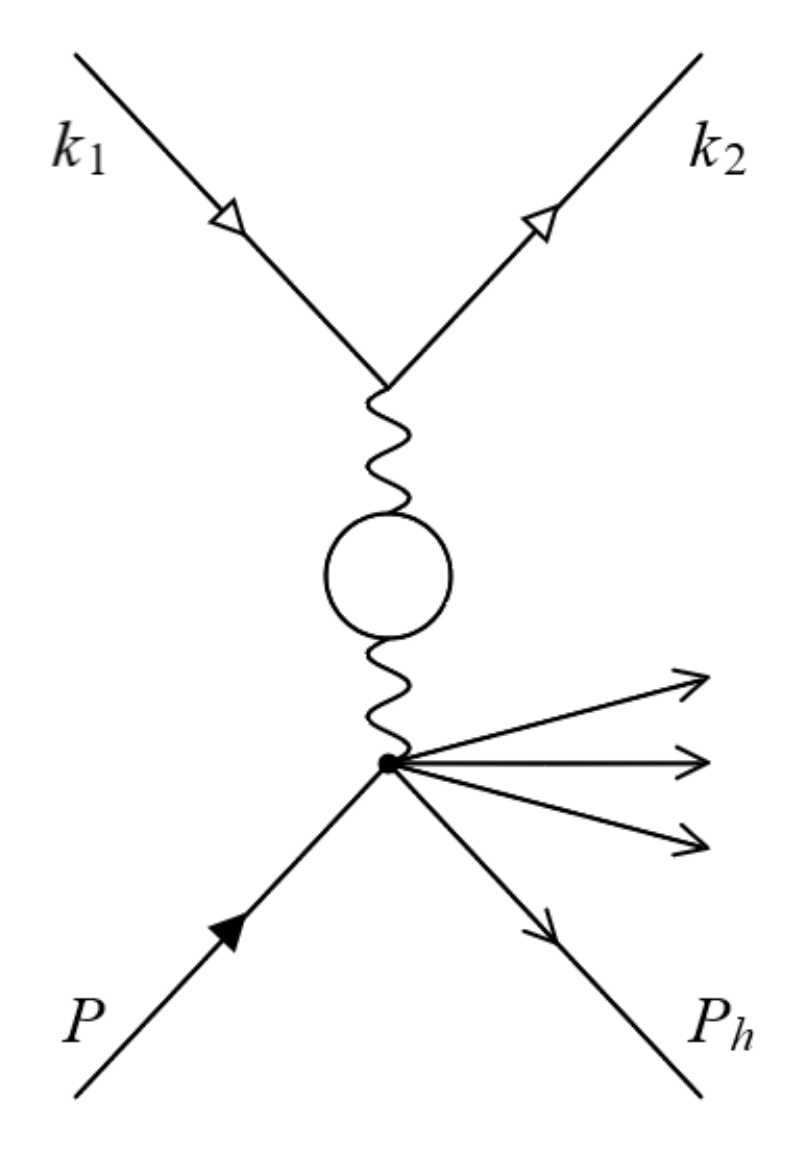}\label{fig:fig_SIDIS_RC4}
\\[-0.1cm]
{\bf (c) \hspace{4.5cm} (d) \hspace{1.5cm}}
\\[-0.1cm]
\caption{Feynman diagrams from (a) to (d) describing the NLO QED RC contributions to SIDIS scattering.} 
\label{fig:fig_SIDIS_RC}
\end{figure}

We can now discuss the real photon emission in the SIDIS scattering process, given by
\beq
\ell(k_{1},\xi) + N(P,\bm{\eta}) \rightarrow \ell^{\prime}(k_{2}) + h(P_{h}) + X(\tilde{P}_{X}) + \gamma(k) ,
\label{eqn_eq:SIDIS_RC12}
\eeq
where $k$ is the four-momentum of the radiated real photon $\gamma$. The four diagrams depicted in Figs.~\ref{fig:fig_SIDIS_RC}(a)-\ref{fig:fig_SIDIS_RC}(d) are the 
lowest-order QED RC contributions to the cross section of the base SIDIS process shown in Fig.~\ref{fig:fig_Born}. The real photon is shown in the figures 
(a) and (b). The considered process is described by all kinematic variables of Eq.~(\ref{eqn_eq:SIDIS_RC1}), plus three additional photonic variables that are
\beq
R = 2k \!\cdot\! P ,~~~~~\tau = \frac{k \!\cdot\! q}{k \!\cdot\! P} ,~~~~~\phi_{k} ,
\label{eqn_eq:SIDIS_RC13}
\eeq
where $\phi_{k}$ is the angle between the $(\mathbf{k_{1}}, \mathbf{k_{2}})$ and $(\mathbf{k}, \mathbf{q})$ planes.
This angle is given by
\beq
\sin{\!(\phi_{k})} = \frac{2\varepsilon^{\mu\nu\rho\sigma}k_{\mu}P_{\nu}k_{1\rho}q_{\sigma}\sqrt{\lambda_{Y}}}
{R \sqrt{ \lambda_{1} \Lb Q^{2} + \tau(S_{x} - \tau M_{N}^{2}) \Rb } } ,
\label{eqn_eq:SIDIS_RC14}
\eeq
where 
\beq
\lambda_{1} = Q^{2} \Lb SX - M_{N}^{2}Q^{2} \Rb - m_{l}^{2}\lambda_{Y} ,~~~~~\mbox{with}\,\,\,X = 2P \!\cdot\! k_{2} ,
\label{eqn_eq:SIDIS_RC15}
\eeq
and $S$, $S_{x}$, and $\lambda_{Y}$ are defined in Eq.~(\ref{eqn_eq:SIDIS_RC5}). The cross-section differential obtained from the real photon radiation 
off the leptonic leg is determined to be
\beq
d\sigma_{R} = \frac{(4\pi\alpha)^{3}}{2\sqrt{\lambda_{S}}\,\tilde{Q}^{4}}\,\tilde{W}_{\mu\nu}L_{R}^{\mu\nu}d\Gamma_{R} .
\label{eqn_eq:SIDIS_RC16}
\eeq
In this case, the phase-space parameterization is the following:
\beq
d\Gamma_{R} = (2\pi)^{4} \frac{d^{3}k}{(2\pi)^{3}2k_{0}} \frac{d^{3}k_{2}}{(2\pi)^{3}2k_{20}} \frac{d^{3}P_{h}}{(2\pi)^{3}2P_{h0}} ,~~~
\mbox{with}~~\frac{d^{3}k}{k_{0}} = \frac{R\,dR\,d\tau\,d\phi_{k}}{2\sqrt{\lambda_{Y}}} .
\label{eqn_eq:SIDIS_RC17}
\eeq
The ``tilde" symbol of $\tilde{W}_{\mu\nu}$ in Eq.~(\ref{eqn_eq:SIDIS_RC16}) means that the arguments of the hadronic tensor are defined via the shifted 
$q \rightarrow q -  k$. With this shift, $\tilde{Q}^{2}$ is expressed by two photonic variables:
\beq
\tilde{Q}^{2} = -(q - k)^{2} = Q^{2} + R\tau . 
\label{eqn_eq:SIDIS_RC17b}
\eeq
The leptonic tensor is separated into two parts: 
\beq
L_{R}^{\mu\nu} = L_{R0}^{\mu\nu} +  L_{R1}^{\mu\nu} .
\label{eqn_eq:SIDIS_RC18}
\eeq
The first term is given by
\beq
L_{R0}^{\mu\nu} = -\frac{1}{2}\,\Tr\!\left[ \Lb \hat{k}_{2} + m_{l} \Rb\!\Gamma_{R}^{\mu\alpha}\!\Lb \hat{k}_{1} + m_{l} \Rb\!\Lb 1 + \gamma_{5}\hat{\xi}_{0} 
\Rb\!\bar{\Gamma}_{R\alpha}^{\nu} \right] ,
\label{eqn_eq:SIDIS_RC19}
\eeq
with
\bea
& &
\Gamma_{R}^{\mu\alpha} = \Lb \frac{k_{1}^{\alpha}}{k \!\cdot\! k_{1}} - \frac{k_{2}^{\alpha}}{k \!\cdot\! k_{2}} \Rb \gamma^{\mu}
- \frac{\gamma^{\mu}\hat{k}\gamma^{\alpha}}{2k \!\cdot\! k_{1}} - \frac{\gamma^{\alpha}\hat{k}\gamma^{\mu}}{2k \!\cdot\! k_{2}} ,
\nonumber\\
& &
\bar{\Gamma}_{R\alpha}^{\nu} = \gamma_{0}\,\Gamma_{R\alpha}^{\nu\dagger}\,\gamma_{0} =
\Lb \frac{k_{1\alpha}}{k \!\cdot\! k_{1}} - \frac{k_{2\alpha}}{k \!\cdot\! k_{2}} \Rb \gamma^{\nu}
- \frac{\gamma^{\nu}\hat{k}\gamma_{\alpha}}{2k \!\cdot\! k_{2}} - \frac{\gamma_{\alpha}\hat{k}\gamma^{\nu}}{2k \!\cdot\! k_{1}} .
\label{eqn_eq:SIDIS_RC20}
\eea
The second term is given by
\beq
L_{R1}^{\mu\nu} = -\frac{1}{2}\,\Tr\!\left[ \Lb \hat{k}_{2} + m_{l} \Rb\!\Gamma_{R}^{\mu\alpha}\!\Lb \hat{k}_{1} + m_{l} \Rb \gamma_{5}\hat{\xi}_{1} 
\bar{\Gamma}_{R\alpha}^{\nu} \right] .
\label{eqn_eq:SIDIS_RC21}
\eeq
$\xi_{0}$ and $\xi_{1}$ are the components of the polarization vector $\xi$ shown in Eq.~(\ref{eqn_eq:SIDIS_RC6}).

The convolutions of both separated leptonic tensors with the shifted hadronic tensor are given by
\bea
& &
\tilde{W}_{\mu\nu}L_{R0}^{\mu\nu} = -2\sum_{i = 1}^{9} \sum_{j = 1}^{k_{i}} \tilde{\mathcal{H}}_{i}\,\theta_{ij}^{0}\,R^{j - 3} ,
\nonumber\\
& &
\tilde{W}_{\mu\nu}L_{R1}^{\mu\nu} = -2\sum_{i = 5,7,9} \sum_{j = 1}^{k_{i}} \tilde{\mathcal{H}}_{i}\,\theta_{ij}^{1}\,R^{j - 3} ,
\label{eqn_eq:SIDIS_RC22}
\eea
where the quantities $\theta_{ij}^{0}$ and $\theta_{ij}^{1}$ are not functions of $R$ but are composite functions of the other kinematic variables, 
and are explicitly demonstrated in Appendix B of \cite{Akushevich:2019mbz}. 

As was mentioned in the Introduction, for the extraction and cancellation of the infrared divergence in these RC calculations, the Bardin-Shumeiko method 
\cite{Bardin:1976qa,Shumeiko:1978cn} is applied. 
And here we represent the final formulas that we have implemented into \texttt{SIDIS-RC EvGen}. For convenience, let us  designate the six-fold differential 
cross section in the form of
\beq
\sigma \equiv \frac{d\sigma}{dx_{{\!}_{Bj}}\,dy\,dz_{h}\,dP_{hT}^{2}\,d\phi_{h}\,d\phi_{{\!}_{S}}} ,
\label{eqn_eq:SIDIS_RC23}
\eeq
and use this notation henceforth.
In this case, our master formula is the inelastic tail of the SIDIS total cross section (with several RC components included), expressed as
\beq
\sigma_{\rm SIDIS}^{in} = \frac{\alpha}{\pi} \Lb \delta_{VR} + \delta_{\rm vac}^{l} + \delta_{\rm vac}^{h} \Rb \sigma_{\rm SIDIS}^{B} + 
\sigma_{R}^{F} + \sigma^{AMM} ,
\label{eqn_eq:SIDIS_RC24}
\eeq
where $\sigma_{\rm SIDIS}^{B}$ is shown in Eq.~(\ref{eqn_eq:SIDIS_RC9}) or Eq.~(\ref{eqn_eq:SIDIS_kin3}). 

$\bullet$ $\delta_{VR}$ is the sum of the infrared divergent terms and is finite:
\begin{eqnarray}
& & \!\!\!\!\!\!\!\!\!\!
\delta_{VR} = 2(Q_{m}^{2}L_{m} - 1)\,\log{\!\Lb \frac{P_{X}^{2} - M_{th}^{2}}{m_{l}\sqrt{P_{X}^{2}}} \Rb} + \frac{1}{2}S^{\prime}L_{S^{\prime}} +
\nonumber\\
& & ~~~
+ \frac{1}{2}X^{\prime}L_{X^{\prime}} + S_{\phi} -2 + \Lb \frac{3}{2}\,Q^{2} + 4m_{l}^{2} \Rb L_{m} -
\nonumber\\
& & ~~~
- \frac{Q_{m}^{2}}{\sqrt{\lambda_{m}}} \Lb \frac{1}{2}\lambda_{m}L_{m}^{2} + 2{\rm Li_{2}}\!\!\left[ \frac{2\sqrt{\lambda_{m}}}{Q^{2} + \sqrt{\lambda_{m}}} \right] 
-\frac{\pi^{2}}{2} \Rb ,
\label{eqn_eq:SIDIS_RC25}
\end{eqnarray}
where $M_{th} = M_{N} + M_{h}$ is the minimum invariant mass value of the undetected hadron states $P_{X}$. ${\rm Li_{2}}$ is Spence's dilogarithm; 
the variables $Q_{m}^{2}$, $P_{x}$, $S^{\prime}$, $X^{\prime}$, and $\lambda_{m}$, are given by eq.~(3); the functions $L_{m}$, $L_{S^{\prime}}$, and 
$L_{X^{\prime}}$, are given by eq.~(C10); the function $S_{\phi}$ is given by eqs.~(40)-(42)~and~(3); all in Ref.~\cite{Akushevich:2019mbz}.

$\bullet$ $\delta_{\rm vac}^{l}$ is the contribution of vacuum polarization by leptons:
\begin{eqnarray}
& & \!\!\!\!\!\!\!\!\!\!\!
\delta_{\rm vac}^{l} = \sum_{i = e,\mu,\tau} \delta_{\rm vac}^{l,i} =
\nonumber\\
& &
= \mathlarger{\sum}_{i = e,\mu,\tau} \Lb \frac{2}{3}(Q^{2} + 2m_{i}^{2})L_{m}^{i} - \frac{10}{9} + \frac{8m_{i}^{2}}{3Q^{2}}(1 - 2m_{i}^{2} L_{m}^{i}) \Rb , 
\label{eqn_eq:SIDIS_RC26}
\end{eqnarray}
where $L_{m}^{i}$ is defined in eq.~(D3) of Ref.~\cite{Akushevich:2019mbz}.

$\bullet$ $\delta_{\rm vac}^{h}$ is the contribution of vacuum polarization by hadrons, taken as a fit to hadronic cross section data from \cite{Burkhardt:1995tt}:
\beq
\delta_{\rm vac}^{h} = -\frac{2\pi}{\alpha} \left[ A + B \log\!{(1 + C|t_{h}|)} \right] ,
\label{eqn_eq:SIDIS_RC26b}
\eeq
where the constants $A$, $B$ and $C$, are shown in Table~\ref{table}.
\begin{table}[h!]
\centering
  \begin{tabularx}{0.85\textwidth}{|X|X|X|X|}
      \hline
      ~$|t_{h}|,~{\rm (GeV/c)^{2}}$ &~~~~~~~$A$ &~~~~~~~$B$ &~~~~~~~$C$ \\
      \hline
    ~~~~~0 - 1       & $-1.345 \times 10^{-9}$     &      $-2.302 \times 10^{-3}$     & ~~~~~4.091\\
      \hline
    ~~~~~1 - 64     & $-1.512 \times 10^{-3}$     &      $-2.822 \times 10^{-3}$     & ~~~~~1.218 \\
      \hline
    ~~~~~$> 64$   & $-1.344 \times 10^{-3}$     &      $-3.068 \times 10^{-3}$     & ~~~~~0.999 \\
     \hline
  \end{tabularx}
      \caption{The values of the three parameters in Eq.~(\ref{eqn_eq:SIDIS_RC26b}), shown in three consecutive ranges of $t_{h} = q^{2}$.}
\label{table}
\end{table}

$\bullet$ $\sigma_{R}^{F}$ is the infrared free contribution, integrated over the three photonic variables of  Eq.~(\ref{eqn_eq:SIDIS_RC13}):
\begin{eqnarray}
& & \!\!\!\!\!\!\!\!\!\!\!
\sigma_{R}^{F} = - \frac{\alpha^{3} S S_{x}^{2}}{64\pi^{2} M_{N} P_{hL} \lambda_{S} \sqrt{\lambda_{Y}}} \mathlarger{\int}_{\tau_{\rm min}}^{\tau_{\rm max}} d\tau
\mathlarger{\int}_{0}^{2\pi} d\phi_{k} \mathlarger{\int}_{0}^{R_{\rm max}} dR \times
\nonumber\\
& & ~~
\times \mathlarger{\sum}_{i = 1}^{9} \Lb \frac{\theta_{i1}}{R} \Lb \frac{\tilde{\mathcal{H}}_{i}}{\tilde{Q}^{4}} -\frac{\mathcal{H}_{i}}{Q^{4}} \Rb 
+ \sum_{j = 2}^{k_{i}} \tilde{\mathcal{H}}_{i} \theta_{ij} \frac{R^{j - 2}}{\tilde{Q}^{4}} \Rb ,
\label{eqn_eq:SIDIS_RC27}
\end{eqnarray}
where $S$, $S_{x}$, $P_{hL}$ and $\lambda_{Y}$ are shown in Eq.~(\ref{eqn_eq:SIDIS_RC5}), $\lambda_{S}$ in Eq.~(\ref{eqn_eq:SIDIS_RC3}),
$\tilde{Q}$ in Eq.~(\ref{eqn_eq:SIDIS_RC17b}). Also, the composite functions $\theta_{i1}$ and $\theta_{ij}$ are explicitly shown in Appendix B, and
$\mathcal{H}_{i}$ is shown in eq.~(14), all in Ref.~\cite{Akushevich:2019mbz}. The the integration limits are
\beq
R_{\rm max} = \frac{P_{x}^{2} - M_{th}^{2}}{1 + \tau -\mu} ,~~~~~\tau_{\rm max/\rm min} = \frac{S_{x} \pm \sqrt{\lambda_{Y}}}{2M_{N}^{2}} ,
\label{eqn_eq:SIDIS_RC28}
\eeq
with $\mu$ defined as $(k \!\cdot\! P_{h})/(k \!\cdot\! P)$  (see eq.~(B3)  of \cite{Akushevich:2019mbz}).

$\bullet$ $\sigma^{AMM}$ is the contribution of the anomalous magnetic moment:
\beq
\sigma^{AMM} = \frac{\alpha^{3}m_{l}^{2}S S_{x}^{2}}{16\pi M_{N}Q^{2}P_{hL}\lambda_{S}} L_{m} \sum_{i = 1}^{9} \theta_{i}^{AMM}\mathcal{H}_{i} ,
\label{eqn_eq:SIDIS_RC29}
\eeq
where the function $L_{m}$ is given by eq.~(C10), and $\theta_{i}^{AMM}$ is defined in eq.~(54)  of \cite{Akushevich:2019mbz}.

\section{\label{sec:MC_gen} MC event generator {\texttt\textbf{SIDIS-RC EvGen}} for studying SIDIS processes with lowest-order 
radiative effects}

This section describes the structure and functionality of the generator \texttt{SIDIS-RC EvGen}. The core of the generator is the \texttt{sidis} package, 
which is divided into a library component, called \texttt{libsidis}, and a generator (binary) component, called \texttt{sidisgen}. The library component 
\texttt{libsidis} is used to directly compute the SIDIS cross sections including NLO RCs, and is combined with a Monte-Carlo (MC) algorithm in the generator 
component \texttt{sidisgen} to produce random events. In particular, in the discussion of \texttt{libsidis}, we provide details on how cross-section calculations 
are fulfilled, along with describing two computational modes for doing it. In the discussion 
of \texttt{sidisgen}, we provide details on how the MC event generation is carried out by using a spatial partitioning method with hyper-cubical 
``foam of cells" based on the FOAM library \cite{FOAM} from ROOT \cite{ROOT}.

\subsection{\label{sec:soft_hard} Dividing the SIDIS inelastic cross section into soft and hard parts}

For the purpose of efficiently generating events and calculating cross sections, we make use of the inelastic cross-section formula in 
Eq.~(\ref{eqn_eq:SIDIS_RC24}) in a slightly modified form, which is obtained as follows.

According to the Bardin-Shumeiko approach \cite{Bardin:1976qa,Shumeiko:1978cn}, the real photon radiation in Eq.~(\ref{eqn_eq:SIDIS_RC16}) 
is split into infrared and infrared-free components:
\beq
d\sigma_{R} = d\sigma_{R}^{IR} + d\sigma_{R}^{F} ,
\label{eqn_eq:SIDIS_MC1}
\eeq
where the infrared contribution is in turn separated into soft $\delta_{S}$ and hard $\delta_{H}$ photon emission parts:
\beq
\sigma_{R}^{IR} = \frac{\alpha}{\pi} (\delta_{S} + \delta_{H}) \sigma^{B} ,
\label{eqn_eq:SIDIS_MC2}
\eeq
(see  eq.~(37) and appendix~C of \cite{Akushevich:2019mbz} for their explicit derivations). Ultimately, they are expressed as
\begin{eqnarray}
& & \delta_{S} = 2(Q_{m}^{2}L_{m} - 1) \Lb P_{IR} + \log\!\Lb \frac{2\bar{k}_{0}}{\nu} \Rb \Rb + \frac{1}{2}\,S^{\prime}L_{S^{\prime}} + 
\frac{1}{2}\,X^{\prime}L_{X^{\prime}} + S_{\phi}, 
\nonumber \\
& & \delta_{H} = 2(Q_{m}^{2}L_{m} - 1)  \log\!\Lb \frac{P_{X}^{2} - M_{th}^{2}}{2\bar{k}_{0} \sqrt{P_{x}^{2}}} \Rb .
\label{eqn_eq:SIDIS_MC3}
\end{eqnarray}
Let us again state that the variables $Q_{m}^{2}$, $P_{X}$, $S^{\prime}$, and $X^{\prime}$, are given by eq.~(3); the functions 
$L_{m}$, $L_{S^{\prime}}$, and $L_{X^{\prime}}$, are given by eq.~(C10); the function $S_{\phi}$ is given by eqs.~(40)-(42)~and~(3); 
all shown in \cite{Akushevich:2019mbz}. $\bar{k}_{0}$ is a small photon energy (soft cutoff) defined in the system 
$\mathbf{P} + \mathbf{q} - \mathbf{P_{h}} = 0$ that divides the soft and hard parts. The $\delta_{S}$ and $\delta_{H}$ contributions 
are $\bar{k}_{0}$-dependent. However, their sum does not depend on $\bar{k}_{0}$ but includes the infrared divergence
\beq
P_{IR} = \frac{1}{n - 4} + \frac{1}{2}\,\gamma_{E} + \log\!{\Lb \frac{1}{2\sqrt{\pi}} \Rb} ,
\label{eqn_eq:SIDIS_MC4}
\eeq
plus the arbitrary mass scale of dimensional regularization $\nu$. Using an additional regularization with the parameter $\bar{k}_{0}$,
one can calculate $\delta_{H}$ at $n = 4$. The $P_{IR}$ and $\nu$ contributions to $\delta_{S} + \delta_{H}$ are canceled by a contribution 
from the leptonic vertex correction given by
\begin{eqnarray}
& & \!\!\!\!\! \delta_{vert} = -2\Lb Q_{m}^{2}L_{m} - 1 \Rb \Lb P_{IR} + \log\!\Lb \frac{m_{l}}{\nu} \Rb \Rb + \Lb \frac{3}{2}\,Q^{2} + 4m_{l}^{2} \Rb L_{m} -
\nonumber \\
& & ~~~~~~ - 2 - \frac{Q_{m}^{2}}{\sqrt{\lambda_{m}}} \Lb \frac{1}{2}\lambda_{m}L_{m}^{2} + 2{\rm Li_{2}}\!\!\left[ \frac{2\sqrt{\lambda_{m}}}{Q^{2} 
+ \sqrt{\lambda_{m}}} \right] -\frac{\pi^{2}}{2} \Rb .
\label{eqn_eq:SIDIS_MC5}
\end{eqnarray}
Thereby, the sum of the three terms, $\delta_{S} + \delta_{H} + \delta_{vert}$, is free from the infrared-divergent term $P_{IR}$ 
and the arbitrary mass scale $\nu$. This sum is $\delta_{VR}$ in Eq.~(\ref{eqn_eq:SIDIS_RC25}).

We can rewrite the cross-section formula from Eq.~(\ref{eqn_eq:SIDIS_RC24}) as
\beq
\sigma_{\rm SIDIS}^{in} = \frac{\alpha}{\pi} \Lb \delta_{vert} + \delta_{\rm vac}^{l} + \delta_{\rm vac}^{h} \Rb \sigma_{\rm SIDIS}^{B} + \sigma^{AMM} + 
\int_{0}^{\infty} \bar{\sigma}_{R}\,d^{3}{\mathbf{k}} ,
\label{eqn_eq:SIDIS_MC6}
\eeq
by using the notation $\sigma_{R} \equiv\int_{0}^{\infty} \bar{\sigma}_{R}\,d^{3}{\mathbf{k}}$ for the radiative part. For this 
integral we have the following terms:
\bea
\int_{0}^{\infty} \bar{\sigma}_{R}\,d^{3}{\mathbf{k}} & = & \int_{0}^{\bar{k}_0} \bar{\sigma}^{IR}_{R}\,d^{3}{\mathbf{k}} + \int_{0}^{\bar{k}_0} \bar{\sigma}^{F}_{R}\,d^{3}{\mathbf{k}} + \int_{\bar{k}_0}^{\infty} \bar{\sigma}_{R}\,d^{3}{\mathbf{k}}  \nonumber \\
& = & \frac{\alpha}{\pi} \delta_{S} \sigma^{B} + \int_{0}^{\bar{k}_0} \bar{\sigma}^{F}_{R}\,d^{3}{\mathbf{k}} + \int_{\bar{k}_0}^{\infty} \bar{\sigma}_{R}\,d^{3}{\mathbf{k}} .
\label{eqn_eq:SIDIS_MC7}
\eea
We can also regroup the terms in Eq.~(\ref{eqn_eq:SIDIS_MC6}) to arrive at the following form:
\bea
\sigma_{\rm SIDIS}^{in} & = & \left[ \frac{\alpha}{\pi} \Lb \delta_{VS} + \delta_{\rm vac}^{l} + \delta_{\rm vac}^{h} \Rb \sigma_{\rm SIDIS}^{B} + 
\sigma^{AMM} + \int_{0}^{\bar{k}_0} \bar{\sigma}^{F}_{R}\,d^{3}{\mathbf{k}} \right]_{\rm non \mbox{-} rad.\,part} +
\nonumber \\
& & \!\!\!\!\!\!\!\qquad + \left[ \int_{\bar{k}_0}^{\infty} \bar{\sigma}_{R}\,d^{3}{\mathbf{k}} \right]_{\rm rad.\,part} \equiv
\nonumber \\
& \equiv & \sigma_{\rm SIDIS}^{nrad} + \sigma_{\rm SIDIS}^{rad} ,
\label{eqn_eq:SIDIS_MC8}
\eea
with $\delta_{VS} = \delta_{S} + \delta_{vert}$.

In \texttt{SIDIS-RC EvGen}, the events are randomly selected to be non-radiative or radiative, based upon the total cross 
sections shown in Eq.~(\ref{eqn_eq:SIDIS_MC8}). The radiative cross section is a nine-fold differential cross section 
(six SIDIS degrees of freedom + three photon degrees of freedom). The non-radiative cross section is a six-fold differential 
cross section, but has a computationally expensive integral over $\bar{\sigma}_{R}^{F}$ included. However, since 
$\bar{\sigma}_{R}^{F}$ is finite even to low photon energies, this integral can be neglected so long as the cutoff 
$\bar{k}_{0}$ is chosen small enough. Neglecting this integral introduces a weak cutoff dependence into the non-radiative 
cross section.

\begin{figure}[h!]
\centering
\includegraphics[width=7.0cm]{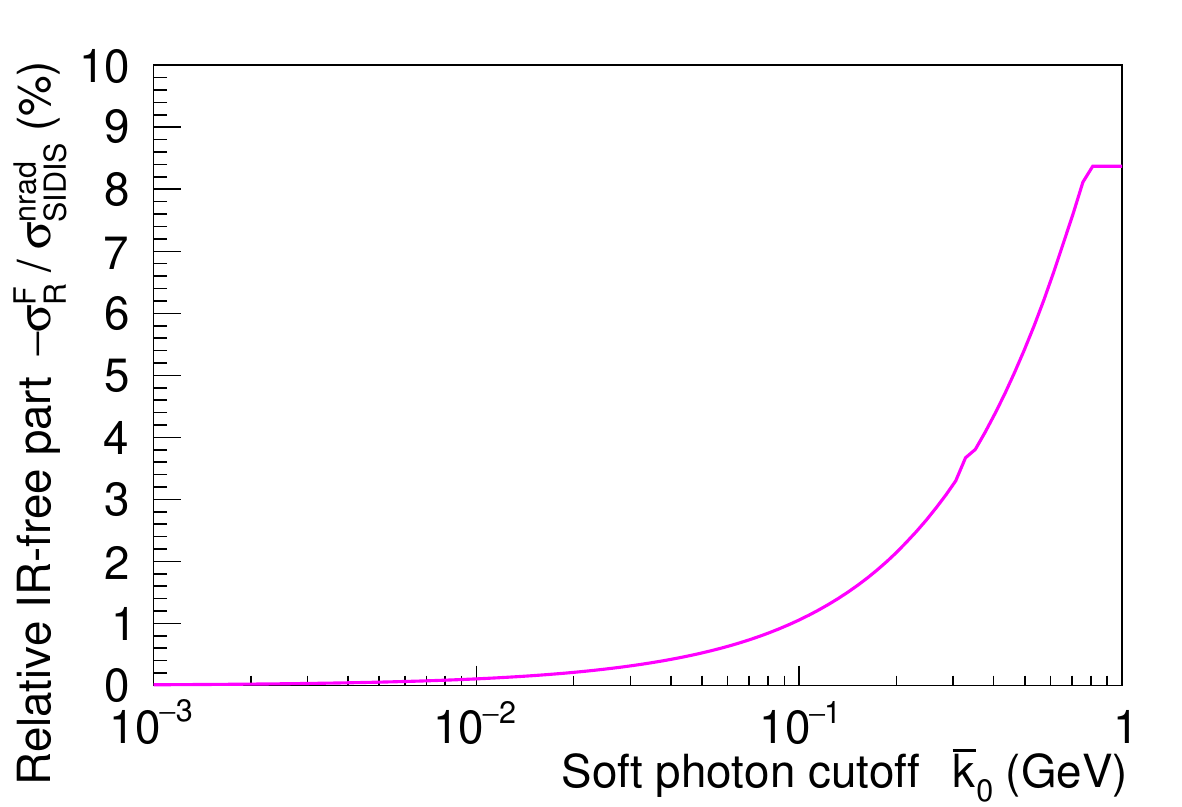}
\caption{(Color online) Numerical estimate of the percentage of the total cross section $\sigma_{\rm SIDIS}^{nrad}$ that stems 
from the non-radiative infrared-divergent-free part $\sigma^{F}_{R}$ (taken with the minus sign), as a function of the soft photon 
cutoff $\bar{k}_0$. As $\bar{k}_0$ goes to zero, $\sigma_{R}^{F}$ contributes less and less to the total cross section. Selected 
kinematics: $E_{b} = 10.6~{\rm GeV}$, $x_{Bj} = 0.2$, $Q^{2} = 1.5~{\rm GeV^{2}}$, $z_{h} = 0.35$, $P_{hT} = 0.3~{\rm GeV}$, 
$\phi_{h} = 0$.}
\label{fig:fig_soft_photon_cutoff}
\end{figure}

In other words, since the generator neglects the infrared-divergent-free part of the cross-section, the resulting cross sections 
unphysically depend on the soft photon cutoff $\bar{k}_{0}$. At $\bar{k}_{0}\rightarrow 0$, this approximation approaches the true 
cross section. However, by selecting too small a value for $\bar{k}_{0}$ will substantially reduce the generator's performance, as a 
greater fraction of the events will need to be drawn from the inefficient radiative part of the cross section. To choose a good value 
for $\bar{k}_0$, a program at

\url{https://github.com/duanebyer/sidis/blob/master/example/plot_cutoff.cpp}
has been provided, which demonstrates the user how the error in the cross-section changes with $\bar{k}_0$. Fig.~\ref{fig:fig_soft_photon_cutoff} 
shows what the output of that program looks like. Then $\bar{k}_0$ can be selected to be as large as possible while meeting whatever 
error requirements users have.

The structure functions can be decomposed further using the beam polarization $\xi$ and target polarization $\bm{\eta}$ 
in the hadron frame (see Eq.~(\ref{eqn_eq:SIDIS_RC1})). Both $\sigma^{B}_{\mathrm{SIDIS}}$ and $\sigma^{AMM}$ have the 
following form
\begin{eqnarray}
\sigma & = & \sigma^{UU} + \bm{\eta}\cdot\bm{\sigma}^{UP} + \lambda \left( \sigma^{LU} + \bm{\eta}\cdot\bm{\sigma}^{LP} \right) ,
\nonumber \\
\bm{\sigma}^{UP} & = & \sigma^{UT_{1}} \mathbf{i} + \sigma^{UT_{2}} \mathbf{j} + \sigma^{UL} \mathbf{k} ,
\nonumber \\
\bm{\sigma}^{LP} & = & \sigma^{LT_{1}} \mathbf{i} + \sigma^{LT_{2}} \mathbf{j} + \sigma^{LL} \mathbf{k} .
\label{eqn_eq:SIDIS_pol}
\end{eqnarray}
The eight polarized parts of the cross section $\sigma^{UU}$, $\sigma^{UL}$, etc., all depend on independent subsets of the SIDIS 
structure functions, so that they can be computed separately and then combined. For the radiative part ($\sigma_{R}$) of the cross section, 
the kinematic shift from the emitted photon causes the polarized parts $\sigma^{UL}$, $\sigma^{UT_{1}}$, and $\sigma^{UT_{2}}$ to mix 
with each other (the same is also the case for $\sigma^{LL}$, $\sigma^{LT_{1}}$, and $\sigma^{LT_{2}}$). That means that there are only 
four independent polarized parts of the radiative cross section: $\sigma_{R}^{UU}$, $\bm{\sigma}_{R}^{UP}$, $\sigma_{R}^{LU}$, and 
$\bm{\sigma}_{R}^{LP}$.

\subsection{\label{sec:lib} Library component}

\url{https://github.com/duanebyer/sidis/tree/master/src}

\url{https://github.com/duanebyer/sidis/tree/master/include/sidis} \\

The C\texttt{++} library \texttt{libsidis} provides a set of functions for computing the six-fold Born cross section 
($\sigma_{\rm SIDIS}^{B}$), the six-fold non-radiative cross section ($\sigma_{\rm SIDIS}^{nrad}$), and the nine-fold radiative 
cross section ($\sigma_{\rm SIDIS}^{rad}$). Additionally, there are auxilliary functions for computing related 
cross sections/corrections (e.g., $\sigma^{AMM}$ or $\delta^{VS}$), kinematic variables, and structure functions.

The top-level \texttt{libsidis} namespace is \texttt{sidis}. It is further organized into several nested namespaces listed as follows:
\begin{itemize}
\item \texttt{xs}: functions for various cross sections;
\item \texttt{kin}: kinematic calculations for the base SIDIS process (6D phase space) and for the radiative SIDIS processes 
(9D phase space)\footnote{Had exclusive structure functions \cite{Akushevich:2022} been included in our framework, 
the radiative exclusive tail for SIDIS would be in a 8D phase space \cite{Akushevich:2019mbz}.};
\item \texttt{lep}: ``leptonic coefficients'', like $\theta_{i}^{B}$, $\theta_{i1}$, $\theta_{ij}$ and $\theta_{i}^{AMM}$, employed for 
various cross sections (see Eqs.~(\ref{eqn_eq:SIDIS_RC9})~(\ref{eqn_eq:SIDIS_RC27})~and~(\ref{eqn_eq:SIDIS_RC29}));
\item \texttt{had}: ``hadronic coefficients'', like $\mathcal{H}$ and $\tilde{\mathcal{H}}$ for various cross sections 
(see again Eqs.~(\ref{eqn_eq:SIDIS_RC9})~(\ref{eqn_eq:SIDIS_RC27})~and~(\ref{eqn_eq:SIDIS_RC29}));
\item \texttt{sf}: parameterizations of the SIDIS structure functions, including TMD PDFs and TMD FFs 
(see Sec.~\ref{sec:SIDIS_struc_func}));
\item \texttt{ps}: particle information, such as masses and charges \cite{Zyla:2020zbs};
\item \texttt{math}: additional convenience functions.
\end{itemize}

Some computations may be duplicated between one cross-section evaluation and the next one, e.g., structure function evaluations when 
integrating over $\phi_{h}$ and/or $\phi_{S}$. In order to reduce unnecessary computations, \texttt{libsidis} provides two main modes 
of use: a streamlined mode that hides the internal data flow, and a flexible mode that provides access to it. The streamlined mode 
should be used in nearly all cases, but the flexible mode \texttt{API} is exported for some occasional situations where it may provide 
a performance advantage.

\subsubsection{\label{sec:flex} Flexible mode}

We start with the flexible mode of \texttt{libsidis}. The dataflow of this mode is diagrammatically shown in Fig.~\ref{fig:fig_diag1}, 
and demonstrated in the program Listing~\ref{lst:libsidis-modes} in Appendix~A.
\begin{figure}[hbt]
\centering
\hspace{0.0cm}
\includegraphics[width=13.5cm]{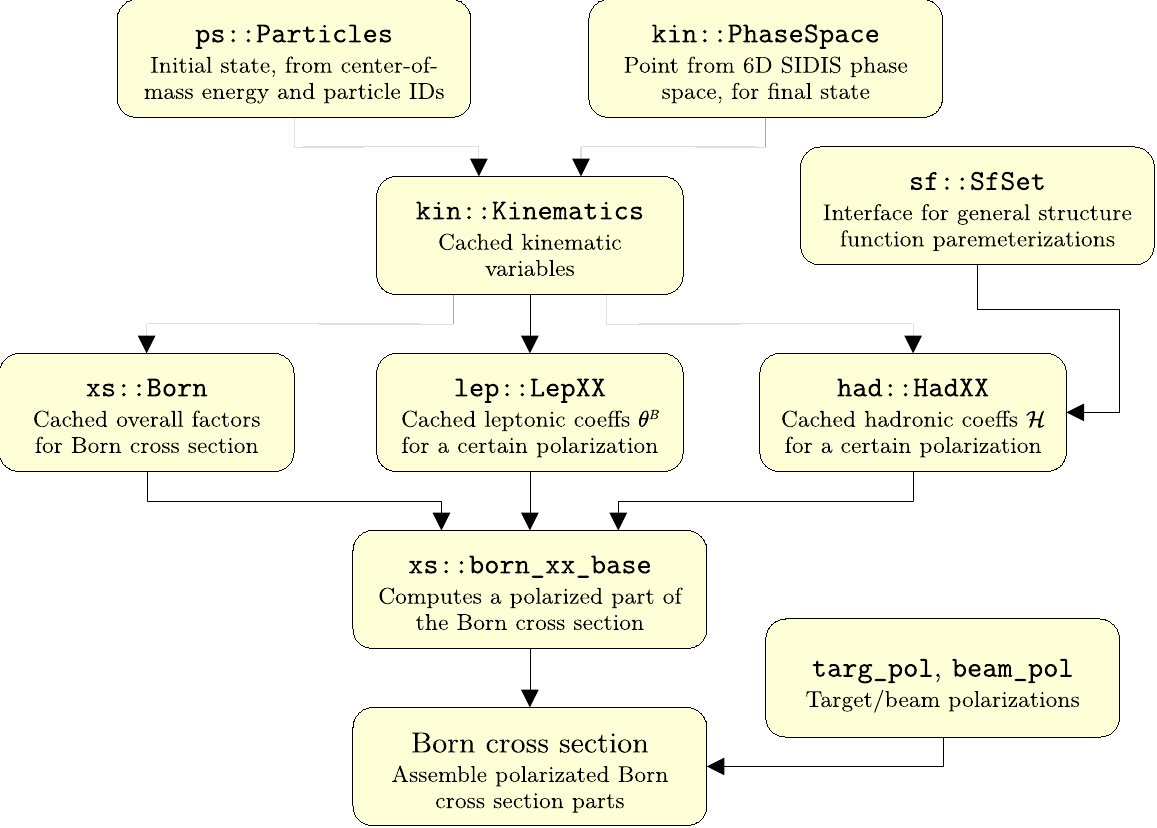}
\caption{(Color online) Dataflow diagram for the flexible mode of \texttt{libsidis}. All type names are given in the \texttt{sidis::}namespace.}
\label{fig:fig_diag1}
\end{figure}
The structure of this diagram is represented as follows:
\begin{itemize}
    \item the particles involved and the six SIDIS phase-space variables are specified using the data structures \texttt{part::Particles} 
    and \texttt{kin::PhaseSpace};
    \item kinematic variables describing the SIDIS process are computed and cached in the \texttt{kin::Kinematics} data structure;
    \item leptonic and hadronic parts of the cross section are computed and cached using the data structures \texttt{lep::LepXX} and 
    \texttt{had::HadXX}; these data structures come in several varieties, both for different polarizations (e.g., \texttt{had::HadUU}, 
    \texttt{had::HadUL}, etc.), and for different types of cross sections (e.g., \texttt{lep::LepBornXX}, \texttt{lep::LepAmmXX}, etc.);
    the hadronic part of the cross section uses the SIDIS structure functions, as provided by the \texttt{sf::SfSet} interface; 
    \item additional kinematic variables are computed for a specific type of cross section; for example, the \texttt{xs::Born} data 
    structure is used for the Born cross section, or the \texttt{xs::Amm} data structure is used for the anomalous magnetic moment 
    cross section, etc.;
    \item both leptonic and hadronic coefficients are combined with those additional kinematic variables to compute one of the polarized 
    parts of the cross section (as shown in Eq.~(\ref{eqn_eq:SIDIS_pol})) using the \texttt{xs::born\_xx\_base} functions; these functions 
    have variations for different polarizations and types of cross sections (e.g., \texttt{amm\_uu\_base}, \texttt{born\_ll\_base}, etc.).
\end{itemize}
The dataflow is the same for radiative cross sections, except that the kinematics-related data structures are substituted by their radiative versions:
\begin{center}
\texttt{kin::PhaseSpace} $\rightarrow$ \texttt{kin::PhaseSpaceRad};

\texttt{kin::Kinematics} $\rightarrow$ \texttt{kin::KinematicsRad}.
\end{center}
The full polarized cross section must be constructed from the polarized parts of the cross section exactly, as described in Eq.~(\ref{eqn_eq:SIDIS_pol}).

\subsubsection{\label{sec:stream} Streamlined mode}

The dataflow diagram for the streamlined mode is shown in Fig.~\ref{fig:fig_diag2}, also demonstrated in the program Listing~\ref{lst:libsidis-modes} 
in Appendix~A.
\begin{figure}[hbt]
\centering
\hspace{0.0cm}
\includegraphics[width=13.4cm]{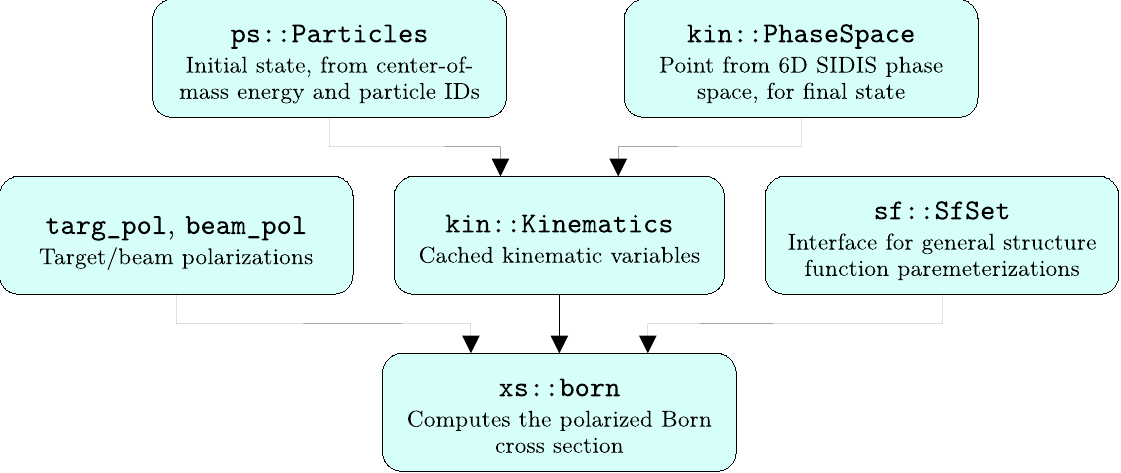}
\caption{(Color online) Dataflow diagram for the streamlined mode of \texttt{libsidis}. All type names are given in the \texttt{sidis::}namespace.
}
\label{fig:fig_diag2}
\end{figure}
The streamlined mode makes use of functions with a simplified interface, such as \texttt{xs::born}, \texttt{xs::nrad}, or \texttt{xs::rad}. 
These streamlined functions compute the cross sections directly from the \texttt{kin::Kinematics} data structure and the \texttt{sf::SfSet} 
structure function set. The streamlined functions will automatically combine the polarized parts to form the full polarized cross section, 
as described in Eq.~(\ref{eqn_eq:SIDIS_pol}).

\subsubsection{\label{sec:SF} Structure function implementation}

\url{https://github.com/duanebyer/sidis/tree/master/share/sidis/sf_set/prokudin} \\

In this section, we continue the discussion of Sec.~\ref{sec:SIDIS_struc_func} but in the context of the event generator. The 
interface \texttt{sf::SfSet} includes all eighteen leading- and subleading-twist SIDIS structure functions. These structure 
functions are often computed using the TMD factorization, which is valid at $P_{hT} \sim k_{\perp} \ll Q$, as mentioned in 
Sec.~\ref{sec:SIDIS_kinematics}. For this situation, the library \texttt{libsidis} provides an implementation 
\texttt{sf::TmdSfSet}. The user-provided TMDs and FFs must then be given via another interface \texttt{sf::TmdSet}. 

A 2D convolution needs to be performed when calculating the structure functions from these TMDs and FFs, based on which the 
type $\texttt{sf::SfSet}$ can compute the structure functions directly from the integral in Eq.~(\ref{eqn_eq:SIDIS_kin11}).
In order to simplify this convolution and further improve the generator's performance, the TMDs and FFs are often taken to be 
Gaussian as in Eq.~(\ref{eqn_eq:SIDIS_kin12}), such that the convolution can be evaluated analytically, which leads, e.g., to
Eq.~(\ref{eqn_eq:SIDIS_kin13}),~(\ref{eqn_eq:SIDIS_kin14}),~or~(\ref{eqn_eq:SIDIS_kin15}). If the Gaussian approximation is 
turned on, then the user needs to provide only the 2D TMDs and FFs computed on $(x_{{\!}_{Bj}}, Q^{2})$ grids 
\cite{WW-SIDIS:2018,Martin:2009iq,mstwpdf,LHAPDF,Liu:2019}, with $\bm{k}_{\perp}$- and $\bm{p}_{\perp}$-dependence coming from 
the Gaussian width parameters. For this situation, the implementation \texttt{sf::GaussianTmdSfSet} can be used to provide the 
structure functions, combined with the Gaussian TMDs and FFs provided through another interface \texttt{sf::GaussianTmdSet}.

We already mentioned in Sec.~\ref{sec:SIDIS_struc_func} that section~3.2 of \cite{Bastami:2018xqd} has a detailed discussion of the 
WW-type approximations for TMDs/FFs, along with sections~4.1~and~4.2 of the same reference showing how to treat the twist-2 and all 
the twist-3 structure functions within these approximations. To twist-3, there are twenty-four TMDs and six FFs that must be provided. 
The twist-3 TMDs and FFs can be estimated in terms of the twist-2 TMDs and FFs, which reduce the independent TMDs to six and independent 
FFs to two. To use these approximations, the implementation \texttt{sf::WwTmdSfSet} can be used, combined with the WW-type basis TMDs 
and FFs given through another interface \texttt{sf::WwTmdSet}. Thereby, \texttt{SIDIS-RC EvGen} can apply WW-type approximations 
to reduce how many TMDs and FFs need to be provided by the user to improve performance.

While several WW-type approximations are compatible with the Gaussian approximation, the others are not. It happens because these 
latter WW-type approximations are not valid at $\bm{k}_{\perp} \rightarrow 0$ (see section~4.4 of \cite{Bastami:2018xqd} for more 
details). In this case, it is possible to combine the two types of approximations by first solving the convolution integrals 
(using the Gaussian approximation), and then applying the integrated WW-type approximations to simplify the obtained results for 
the structure functions. One can combine them using the implementation \texttt{sf::GaussianWwTmdSfSet} and the associated interface 
\texttt{sf::GaussianWwTmdSet}. Thus, in summary we have the following implementations that can be used for parameterizing the SIDIS 
structure functions provided by
\begin{itemize}
\item[(i)] a user-defined implementation of \texttt{sf::SfSet};
\item[(ii)] \texttt{sf::TmdSfSet}, with the TMDs/FFs specified as functions or on grids through the \texttt{sf::TmdSet} 
interface, and then numerically convoluted;
\item[(iii)] \texttt{sf::GaussianTmdSfSet}, with the TMDs/FFs specified as Gaussian approximations through the 
\texttt{sf::GaussianTmdSet} interface, and then analytically convoluted;
\item[(iv)] \texttt{sf::WwTmdSfSet}, with the WW basis TMDs/FFs specified through the \texttt{sf::WwTmdSet} interface 
(and unspecified TMDs/FFs computed through WW-type approximation) and then numerically convoluted;
\item[(v)] \texttt{sf::GaussianWwTmdSfSet}, with the WW basis TMDs/FFs specified as Gaussian approximations through the 
\texttt{sf::GaussianWwTmdSet} interface, and then being analytically convoluted.
\end{itemize}

\subsubsection{\label{sec:LH_coeff} Leptonic/Hadronic coefficients}

As mentioned in the discussion of the flexible mode in Sec.~\ref{sec:flex}, the coefficients $\theta_{i}$ and $\mathcal{H}_{i}$, 
referred to as the leptonic coefficients and hadronic coefficients respectively, are packaged into two families of data structures 
$\texttt{lep::LepXX}$ and $\texttt{had::HadXX}$. These data structures come in many varieties, both for different types of cross 
sections (Born, AMM, radiative), and different polarizations. The reason to separate these coefficients by polarization is to 
reduce the computational cost when only certain polarization parts of a cross section are used. For example, when the beam or 
target is unpolarized, the polarized parts of the cross section should not be computed at all. Or, when computing a transverse single 
spin asymmetry, only the $\sigma^{UT}$ part of the cross section may be needed. 

Here, we provide a taxonomy of the leptonic/hadronic coefficient structures. The polarizations are notated with two symbols $XX$. 
The first symbol denotes the lepton polarization, which can be $U$ for unpolarized, $L$ for longitudinally polarized, or $X$ for 
arbitrary polarization (lepton transverse polarization is not allowed). The second symbol denotes the hadron polarization, which 
can be $U$ for unpolarized, $L$ for longitudinally polarized, $T$ for transversely polarized, $P$ for fully polarized in an arbitrary 
direction, or $X$ for arbitrary polarization.

\begin{description}
\item[]\underline{\texttt{had::HadXX} (lepton pol.: $U$, $L$, $X$; hadron pol.: $U$, $L$, $T$, $P$, $X$)} \\
Standard set of structure functions $\mathcal{H}$, as explicitly shown in Eq.~(14) of \cite{Akushevich:2019mbz}, used for all 
non-radiative cross sections.
\item[]\underline{\texttt{had::HadRadXX} (lepton pol.: $U$, $L$, $X$; hadron pol.: $U$, $P$, $X$)} \\
Shifted structure functions $\tilde{\mathcal{H}}$, corresponding to the kinematic shift introduced by real photon emission. 
The hadron polarization cannot be $L$ or $T$ alone, because $\sigma^{(U/L)L}$, $\sigma^{(U/L)T_{1}}$, and $\sigma^{(U/L)T_{2}}$ parts 
mix together in the shifted hadronic frame. $T_{2}$ refers to the part of the transverse polarization  perpendicular to both $\bm{q}$ 
and $\bm{P_{hT}}$ in the target rest frame, while $T_{1}$ is the remaining part of the transverse polarization with a component along 
$\bm{P_{hT}}$. Used for the radiative cross section.
\item[]\underline{\texttt{had::HadRadFXX} (lepton pol.: $U$, $L$, $X$; hadron pol.: $U$, $P$, $X$)} \\
Subtracted shifted structure functions, used in the calculation of the infrared-divergent-free cross section $\sigma_{R}^{F}$ 
(see Eq.~(\ref{eqn_eq:SIDIS_RC27})). This data structure stores both the shifted structure functions $\tilde{\mathcal{H}}$, and 
the difference $(\mathcal{H} - \tilde{\mathcal{H}})/R$, using the derivative to approximate when $R$ is small.
\end{description}

The leptonic coefficients use a different set of hadron polarization symbols. The same subset of coefficients $\theta_{i}$ is used for 
``out-of-plane'' hadron polarizations $U$ and $T_{2}$, and another subset is used for ``in-plane'' hadron polarizations $L$ and $T_{1}$. 
So, for the lepton coefficients only, the hadron polarization symbol $U$ refers to both $U$ and $T_{2}$ polarizations, while the hadron 
polarization symbol $P$ refers to the others $L$ and $T_{1}$. The lepton polarization symbols remain unchanged.
\begin{description}
\item[]\underline{\texttt{lep::LepBornXX} (lepton pol.: $U$, $L$, $X$; hadron pol.: $U$, $P$, $X$)} \\
Leptonic coefficients $\theta^{B}$, as shown in Eq.~(16) of \cite{Akushevich:2019mbz}, used for the Born cross section $\sigma_{\rm SIDIS}^{B}$.
\item[]\underline{\texttt{lep::LepAmmXX} (lepton pol.: $U$, $L$, $X$; hadron pol.: $U$, $P$, $X$)} \\
Leptonic coefficients $\theta^{AMM}$, as shown in Eq.~(54) of \cite{Akushevich:2019mbz}, used for the anomalous magnetic moment 
(AMM) cross section $\sigma^{AMM}$.
\item[]\underline{\texttt{lep::LepNradXX} (lepton pol.: $U$, $L$, $X$; hadron pol.: $U$, $P$, $X$)} \\
Combination of \texttt{lep::LepBornXX} and \texttt{lep::LepAmmXX} coefficients that, combined with various correction factors, 
give the complete non-radiative cross section.
\item[]\underline{\texttt{lep::LepRadXX} (lepton pol.: $U$, $L$, $X$; hadron pol.: $U$, $P$, $X$)} \\
Leptonic coefficients $\theta$, as shown in Eq.~(B1) and Eq.(B2) of \cite{Akushevich:2019mbz}, used for the radiative part of 
the cross section.
\end{description}
Because of a large number of coefficient structures and the polarized variants, they are produced automatically using the \texttt{cog}
code-generation tool \cite{Cog}.

\subsection{\label{sec:gen_comp} Binary (generator) component}

\url{https://github.com/duanebyer/sidis/tree/master/app/sidisgen} \\

The generator component, \texttt{sidisgen}, uses MC generation to produce points (these are the ``events'') in the SIDIS phase space
for numerical integration weighted by the SIDIS cross section $\sigma_{\mathrm{SIDIS}}^{in}$ (from \texttt{libsidis}). The events are 
drawn from a probability distribution $\Gamma\sim P(\gamma)$\footnote{The variable $\gamma$ describes a point in the phase space.} 
and weighted by $w(\gamma)=\sigma_{\mathrm{SIDIS}}^{in}(\gamma)/P(\gamma)$ (we define $W \equiv w(\Gamma)$). Then, the expectation
value can be used to evaluate cross-section-weighted integrals over the SIDIS phase space:
\beq
\left< O \,W \right> = \int O(\gamma)\,\sigma_{\mathrm{SIDIS}}^{in}(\gamma) \, d\gamma .
\label{eqn_eq:aveO}
\eeq
The MC generator used by \texttt{sidisgen} attempts to minimize the variance in the weights $W$. If all of the weights are equal, 
then $P(\gamma)\propto\sigma_{\mathrm{SIDIS}}^{in}(\gamma)$, so that the events produced by \texttt{sidisgen} can be used to mock 
data measured by SIDIS experiments at HERMES, COMPASS, and JLab.

The quality of the MC generator can be described by the ratio of the effective sample size to the true sample size \cite{Martino:2017,Kong:1992}. 
The square root of this quantity is labeled by $\varepsilon$, which we call the efficiency of the generator:
\beq
\varepsilon = \sqrt{\frac{N_{\rm eff}}{N_{\rm tot}}} = \sqrt{\frac{\left< W \right>^{2}}{\left <W^{2} \right>}} = \sqrt{\frac{1}{1 + {\sigma_{W}^{2}}/{{\left< W \right>}^{2}}}}.
\label{eqn_eq:SIDIS_MC10}
\eeq
The efficiency varies in the range of $[0, 1]$. An efficiency of $1$ corresponds to completely uniformly weighted events. The
efficiency is directly related to the variance in the ratio $\sigma_{\mathrm{SIDIS}}^{in}(\gamma)/P(\gamma)$. 
For defining $\sqrt{N_{\rm eff}/N_{\rm tot}}$, we take inspiration from \cite{Kong:1992}, which can be motivated as the ratio 
of standard deviations of an observable $\mathcal{O}$ between the importance distribution $P(\gamma)$ and the true cross section 
$\sigma_{\textrm{SIDIS}}^{in}(\gamma)$. The better $P(\gamma)$ approximates the cross section, the better the efficiency will be. 
The notation $\sigma_{W}$ in Eq.~(\ref{eqn_eq:SIDIS_MC10}) denotes the standard deviation (not to be confused with the cross-section 
notation).

Because non-radiative and radiative weighted events reside in different phase spaces (6D and 9D, respectively), \texttt{sidisgen} 
has a distinct MC sub-generator for each event type. Each time a new event is requested, it is selected at random to be either
non-radiative or radiative. The ratio of non-radiative to radiative events is chosen to maximize the overall efficiency of the
generator, corresponding to a relative probability of drawing events from each sub-generator of
\bea
P_{(n)rad} \propto \sqrt{\left< {W_{(n)rad}}^{2} \right>} .
\label{eqn_eq:SIDIS_MC_P}
\eea

\subsubsection{\label{sec:FOAM1} MC generation with the FOAM library}

The FOAM library from ROOT is used by \texttt{sidisgen} as the underlying MC engine for both non-radiative and radiative sub-generators,
and approximates the (true) cross sections with a spatial indexing tree, called a foam. Under default settings, the foam is a 
$k$-d tree\footnote{A $k$-d tree subdivides a rectangular $k$-dimensional region through recursive axis-aligned bisections.}, 
constructed through a recursive process in which the cross section is sampled randomly within each foam cell, and then the cells 
are divided to minimize the variance within each daughter cell. This process makes a nested tree structure of hyper-rectangles. 

The foam is initialized once, and then used many times to produce events. For the event production, the FOAM library chooses a 
cell at random, with a probability chosen such that the overall efficiency is maximized:
\bea
P_{k} \propto V_{k} \sqrt{\int_{V_{k}} \Lb \sigma_{\mathrm{SIDIS}}^{in}(\gamma) \Rb^{2}\,d\gamma } .
\eea
Then, a point $\gamma_{i}$ is chosen randomly within the cell, and a weight $w_{i}=w(\gamma_i)$ is assigned to that point. As long 
as the foam is sufficiently dense such that the cross section does not vary significantly over each cell, the events will be weighted 
nearly uniformly.

Some kinematic cuts can be provided on the six-fold cross-section variables, $x_{{\!}_{Bj}}$, $y$, $z_{h}$, $P_{hT}$, $\phi_{h}$,
$\phi_{{\!}_{S}}$. These cuts are accounted for during the FOAM tree initialization, and the library ensures that out-of-bound 
events are not produced outside of the cuts (meaning that no need to generate and then discard those events). The radiative
contribution $\sigma_{\rm SIDIS}^{rad}$ to the SIDIS cross section needs to be integrated over photon degrees of freedom 
to be compared to the non-radiative contribution $\sigma_{\rm SIDIS}^{nrad}$. For ease of use, $\texttt{libsidis}$ provides
photon-integrated cross-section functions $\texttt{xs::rad\_integ}$ and $\texttt{xs::rad\_f\_integ}$ that utilizes the VEGAS Monte-Carlo 
integrator \cite{VEGAS} from the GNU Scientific Library (GSL) \cite{GSL}. The Cubature package \cite{Cubature} for the multi-dimensional 
adaptive quadrature method is used as a backup integrator. The photon-integrated cross sections make use of the variable transforms discussed 
in the next Section~\ref{sec:FOAM2}.

\subsubsection{\label{sec:FOAM2} Improvement of the foam efficiency}

\begin{figure}[hbt!]
\centering
\includegraphics[width=6.75cm]{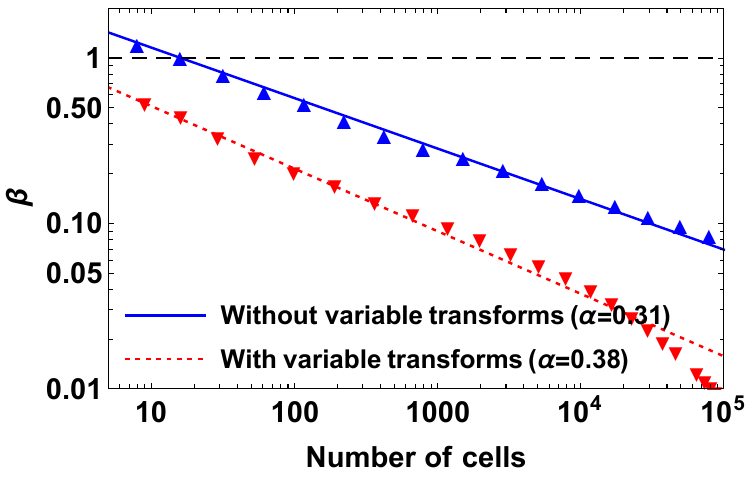}\label{fig:fig_non-rad_eff}
\includegraphics[width=6.75cm]{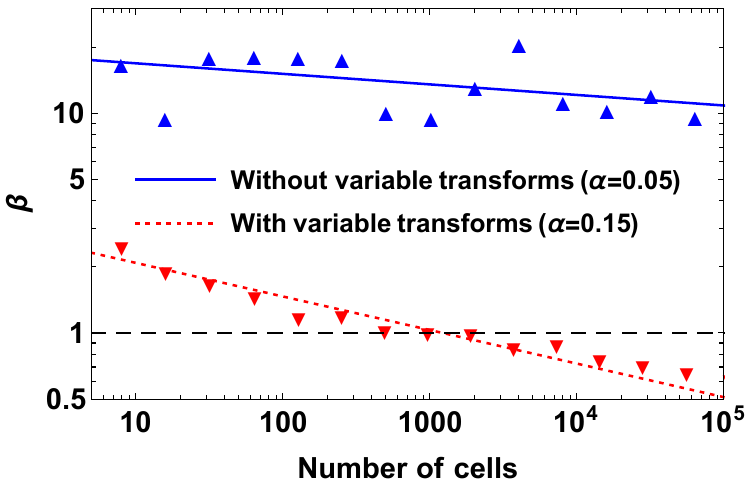}\label{fig:fig_rad_eff}
\caption{(Color online) The improvement of the foam efficiency as the number of cells in the foam increases, for the non-radiative 
(left plot) and radiative (right plot) cross sections. Smaller $\beta$ corresponds to better efficiency, dashed line at $50\%$
efficiency. The fluctuations of the points in the right plot are caused by a poor estimate of the foam's efficiency at high $\beta$. 
The use of variable transforms (see later in Eq.~\ref{eq:rad_var_transform}) substantially
improve the scaling parameter $\alpha$ in both cases.}
\label{fig:fig_eff}
\end{figure}

As more cells are added to the foam, the efficiency improves. In order to describe this process, another useful quantity, 
$\beta = (1/\varepsilon) - 1$, is introduced, which ranges from $0$ (being $100\%$ efficient) to $+\infty$ (being $0\%$ efficient).
Empirically, $\beta$ scales like a negative power law with increasing number of cells in the $k$-d tree: 
$\beta \propto N_{\mathrm{cells}}^{-\alpha}$, as shown in Fig.~\ref{fig:fig_eff}. For most distributions, the power $\alpha$ is
observed to vary between $0$ and $2$. If $\alpha$ is small, it takes far longer to reach the desired efficiency.

When $\sigma_{\rm SIDIS}^{rad}$ is expressed in terms of the photon variables (see Eq.~(\ref{eqn_eq:SIDIS_RC13})) $R$ 
- proportional to photon energy, $\tau$ - related to photon polar angle, and $\phi_{k}$ - photon azimuthal angle, there 
are two radiative peaks in the $(\tau, \phi_{k})$ distribution, corresponding to photon emission collinear with the incoming and 
outgoing electrons. The locations of these peaks alter further with the other phase-space variables, leading to ridges in 
the full 9D distribution. The ridges are not well aligned with the hyper-rectangle axes, making it difficult for 
the foam to approximate them, and consequently resulting in a very poor scaling factor $\alpha_{rad}\sim0.05$, 
corresponding to the upper triangle points and the solid line in the right plot of Fig.~\ref{fig:fig_eff}. With such 
a poor scaling factor, it would take on the order of $10^{10}$ cells to reach an efficiency of only $75\%$. We mitigated 
this problem in several ways.

$\bullet$ First, the FOAM library is extended to improve performance, add support to its original multi-threading, 
and use more accurate estimators for the cell division. This makes it feasible to use an order more cells in the foam 
construction.

\begin{figure}[hbt!]
\centering
\includegraphics[width=6.5cm]{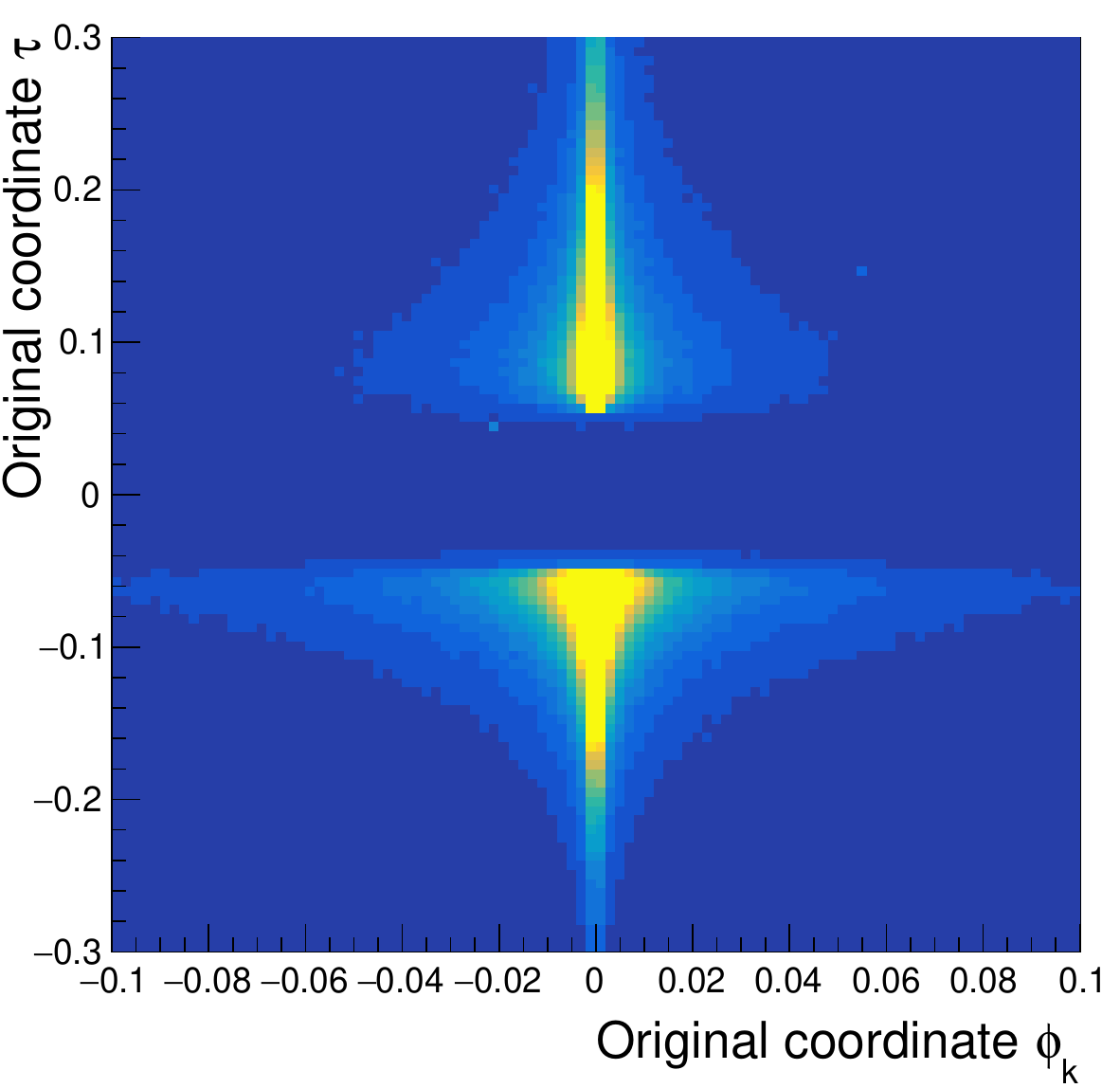}\label{fig:fig_rad_dist}
\hspace{0.0in}
\includegraphics[width=6.5cm]{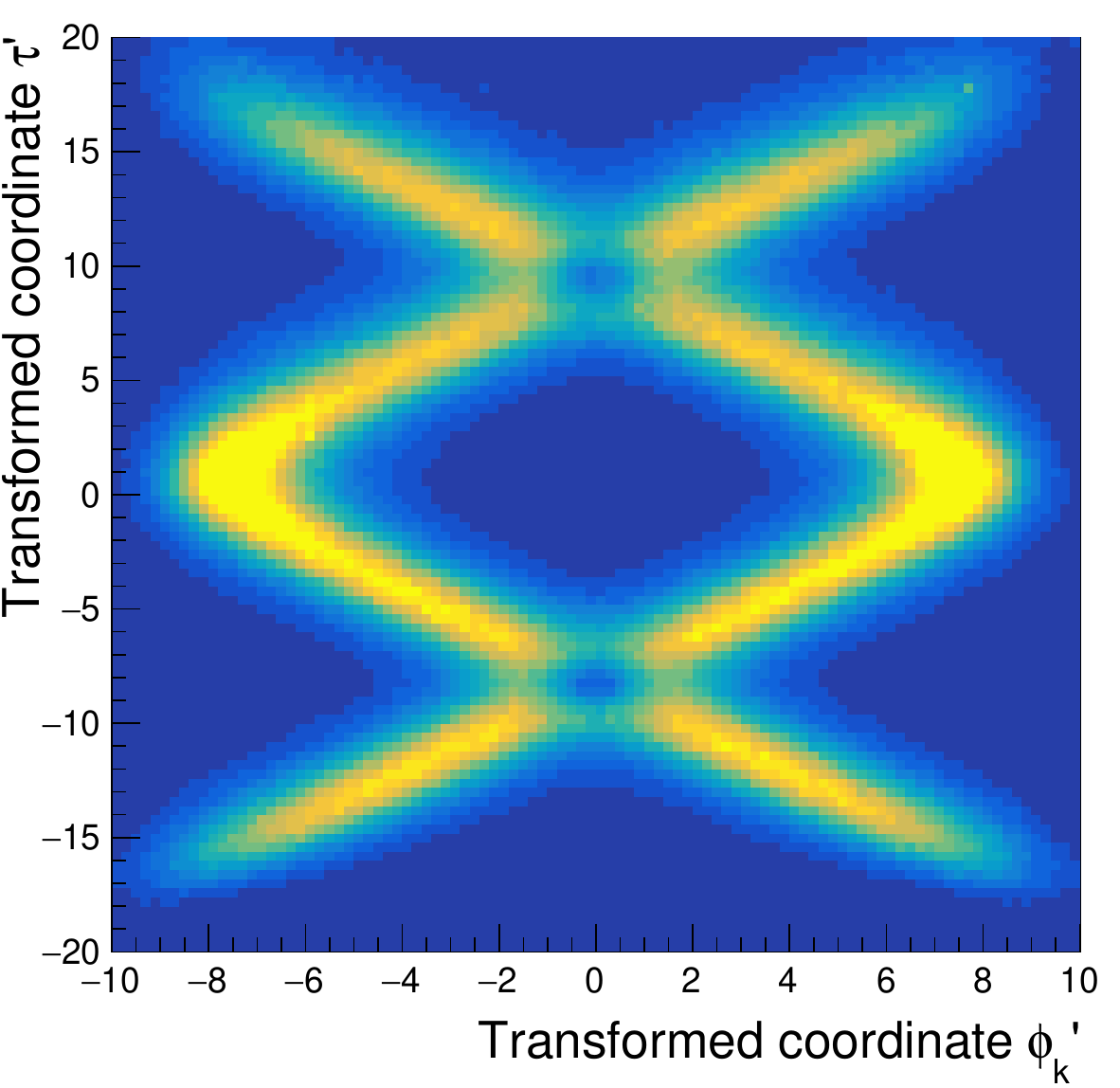}\label{fig:fig_rad_dist_t}
\caption{\label{fig:fig_rad_comp} (Color online) The change in the radiative part of the distribution as a result of the variable
transforms from Eq.~\ref{eq:rad_var_transform}. The left plot shows the two sharp peaks in the radiative distribution, which are 
difficult for the foam to approximate. The right plot shows much broader transformed peaks, allowing the foam to more accurately
approximate it.}
\end{figure}

$\bullet$ Second, two variable transformations are used to smooth out the radiative peaks. The locations and widths of the 
peaks in $(\tau,\phi_{k})$ around the incoming and outgoing electron directions can be calculated from the electron kinematics. 
Then, a sigmoid transformation in $\phi_{k}$ and a double-sigmoid transformation in $\tau$ can be used to smooth the peaks, 
choosing the steepest parts of such transformations to coincide with the peak locations. We use
\bea
\phi_{k}^{\prime} & = & \arcsinh{\!\Lb \frac{\phi_{k}}{w_{\phi_{k}}} \Rb} ,
\nonumber \\
\tau^{\prime} & = & \arcsinh{\!\Lb \frac{\tau - \tau_{p,1}}{w_{\tau,1}} \Rb} + 
\arcsinh{\!\Lb \frac{\tau - \tau_{p,2}}{w_{\tau,2}} \Rb} ,
\label{eq:rad_var_transform}
\eea
where $w_{\phi_{k}}$, $w_{\tau,1}$, $w_{\tau,2}$ are the peak widths, and $(0,\tau_{p,1})$, $(0,\tau_{p,2})$ are the peak locations.
The result of these transformations is shown in Fig.~\ref{fig:fig_rad_comp}. Additional variable transformations are applied to some 
of the other SIDIS variables as well, such as a $(\cdot)^{-1}$ transform on the Bjorken $x_{{\!}_{Bj}}$, to roughly
match the expected cross-section behavior. The transformations substantially improve the foam construction process, as shown in
Fig.~\ref{fig:fig_eff}.

$\bullet$ Third, rejection sampling is used to reduce variance of events produced by the non-radiative and radiative sub-generators. 
The radiative sub-generator can often reach an efficiency of $75\%$, and the rejection sampling can push this number up further towards
$95\%$ or higher, at a cost to event generation performance.

\section{\label{sec:programs} Numerical results obtained from {\texttt\textbf{SIDIS-RC EvGen}}}

By having at our disposal all the discussion in Sec.~\ref{sec:MC_gen}, in this section we demonstrate a number of results obtained 
after running \texttt{SIDIS-RC EvGen}: namely on the SIDIS Collins and Sivers azimuthal transverse SSAs. 
Fast computations of the asymmetries in general are provided in the $\texttt{asym}$ namespace, combining the GSL VEGAS integrator 
with the flexible mode for efficient integration.

\subsection{\label{sec:asym} Charged-hadron Collins and Sivers azimuthal transverse single-spin asymmetries}

Transverse SSAs are among the most important and fascinating observables in electron scattering and hadronic physics. During the last fifteen years or so, 
the transverse SSAs in SIDIS have been studied at HERMES, COMPASS, and JLab
\cite{HERMES:2009lmz,HERMES:2010mmo,HERMES:2012kpt,COMPASS:2012dmt,COMPASS:2014bze,COMPASS:2014kcy,Qian:2011py}.
Their studies are crucial for our cardinal understanding of the 3D momentum structure of the nucleon.

The SIDIS differential cross section in Eq.~(\ref{eqn_eq:SIDIS_kin3}) (or Eq.~(\ref{eqn_eq:SIDIS_RC9})), including all leading-twist (twist-2) and 
subleading-twist (twist-3) structure functions in the $1/Q$ expansion, can also be expressed in terms of SSAs in the following way \cite{Bastami:2018xqd}:
\begin{eqnarray*}
& & \!\!\!\!\!\!\!\!\!\!\!\!\!\!\!\!\!\!\!\!\!\!\!\!\!\!\!\!\!\!\!\!
\sigma_{\rm SIDIS}^{B} = \frac{\alpha^{2}}{x_{{\!}_{Bj}} y Q^{2}} \Lb 1 + \frac{\gamma^{2}}{2x_{{\!}_{Bj}}} \Rb c_{1}\,F_{UU} \times
\nonumber\\
& & \!\!\!\!\!\!\!\!\!\!\!\!\!\!\!\!\!\!\!\!\!\!\!
\times \Bigg\{ \bigg[ 1 + \Lb \frac{c_{3}}{c_{1}} \Rb \cos{\!(\phi_{h})}\,A_{UU}^{\cos{\!(\phi_{h})}} +
\nonumber\\
& & \!\!\!\!\!\!\!\!\!\!
+ \Lb \frac{c_{2}}{c_{1}} \Rb \cos{\!(2\phi_{h})}\,A_{UU}^{\cos{\!(2\phi_{h})}} + \lambda_{e} \Lb \frac{c_{4}}{c_{1}} \Rb \sin{\!(\phi_{h})}\,A_{LU}^{\sin{\!(\phi_{h})}} \bigg] +
\end{eqnarray*}
\vskip 0.10truecm
\begin{displaymath}
\!\!\!\!\!\!\!\!\!\!\!\!\!\!\!\!\!\!\!\!\!\!\!\!\!\!
+ S_{L} \bigg[ \Lb \frac{c_{3}}{c_{1}} \Rb \sin{\!(\phi_{h})}\,A_{UL}^{\sin{\!(\phi_{h})}} + \Lb \frac{c_{2}}{c_{1}} \Rb \sin{\!(2\phi_{h})}\,A_{UL}^{\sin{\!(2\phi_{h})}} \bigg] +
\end{displaymath}
\vskip 0.35truecm
\begin{displaymath}
\!\!\!\!\!\!\!\!\!\!\!\!\!\!\!\!\!\!\!\!\!\!\!\!\!\!\!\!\!\!\!\!\!\!\!\!\!\!\!\!\!\!\!\!\!\!\!\!\!\!\!\!\!\!\!\!\!\!\!\!\!\!\!
+ S_{L}\lambda_{e} \bigg[ \Lb \frac{c_{5}}{c_{1}} \Rb A_{LL} + \Lb \frac{c_{4}}{c_{1}} \Rb \cos{\!(\phi_{h})}\,A_{LL}^{\cos{\!(\phi_{h})}} \bigg] +
\end{displaymath}
\vskip -0.05truecm
\begin{eqnarray*}
& &
+ S_{T} \bigg[ \sin{\!(\phi_{h} - \phi_{{\!}_{S}})} A_{UT,T}^{\sin{\!(\phi_{h} - \phi_{{\!}_{S}})}} +
\nonumber\\
& &~~
+ \Lb \frac{c_{2}}{c_{1}} \Rb \sin{\!(\phi_{h} + \phi_{{\!}_{S}})}\,A_{UT}^{\sin{\!(\phi_{h} + \phi_{{\!}_{S}})}} + 
\Lb \frac{c_{2}}{c_{1}} \Rb \sin{\!(3\phi_{h} - \phi_{{\!}_{S}})}\,A_{UT}^{\sin{\!(3\phi_{h} - \phi_{{\!}_{S}})}} + 
\nonumber\\
& &~~
+ \Lb \frac{c_{3}}{c_{1}} \Rb \sin{\!(\phi_{{\!}_{S}})}\,A_{UT}^{\sin{\!(\phi_{{\!}_{S}})}} + 
\Lb \frac{c_{3}}{c_{1}} \Rb \sin{\!(2\phi_{h} - \phi_{{\!}_{S}})}\,A_{UT}^{\sin{\!(2\phi_{h} - \phi_{{\!}_{S}})}} \bigg] + 
\end{eqnarray*}
\vskip -0.30truecm
\begin{eqnarray}
& & \!\!\!\!\!\!\!\!\!\!\!\!\!\!\!\!\!\!\!\!\!\!\!\!\!
+ S_{T} \lambda_{e} \bigg[ \Lb \frac{c_{5}}{c_{1}} \Rb \cos{\!(\phi_{h} - \phi_{{\!}_{S}})}\,A_{LT}^{\cos{\!(\phi_{h} - \phi_{{\!}_{S}})}} +
\Lb \frac{c_{4}}{c_{1}} \Rb \cos{\!(\phi_{{\!}_{S}})}\,A_{LT}^{\cos{\!(\phi_{{\!}_{S}})}} +
\nonumber\\
& & \!\!\!\!\!
+ \Lb \frac{c_{4}}{c_{1}} \Rb \cos{\!(2\phi_{h} - \phi_{{\!}_{S}})}\,A_{LT}^{\cos{\!(2\phi_{h} - \phi_{{\!}_{S}})}} \bigg] \Bigg\} ,
\label{eqn_eq:SIDIS_SSA1}
\end{eqnarray}
with the asymmetries generally defined as various ratios of polarized and unpolarized cross sections:
\beq
A_{XY}^{\mbox{a.d.}} \equiv A_{XY}^{\mbox{a.d.}}(x_{{\!}_{Bj}},Q^2,z_{h},P_{hT}) = \frac{F_{XY}^{\mbox{a.d.}}(x_{{\!}_{Bj}},Q^2,z_{h},P_{hT})}
{F_{UU}(x_{{\!}_{Bj}},Q^2,z_{h},P_{hT})} ,
\label{eqn_eq:SIDIS_SSA2}
\eeq
where the superscript $\mbox{a.d.}$ means angular dependence (azimuthal modulations), and the subscripts $XY$ are explained in the paragraph below 
Eq.~(\ref{eqn_eq:SIDIS_kin3}). Meanwhile, how these polarization indices are organized in the event generator framework (in somewhat
different way) are shown in details in  Sec.~\ref{sec:LH_coeff}.

It is obvious from Eq.~(\ref{eqn_eq:SIDIS_SSA1}) that at the order of $\mathcal{O}(1/Q)$, there are totally five SSAs with 
an unpolarized beam and a transversely polarized target. Two of these SSAs, due to the Collins effect \cite{Collins:1992kk} and 
Sivers effect \cite{Sivers:1989cc}, are twist-2 observables: $A_{UT}^{\rm Collins}$ and $A_{UT}^{\rm Sivers}$, respectively. 
The Collins effect emerges from the convolution of the transversity TMD and the Collins FF, the Sivers effect stems from the
convolution of the Sivers TMD and the unpolarized FF\footnote{An example showing the power of the TMD factorization (mentioned in
Sec.~\ref{sec:SIDIS_kinematics}), where the structure functions $F_{UT}^{\sin{\!(\phi_{h} + \phi_{{\!}_{S}})}}$ and 
$F_{UT}^{\sin{\!(\phi_{h} - \phi_{{\!}_{S}})}}$ $\Lb \equiv F_{UT,T}^{\sin{\!(\phi_{h} - \phi_{{\!}_{S}})}} \Rb$ are expressed 
by those convolutions, is the agreement between model calculations and COMPASS/HERMES experimental results for the Sivers and
Collins effects \cite{Anselmino:2008jk,Anselmino:2008sga}.}.

The general form of the transverse non-separated SSA can be written with all three leading-twist (or twist-2) and two 
subleading-twist (or twist-3) terms:
\begin{eqnarray}
& & \!\!\!\!\!\!\!\!\!\!\!\!\!\!\! 
A_{UT} = A_{UT}^{\rm Collins}\sin{\!(\phi_{h} + \phi_{{\!}_{S}})} + A_{UT}^{\rm Sivers}\sin{\!(\phi_{h} - \phi_{{\!}_{S}})} + 
\nonumber \\
& & \!\! + A_{UT}^{\rm Pretzelosity}\sin{\!(3\phi_{h} - \phi_{{\!}_{S}})} + A_{UT}^{\rm sl\mbox{-}t1}\sin{\!(\phi_{{\!}_{S}})}
+ A_{UT}^{\rm sl\mbox{-}t2}\sin{\!(2\phi_{h} - \phi_{{\!}_{S}})} , 
\label{eqn_eq:SIDIS_SSA3}
\end{eqnarray}
with the SSAs extracted from the Born cross section (see e.g. \cite{Barone:2015ksa}) given by
\begin{eqnarray}
& & \!\!\!\!\!\!\!\!\!\!\!\!\!
A_{UT}^{\rm Collins} \equiv A_{UT}^{\sin{\!(\phi_{h} + \phi_{{\!}_{S}})}} \equiv 2\langle \sin{\!(\phi_{h} + \phi_{{\!}_{S}})} \rangle =
\nonumber \\
& & ~~ = 2\,\frac{\int\limits_{0}^{2\pi} d\phi_{{\!}_{S}} \int\limits_{0}^{2\pi}d\phi_{h}\,\sin{\!(\phi_{h} + \phi_{{\!}_{S}})}\,\sigma^{\rm B}}
{\int\limits_{0}^{2\pi} d\phi_{{\!}_{S}}  \int\limits_{0}^{2\pi}d\phi_{h}\,\sigma^{\rm B}} =
\frac{c_{2}}{c_{1}} \frac{F_{UT}^{\sin{\!(\phi_{h} + \phi_{{\!}_{S}})}}}{F_{UU}} ,
\label{eqn_eq:SIDIS_SSA4}
\end{eqnarray}
\begin{eqnarray}
& & \!\!\!\!\!\!\!\!\!\!\!\!\!
A_{UT}^{\rm Sivers} \equiv A_{UT}^{\sin{\!(\phi_{h} - \phi_{{\!}_{S}})}} \equiv 2\langle \sin{\!(\phi_{h} - \phi_{{\!}_{S}})} \rangle =
\nonumber \\
& & \,\,\,\, = 2\,\frac{\int\limits_{0}^{2\pi} d\phi_{{\!}_{S}} \int\limits_{0}^{2\pi}d\phi_{h}\,\sin{\!(\phi_{h} - \phi_{{\!}_{S}})}\,\sigma^{\rm B}}
{\int\limits_{0}^{2\pi} d\phi_{{\!}_{S}}  \int\limits_{0}^{2\pi}d\phi_{h}\,\sigma^{\rm B}} =
\frac{1}{c_{1}} \frac{F_{UT}^{\sin{\!(\phi_{h} - \phi_{{\!}_{S}})}}}{F_{UU}} ,
\label{eqn_eq:SIDIS_SSA5}
\end{eqnarray}
where the prefactors are given in Eqs.~(\ref{eqn_eq:SIDIS_kin4}),~(\ref{eqn_eq:SIDIS_kin5}),~(\ref{eqn_eq:SIDIS_kin2})~and~(\ref{eqn_eq:SIDIS_kin9}).
$A_{UT}^{\rm  Pretzelosity}$ is defined similarly. For having more details on these SSAs, we refer to \cite{Bastami:2018xqd}: notably, its section~5.3 
for $A_{UT}^{\rm Sivers}$, section~5.4 for $A_{UT}^{\rm Collins}$, section~7.6 for $A_{UT}^{\rm sl\mbox{-}t1}$, and section~7.7 for $A_{UT}^{\rm sl\mbox{-}t2}$.
Meanwhile, $F_{UU}$, $F_{UT}^{\sin{\!(\phi_{h} + \phi_{{\!}_{S}})}}$, and $F_{UT}^{\sin{\!(\phi_{h} - \phi_{{\!}_{S}})}}$ are given in 
Eqs.~(\ref{eqn_eq:SIDIS_kin13}),~(\ref{eqn_eq:SIDIS_kin14}),~and~(\ref{eqn_eq:SIDIS_kin15}), respectively.
For $F_{UT}^{\sin{\!(\phi_{{\!}_{S}})}}$ and $F_{UT}^{\sin{\!(2\phi_{h} - \phi_{{\!}_{S}})}}$, we refer back to our discussion at the end of 
Sec.~\ref{sec:SIDIS_struc_func}.

When RCs are included, the asymmetries become modified through the substitution $\sigma^{B}\rightarrow\sigma^{in}_{\rm SIDIS}$ in Eq.~(\ref{eqn_eq:SIDIS_SSA4}) 
and Eq.~(\ref{eqn_eq:SIDIS_SSA5}), using Eq.~(\ref{eqn_eq:SIDIS_MC8}) for $\sigma^{in}_{\rm SIDIS}$. However, because the radiated photon causes the hadron 
frame to shift, the RC-modified asymmetries do not have the simple relationship to the structure functions given by the right-hand-side of Eq.~(\ref{eqn_eq:SIDIS_SSA4}) 
and Eq.~(\ref{eqn_eq:SIDIS_SSA5}).

\begin{figure}[hbt!]
\centering
\includegraphics[width=6.5cm]{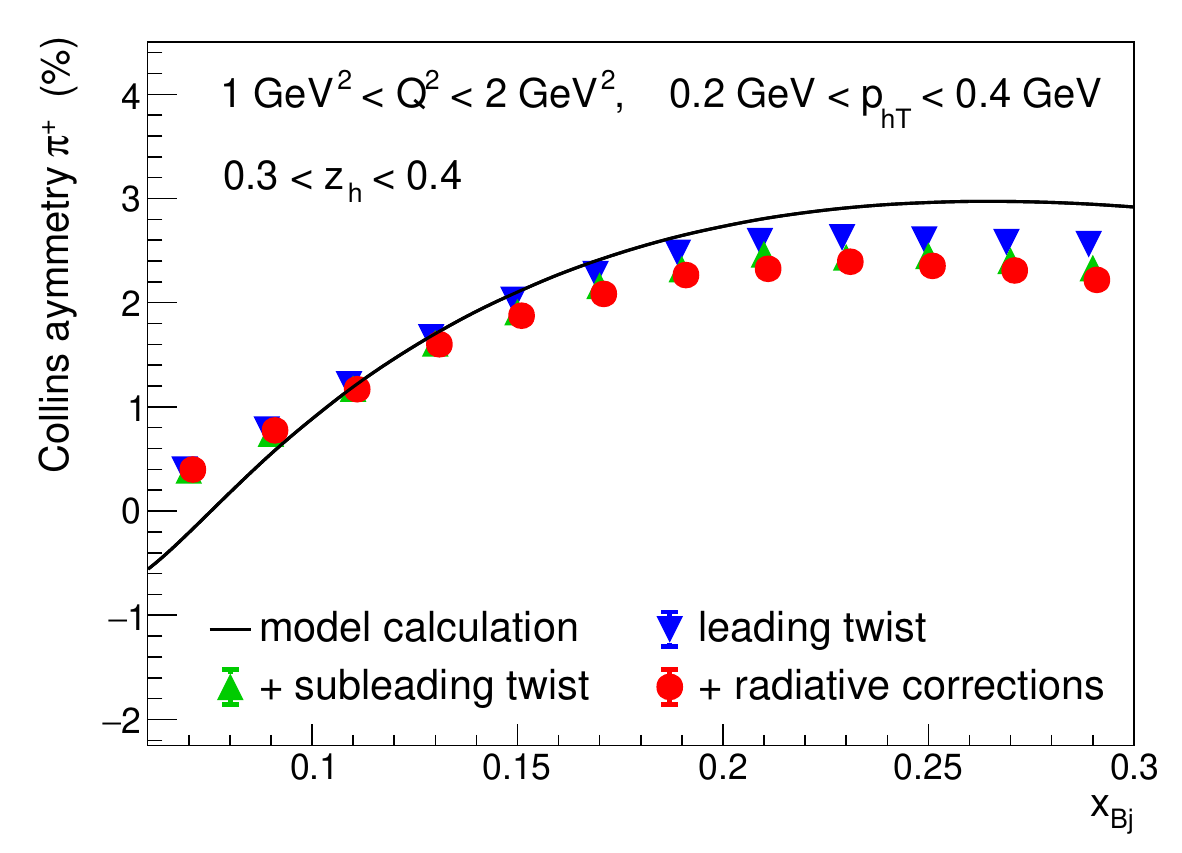}\label{fig:fig_Collins_1}
\includegraphics[width=6.5cm]{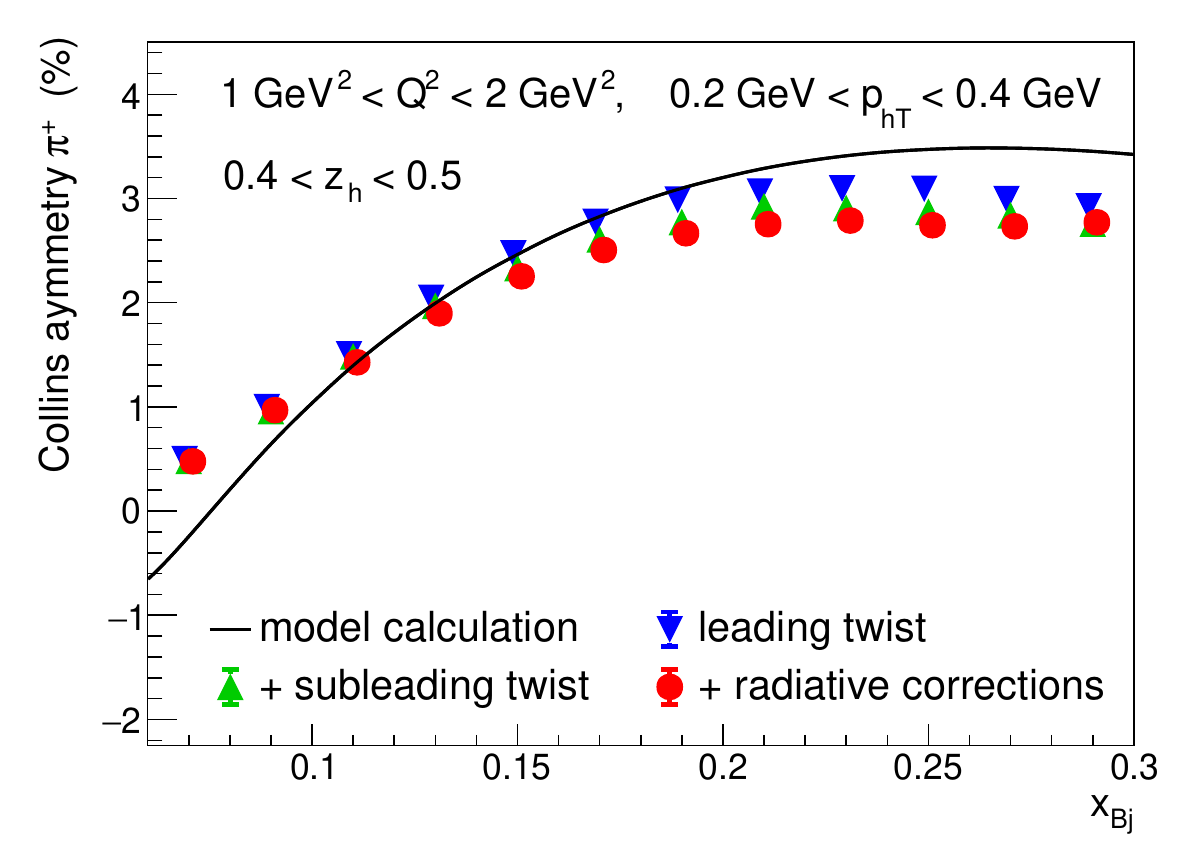}\label{fig:fig_Collins_2}
\\[-0.1cm]
{\bf (a) \hspace{5.5cm} (b) \hspace{5.5cm}}
\hskip 0.5truecm
\includegraphics[width=6.5cm]{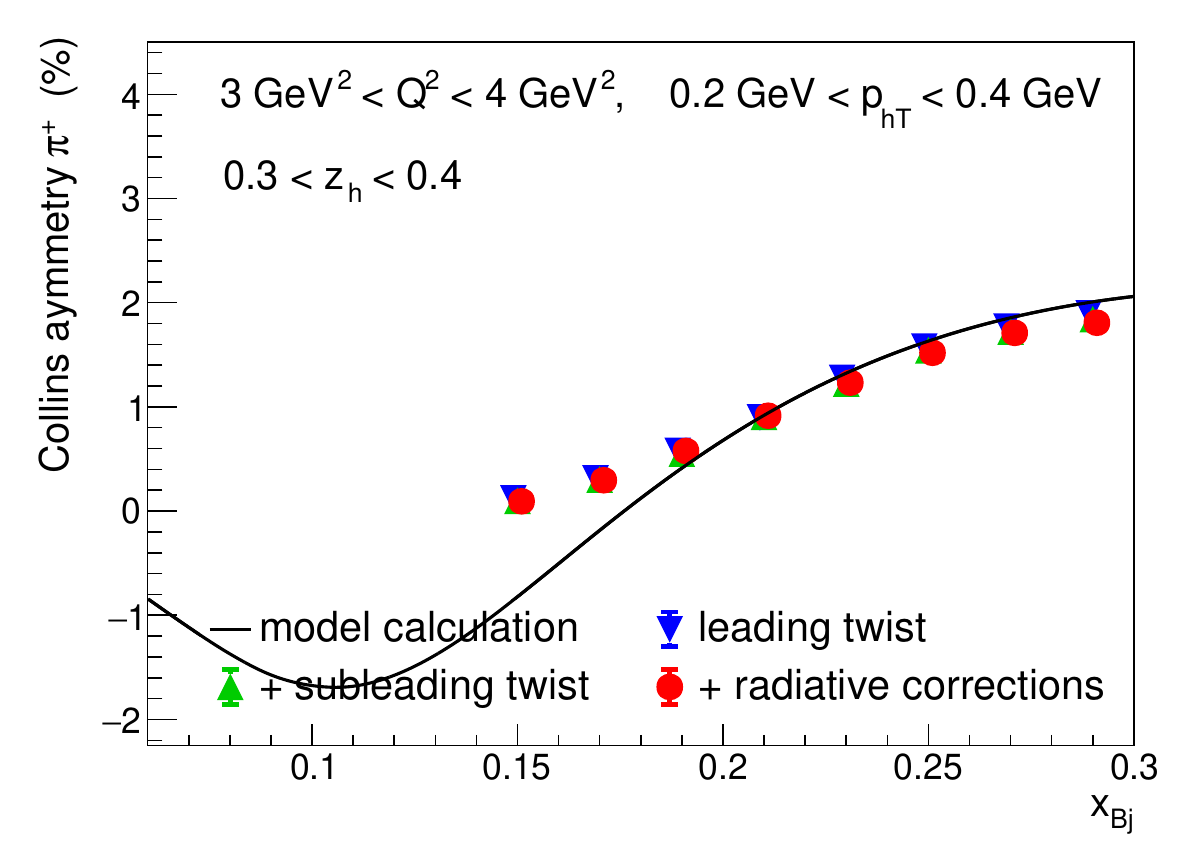}\label{fig:fig_Collins_3}
\includegraphics[width=6.5cm]{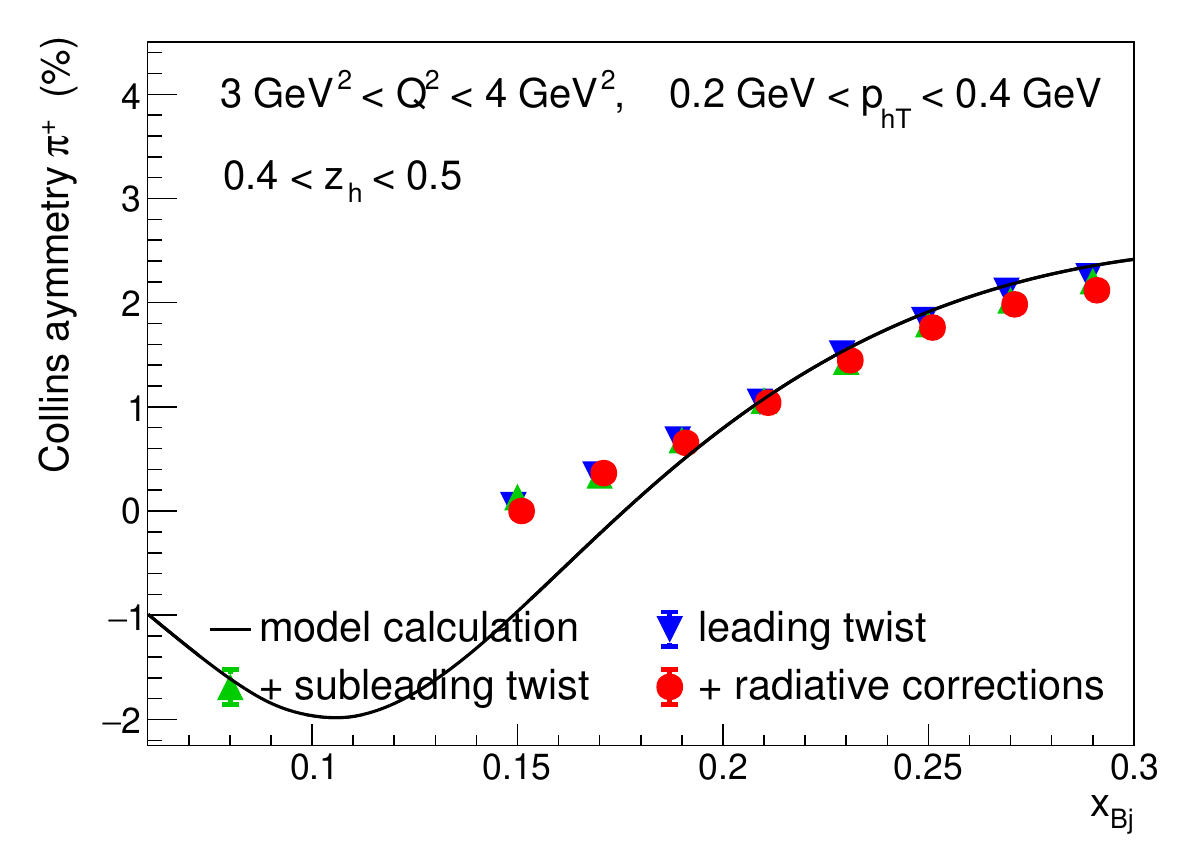}\label{fig:fig_Collins_4}
\\[-0.1cm]
{\bf (c) \hspace{5.5cm} (d) \hspace{5.5cm}}
\\[-0.1cm]
\caption{(Color online) The Collins SSA pseudo-data on $\pi^{+}$ as a function of $x_{{\!}_{Bj}}$, obtained from 
\texttt{SIDIS-RC EvGen} in specific $Q^{2}$, $z_{h}$, and $P_{hT}$ bins that are shown in the plots. The model calculation 
is at the bin center while the pseudo-data is integrated over the entire bin, leading to some additional difference between 
them. For better visualization, the points have been slightly offset horizontally. For producing these pseudo-data sets,
the electron beam energy of 10.6~GeV has been used (for the other kinematic parameters, see Listing~\ref{lst:sidisgen-parameters} 
in Appendix~B.)}
\label{fig:fig_SSA_Collins}
\end{figure}

\begin{figure}[hbt!]
\centering
\includegraphics[width=6.5cm]{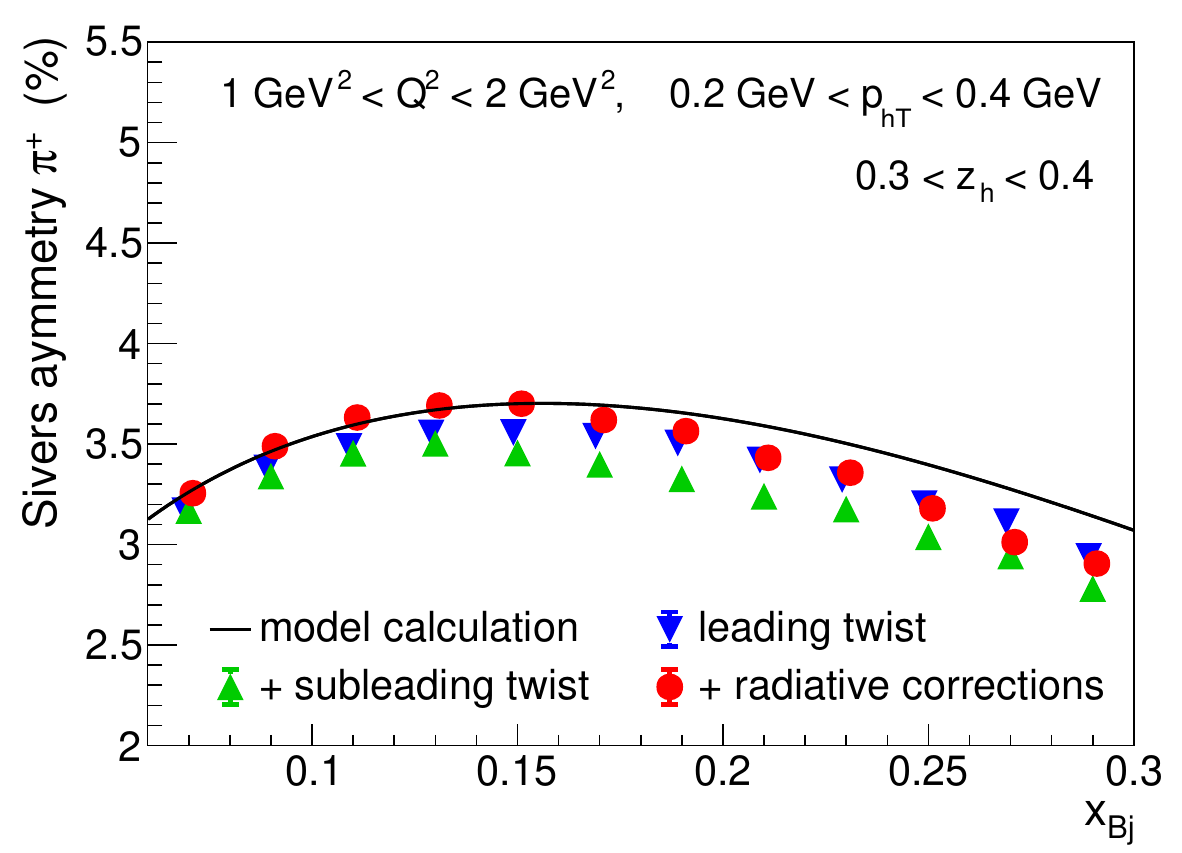}\label{fig:fig_Sivers_1}
\includegraphics[width=6.5cm]{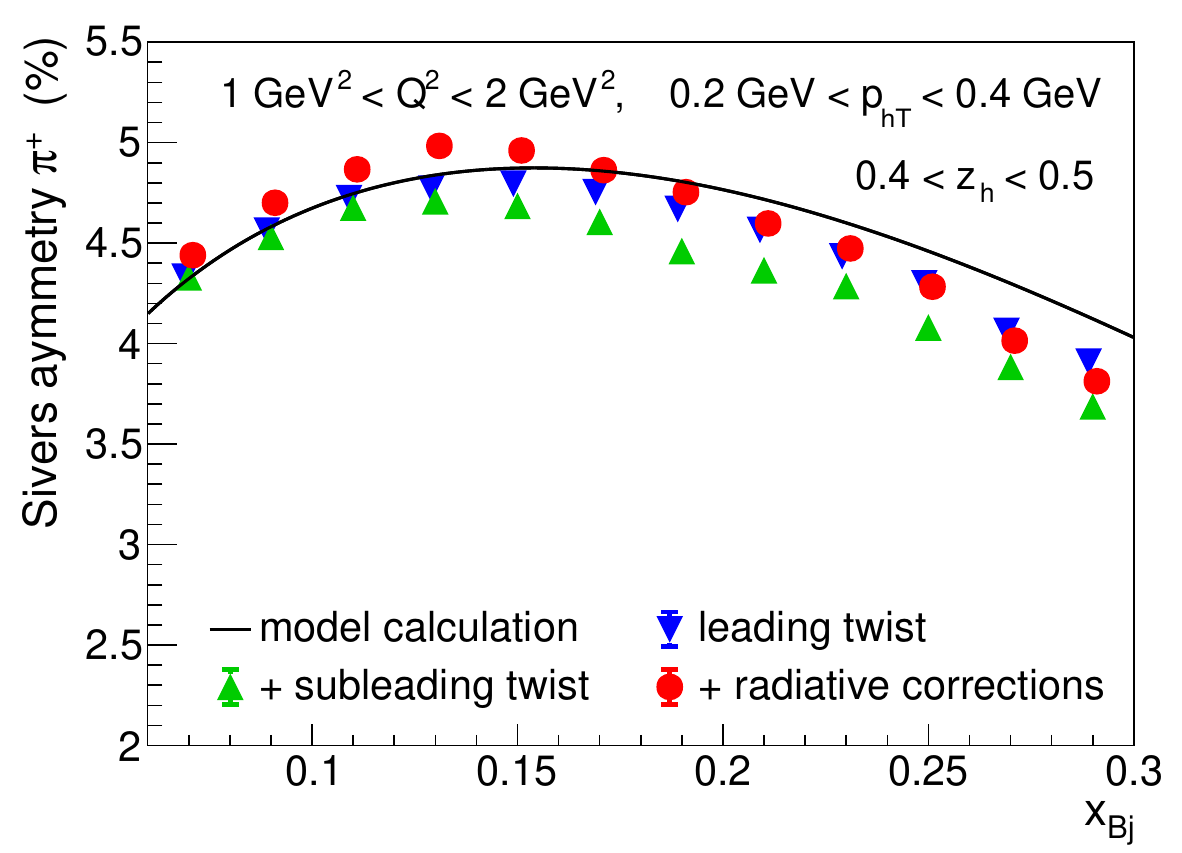}\label{fig:fig_Sivers_2}
\\[-0.1cm]
{\bf (a) \hspace{5.5cm} (b) \hspace{5.5cm}}
\hskip 0.5truecm
\includegraphics[width=6.5cm]{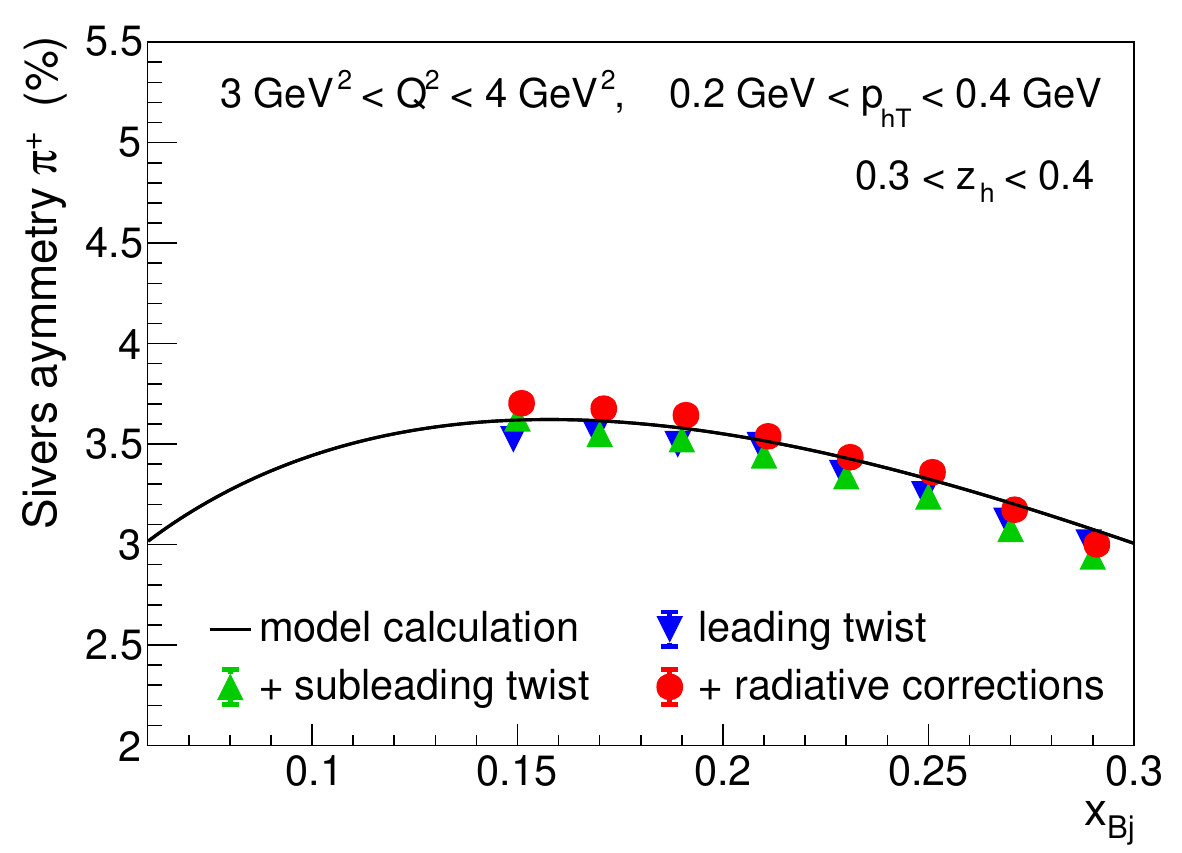}\label{fig:fig_Sivers_3}
\includegraphics[width=6.5cm]{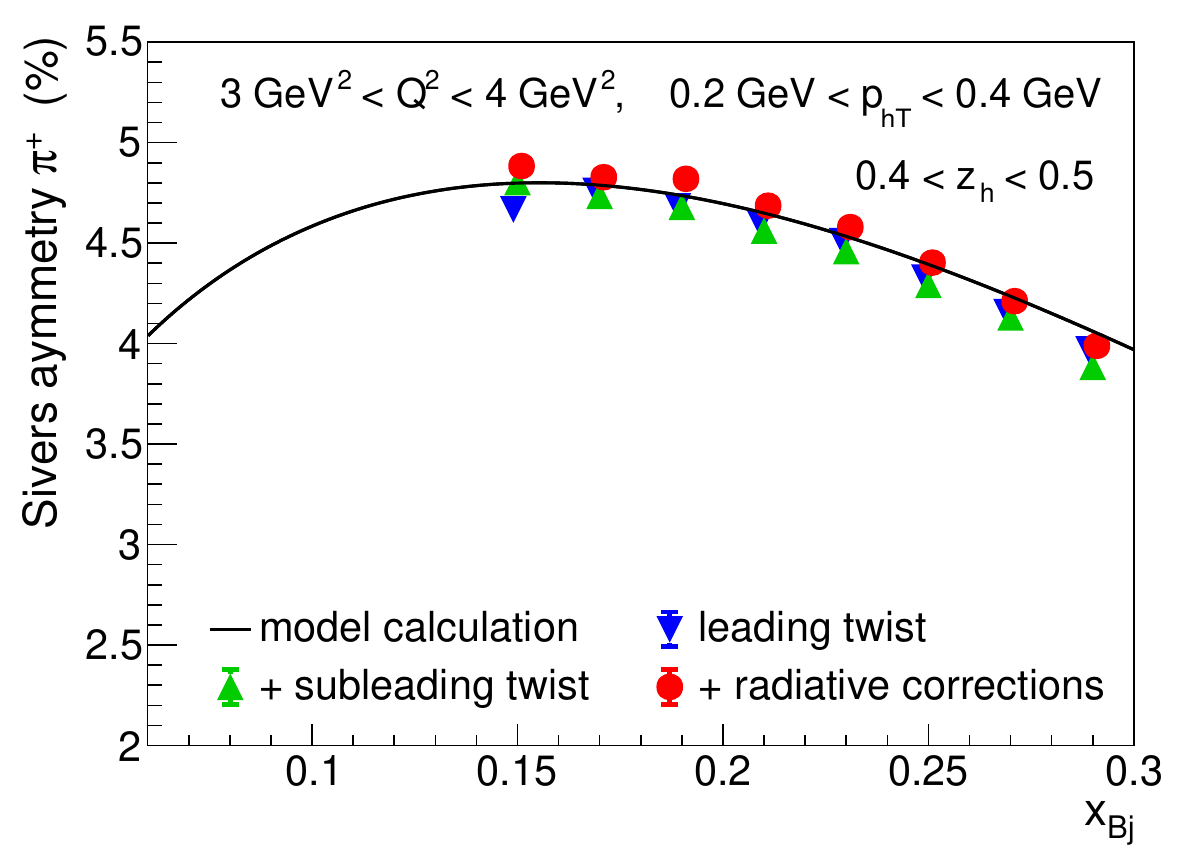}\label{fig:fig_Sivers_4}
\\[-0.1cm]
{\bf (c) \hspace{5.5cm} (d) \hspace{5.5cm}}
\\[-0.1cm]
\caption{(Color online) The Sivers SSA pseudo-data on $\pi^{+}$ as a function of $x_{{\!}_{Bj}}$, obtained from 
\texttt{SIDIS-RC EvGen} in the same $Q^{2}$, $z_{h}$, and $P_{hT}$ binning, as in Fig.~\ref{fig:fig_SSA_Collins}.  
The model calculation is at the bin center while the pseudo-data is integrated over the entire bin, leading to 
some additional difference between them. For better visualization, the points have been slightly offset horizontally.}
\label{fig:fig_SSA_Sivers}
\end{figure}

\begin{figure}[hbt!]
\centering
\includegraphics[width=6.5cm]{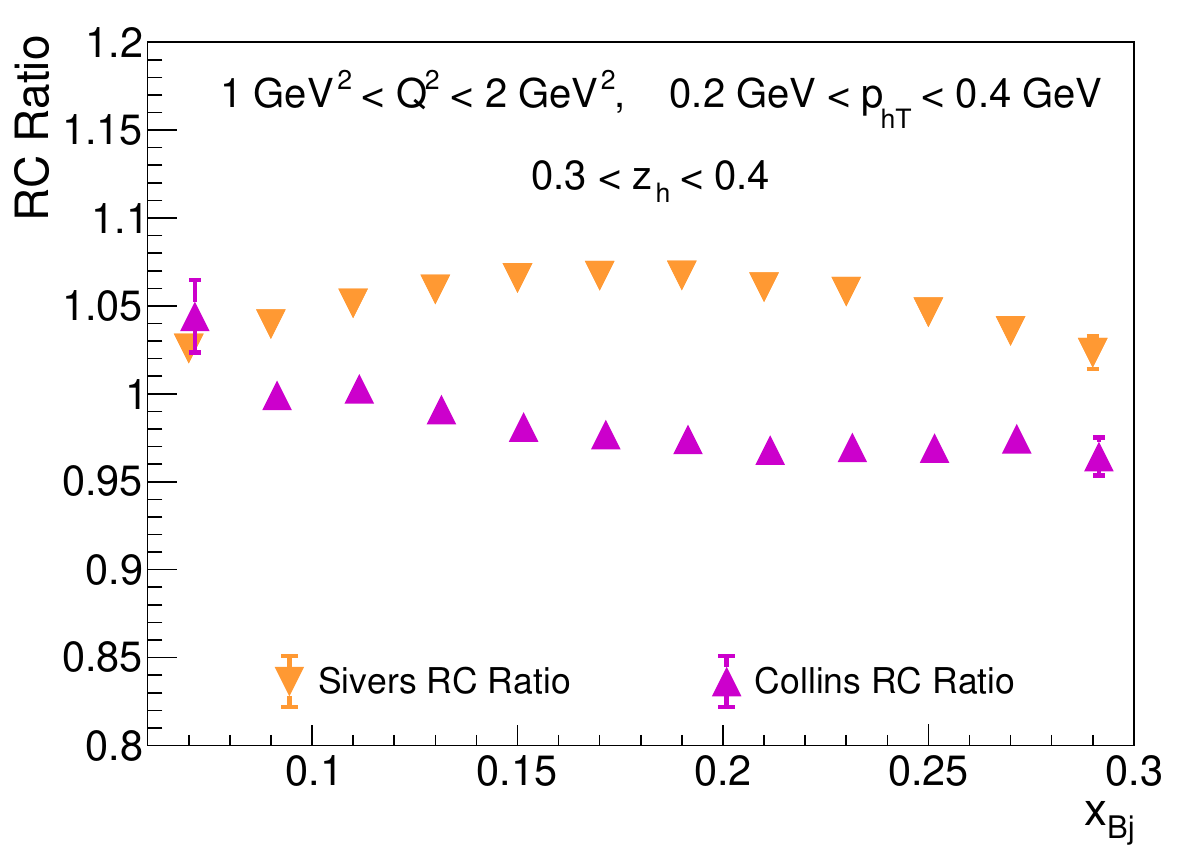}\label{fig:fig_asym-ratio-0-0}
\includegraphics[width=6.5cm]{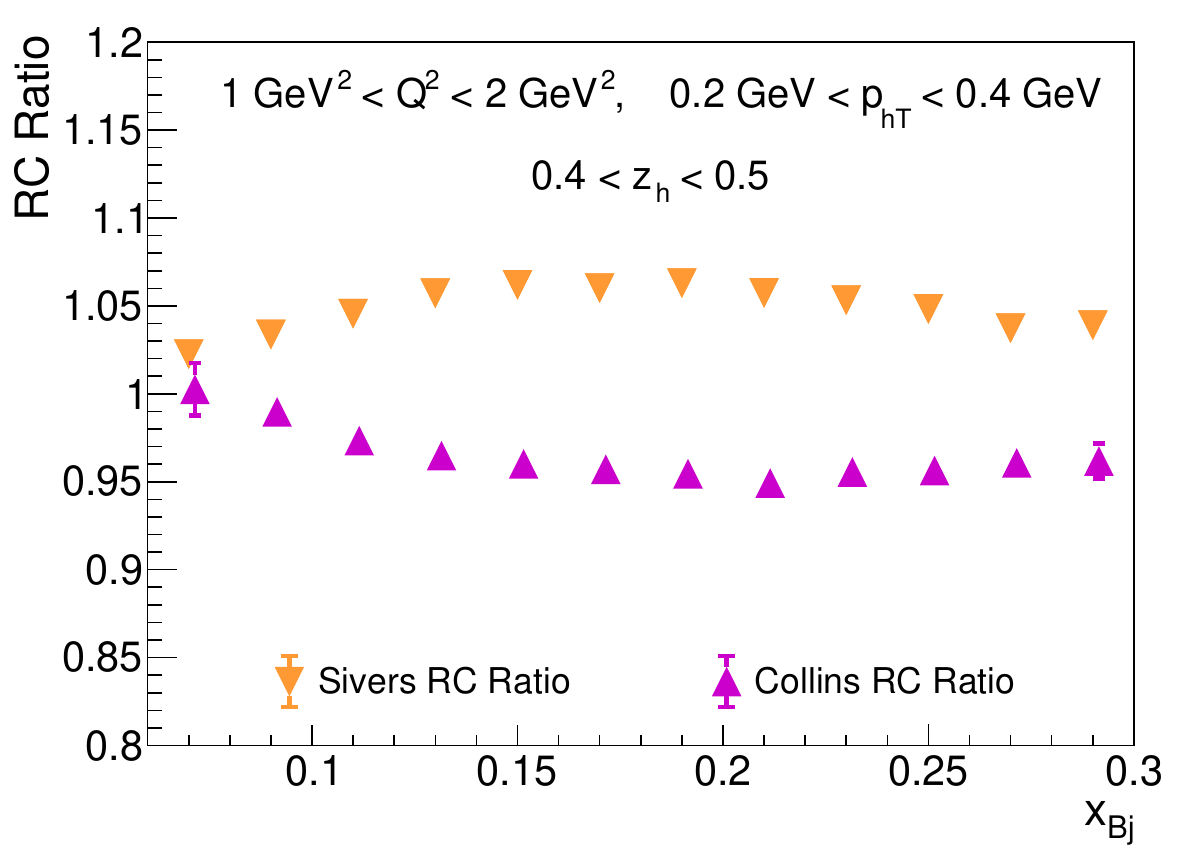}\label{fig:fig_asym-ratio-0-2}
\\[-0.1cm]
{\bf (a) \hspace{5.5cm} (b) \hspace{5.5cm}}
\hskip 0.5truecm
\includegraphics[width=6.5cm]{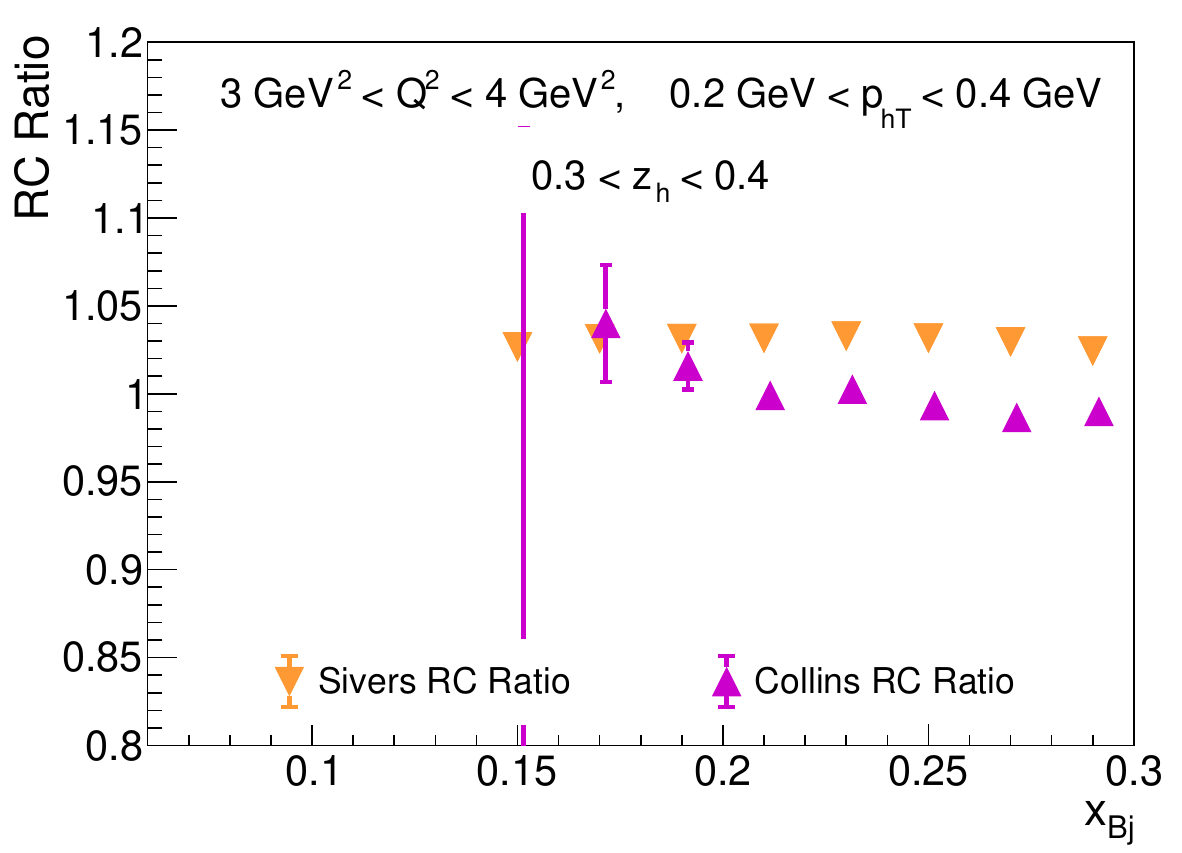}\label{fig:fig_asym-ratio-2-0}
\includegraphics[width=6.5cm]{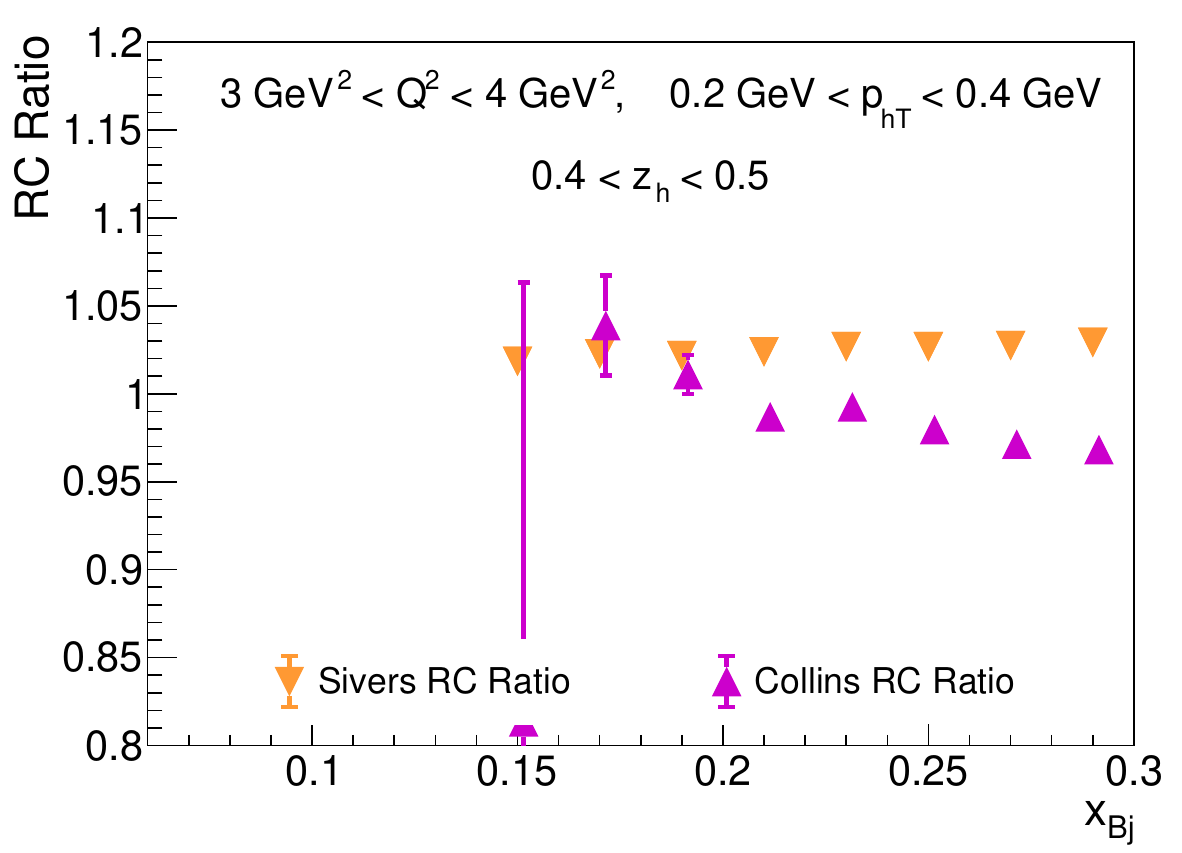}\label{fig:fig_asym-ratio-2-2}
\\[-0.1cm]
{\bf (c) \hspace{5.5cm} (d) \hspace{5.5cm}}
\\[-0.1cm]
\caption{(Color online) Ratios of the red circle pseudo-data to the green upward triangle pseudo-data shown in
Fig.~\ref{fig:fig_SSA_Collins} and Fig.~\ref{fig:fig_SSA_Sivers}. The ratios demonstrate the lowest-order RC 
effects on the Collins and Sivers SSAs when the subleading-twist effects are also taken into account. 
The large error bars of the Collins RC ratio in the lowest bin of $x_{Bj} \sim 0.15$ appear because 
the Collins asymmetry drops to near zero at this point. This figure shows one of the principal results we gain 
from \texttt{SIDIS-RC EvGen}.}
\label{fig:fig_SSA_ratio}
\end{figure}

Figs.~\ref{fig:fig_SSA_Collins}(a)-\ref{fig:fig_SSA_Collins}(d) show the Collins SSA for positively charged pions as a function of $x_{{\!}_{Bj}}$ 
in given kinematic bins of $Q^{2}$, $z_{h}$, and $P_{hT}$. 
\begin{itemize}
\item[$\circ$] The solid curves are obtained from direct calculations of $A_{UT}^{\rm Collins}$ at the bin center. 

\item[$\circ$] The blue downward triangle marker shows the Collins SSA measured from pseudo-data generated by \texttt{sidisgen} using the Born cross 
section $\sigma^{B}$ with structure functions at leading twist only.

\item[$\circ$] The green upward triangle marker shows the Collins SSA measured from pseudo-data using the Born cross section $\sigma^{B}$ with structure 
functions at leading and subleading twists.

\item[$\circ$] The red circle marker shows the Collins SSA measured from pseudo-data using the full RC cross section
$\sigma^{in}_{\rm SIDIS}$ with structure functions at leading and subleading twists. 
\end{itemize}
The pseudo-data errors are statistical uncertainties, for the treatment and propagation of which, we refer back to Sec.~\ref{sec:flex}.
Figs.~\ref{fig:fig_SSA_Sivers}(a)-\ref{fig:fig_SSA_Sivers}(d) show analogous plots made in the same bins but for the Sivers SSA.
In Figs.~\ref{fig:fig_SSA_ratio}(a)-\ref{fig:fig_SSA_ratio}(d), one can see ratios describing the RC effects on both Collins and Sivers asymmetries, 
when $A_{UT}^{\rm Collins|Sivers}|_{\rm RC}$ are compared to $A_{UT}^{\rm Collins|Sivers}$:
\beq
\mbox{RC~Ratio} = \frac{A_{UT}^{\rm Collins|Sivers}|_{\rm RC}}
{A_{UT}^{\rm Collins|Sivers}} .
\label{eqn_eq:SIDIS_Ratio}
\eeq
In these latter figures, it is apparent that the discussed observables can amount up to about 5\% of the Born-level results. 
In the context of upcoming comparisons with experimental data and/or more precise predictions, it will be important that 
\texttt{SIDIS-RC EvGen} also includes the missing higher-order RC effects in the future. One way as a first approximation 
could be employing the so-called (simplest) exponentiation procedure that had been used in \cite{Akushevich:2015toa} for 
calculations of RCs beyond the ultra-relativistic limit in unpolarized elastic $e+p$ scattering.

The exponentiation procedure is based on simply accounting for multiple soft photons and respective loops for canceling the 
infrared divergence. In our case, we need to consider $\sigma_{\rm SIDIS}^{B}$ + (term proportional of $\delta_{VR}$) as 
the first two terms in expansion of the exponent $\exp{\!\left[ (\alpha/\pi) \delta_{VR} \right]}\sigma_{\rm SIDIS}^{B}$ 
in a series over $\alpha$. Here, $\delta_{VR}$ is the sum of the infrared divergent terms in Eq.~(\ref{eqn_eq:SIDIS_RC25}). 
This is not only because one can find a straightforward way to account for multiple soft-photon radiation but also it is a 
regularization of the lowest-order RCs, since $\delta_{VR}$ will be infinitely close to the pion threshold at 
$P_{X}^{2} \rightarrow M_{th}^{2}$. The term $\delta_{VR}$ in Eq.~(\ref{eqn_eq:SIDIS_RC24}) can be later substituted 
by $\delta_{VR} \rightarrow (\pi/\alpha) \Lb \exp{\!\left[ (\alpha/\pi) \delta_{VR} \right]} - 1 \Rb$ in \texttt{SIDIS-RC EvGen}. 
Note that Eq.~(\ref{eqn_eq:SIDIS_MC8}) should be modified accordingly. Such a new implementation in the event generator, 
even given its simple nature, may increase the precision of the pseudu-data in all Figs.~\ref{fig:fig_SSA_Collins}, 
\ref{fig:fig_SSA_Sivers}, and \ref{fig:fig_SSA_ratio}.


With the $A_{UT}^{\rm Collins}$, $A_{UT}^{\rm Sivers}$, $A_{UT}^{\rm Pretzelosity}$ SSAs and $A_{UT}^{\rm sl\mbox{-}t1}$,
$A_{UT}^{\rm sl\mbox{-}t2}$ terms, it is possible to address essential questions in modern nuclear physics such as (i) 
whether one can provide a high precision test of lattice QCD predictions via the tensor charge \cite{DAlesio:2020vtw} 
(a fundamental quantity related to the nucleon spin); (ii) whether there are clear signatures of relativistic effects 
inside the nucleon, (iii) how one can extract quantitative information about a contribution of the quark orbital angular 
momentum to the proton spin; (iv) how to quantify the quark transverse motion inside the nucleon and observe spin-orbit
correlations.

\subsection{\label{sec:xs} Prospects on charged-hadron multiplicity and asymmetry data comparisons}

In the partonic description, the azimuthal asymmetries in Eq.~(\ref{eqn_eq:SIDIS_SSA1}) are applicable for studying spin-independent and 
spin-dependent TMDs and FFs (see Eqs.~(\ref{eqn_eq:SIDIS_RC8}) and (\ref{eqn_eq:SIDIS_RC11})). In this regard, another relevant experimental 
observable to investigate TMDs and FFs is the differential hadron multiplicity \cite{HERMES:2012uyd,COMPASS:2017mvk,CLAS:2008nzy}, defined as
\beq
n_{\rm SIDIS}^{h}(x_{{\!}_{Bj}}, Q^{2}, z_{h}, P_{hT}^{2}) \equiv \Lb \frac{d\sigma^{B}_{\rm SIDIS}}{dx_{{\!}_{Bj}}\,dQ^{2}\,dz_{h}\,dP_{hT}^{2}} \Rb 
\bigg/ 
\Lb \frac{d\sigma_{\rm DIS}}{dx_{{\!}_{Bj}}\,dQ^{2}} \Rb ,
\label{eqn_eq:SIDIS_mult}
\eeq
where the denominator is the DIS cross section.

Comparisons with experimental data on the multiplicity observable from Eq.~(\ref{eqn_eq:SIDIS_mult}) and on other asymmetries from 
Eq.~(\ref{eqn_eq:SIDIS_SSA1}), based on using pseudo-data from \texttt{SIDIS-RC EvGen}, will be topics for investigation in the future, 
after exclusive structure functions are also implemented in \texttt{SIDIS-RC EvGen}.

\section{\label{sec:concl} Summary and outlook}

MC event generators are important constituents of a variety of experimental analyses. They are extensively used to make predictions and preparations for 
designed experiments. Physics models and frameworks used for underlying event generation and their reliability are essential for understanding of various 
systematic effects in studies of different observables. One source of such effects are induced by radiative corrections in semi-inclusive deep inelastic scattering.

In this paper, we presented a C\texttt{++} coded standalone MC event generator called \texttt{SIDIS-RC EvGen}, for a purpose of studying the partonic structure of the 
nucleon in 3D momentum space, by using SIDIS that also includes the lowest-order RCs. We discussed and simultaneously provided various details on
\begin{itemize}
\item[(1)] the twist-2 and twist-3 SIDIS structure functions, which are thoroughly and systematically investigated in the Wandzura-Wilczek-type and Gaussian 
approximations \cite{Bastami:2018xqd,WW-SIDIS:2018}\footnote{These references generally render very useful inputs for MC event generators.} on the
 basis of available data;

\item[(2)] the inelastic tail to the SIDIS six-fold differential cross section, including the lowest-order RC components, for which explicit exact analytical formulas 
(with a longitudinally polarized lepton and an arbitrarily polarized target) are computed in compact and covariant form convenient for numerical 
analyses \cite{Akushevich:2019mbz};

\item[(3)] our created event generator's structure and functionality focusing on its library component for computing cross sections and binary (generator) 
component for event generation \cite{SIDIS-RC_EvGen:2020}, along with showing numerical results on the Collins and Sivers SSAs 
in the SIDIS process. 
\end{itemize}

It is relevant to outline the following potential prospects related to our work with \texttt{SIDIS-RC EvGen}; such as
\begin{itemize}
\item[(i)] a) increasing the efficiency of \texttt{SIDIS-RC EvGen} by improving the foam efficiency, or even replacing the 
               FOAM algorithm by the VEGAS algorithm;
               
           b) providing with other options for including state-of-the-art parameterizations of TMDs, used in most recent and upcoming 
           phenomenological studies; such as of Refs.~\cite{Scimemi:2019cmh,Bertone:2019nxa,Bacchetta:2022awv,Cammarota:2020qcw,Cerutti:2022yya,Gamberg:2022kdb,Echevarria:2020hpy,Bacchetta:2020gko,Bury:2020vhj,Bury:2021sue,Barry:2023qqh}.

\item[(ii)] a) incorporating the exclusive structure functions into our current framework;

            b) including $Q^{2}$ evolution of the SIDIS structure functions as an optional feature;

            c) improving the parameterization that describes the contribution of vacuum polarization by hadrons ($\delta_{\rm vac}^{h}$), 
               making use of the most recent hadronic data for fitting \cite{Davier:2019can,Davier:2017zfy}, the advanced calculations 
               from \cite{Bonciani:2003ai,Jegerlehner:2009ry,Jegerlehner:2017gek}, and the software package \texttt{alphaQED} 
               of F.~Jegerlehner \cite{Jegerlehner:2019};
               
            d) implementing the exponentiation procedure for $\sigma_{\rm SIDIS}^{in}$ already discussed in Sec.~\ref{sec:asym};

            e) planning by some of us to calculate the second-order SIDIS RCs in a more robust way (as a follow-up to Ref.~\cite{Akushevich:2019mbz}), 
               and subsequently incorporating the updated RC framework in the event generator.

\item[(iii)] comparing the event generator's output with the HERMES, COMPASS, and JLab various SSA data sets
\cite{HERMES:2009lmz,HERMES:2010mmo,HERMES:2012kpt,COMPASS:2012dmt,COMPASS:2014bze,COMPASS:2014kcy,Qian:2011py}, with the charged-hadron 
multiplicity data sets \cite{HERMES:2012uyd,COMPASS:2017mvk,CLAS:2008nzy}, as well as making predictions for the future EIC 
\cite{Accardi:2012qut,AbdulKhalek:2021gbh}.

\item[(iv)] developing eventually a versatile framework for nucleon's 3D momentum structure extraction from our standalone generator: an example of how a 
more flexible MC generator can be built by extending existing generators is like the multi-purpose particle physics event generator Herwig\texttt{++} 3.0, built on 
the basis of the ThePEG substructure that underlies the Herwig\texttt{++} versions \cite{Bellm:2015jjp}. 
\end{itemize}

\medskip
\section*{Acknowledgements}
This work has been supported in part by the U.S. Department of Energy, Office of Science, Office of Nuclear Physics under 
contracts No.~DE-FG02-03ER41231 (D.B., V.K., H.G., and Z.Z.), No.~DE-AC02-06CH11357 (C.P.), No.~DE-AC05-06OR23177 (A.P.) 
under which Jefferson Science Associates, LLC, manages and operates Jefferson Lab; and within the framework of the TMD Topical 
Collaboration (A.P.), and by the National Science Foundation Grant No.~PHY-2012002 (A.P.).

\appendix
\renewcommand{\theequation}{A\arabic{equation}}
\setcounter{equation}{0}
\renewcommand{\thefigure}{A\arabic{figure}}
\setcounter{figure}{0}

\section*{\label{sec:appA} Appendix A. Program listings of the flexible/streamlined modes}

\lstset{
    language=C++,
    numbers=left,
    basicstyle=\scriptsize\ttfamily\color{black},
    keywordstyle=\bfseries\color{teal},
    commentstyle=\itshape\color{gray},
    identifierstyle=\color{black},
    stringstyle=\color{red}
}
\begin{lstlisting}[floatplacement=h!,language=C++,caption={\label{lst:libsidis-modes} Demonstration of differences between the \texttt{libsidis} 
flexible mode (Sec.~\ref{sec:flex}) and streamlined mode (Sec.~\ref{sec:stream}).}]
using namespace sidis;
// Initial state.
part::Particles ps(
    part::Nucleus::P, part::Lepton::E, part::Hadron::PI_P,
    MASS_P + MASS_PI_0);
double S = 2 * MASS_P * 10;
// Phase space.
kin::PhaseSpace ph_space { 0.2, 0.9, 0.3, 2., 0.5*PI, 0.2*PI };
kin::Kinematics kin(ps, S, ph_space);
// Structure function parameterization.
sf::set::ProkudinSfSet sf_set;
// Polarization vectors.
double beam_pol = 0.5;
math::Vec3 targ_pol(0.1, 0.4, 0.2);
if (flexible) {
    // Born coefficients.
    xs::Born b(kin);
    // Leptonic coefficients.
    lep::LepBornUU lep_uu(kin);
    lep::LepBornUP lep_up(kin);
    lep::LepBornLU lep_lu(kin);
    lep::LepBornLP lep_lp(kin);
    // Hadronic coefficients.
    had::HadUU had_uu(kin, sf_set);
    had::HadUL had_ul(kin, sf_set);
    had::HadUT had_ut(kin, sf_set);
    had::HadLU had_lu(kin, sf_set);
    had::HadLL had_ll(kin, sf_set);
    had::HadLT had_lt(kin, sf_set);
    // Polarized parts of Born cross section.
    double xs_uu =  xs::born_uu_base (b, lep_uu, had_uu);
    double xs_ul =  xs::born_ul_base (b, lep_up, had_ul);
    double xs_ut1 = xs::born_ut1_base(b, lep_up, had_ut);
    double xs_ut2 = xs::born_ut2_base(b, lep_uu, had_ut);
    double xs_lu =  xs::born_lu_base (b, lep_lu, had_lu);
    double xs_ll =  xs::born_ll_base (b, lep_lp, had_ll);
    double xs_lt1 = xs::born_lt1_base(b, lep_lp, had_lt);
    double xs_lt2 = xs::born_lt2_base(b, lep_lu, had_lt);
    math::Vec3 xs_up(xs_ut1, xs_ut2, xs_ul);
    math::Vec3 xs_lp(xs_lt1, xs_lt2, xs_ll);
    // Assemble to get complete Born cross section.
    double xs = (xs_uu + math::dot(xs_up, targ_pol))
        + beam_pol*(xs_lu + math::dot(xs_lp, targ_pol));
}
if (streamlined) {
    // Compute polarized cross section.
    double xs = xs::born(kin, sf_set, beam_pol, targ_pol);
}
\end{lstlisting}

\appendix
\renewcommand{\theequation}{B\arabic{equation}}
\setcounter{equation}{0}
\renewcommand{\thefigure}{B\arabic{figure}}
\setcounter{figure}{0}
\section*{\label{sec:appB} Appendix B. Quick start and Build of \texttt{SIDIS-RC EvGen}}

This appendix describes the procedure of the quick start of the event generator (for its building see 
\url{https://github.com/duanebyer/sidis}).
But first of all, let us list the following libraries that this software does utilize.
\begin{itemize}
\item[$\bullet$] WW-SIDIS: structure functions for the proton using the WW-type and Gaussian approximations \cite{Bastami:2018xqd,WW-SIDIS:2018};
\item[$\bullet$] GSL: multi-dimensional Monte-Carlo integration \cite{GSL}, including the Monte-Carlo integrator VEGAS \cite{VEGAS};
\item[$\bullet$] FOAM: MC event generation using spatial partitioning \cite{FOAM};
\item[$\bullet$] MSTWPDF: parton distribution functions for the proton \cite{Martin:2009iq,mstwpdf};
\item[$\bullet$] Cog: code generation with Python \cite{Cog};
\item[$\bullet$] ROOT: plotting and data analysis \cite{ROOT};
\item[$\bullet$] Cubature: a backup package for multi-dimensional numerical integration \cite{Cubature}.
\end{itemize}

\noindent \underline{\bf{Quick start (generator):}} An example of the event generator's \texttt{sidisgen} component's user command-line interface is demonstrated 
in Listing~\ref{lst:sidisgen-parameters}, which describes an input file with all parameters needed for event generation. The allowed options can be seen by 
\texttt{sidisgen} $--$\texttt{help}. The generator is run in two steps using the input file. The FOAM library must be initialized before the events can be generated 
themselves. With this initialization, one approximates the SIDIS differential cross section in a specified kinematic region. Thus, we first call
\beq
\texttt{sidisgen~--initialize}~\texttt{<parameter file>}
\label{eqn_eq:sidisgen_in}
\eeq
to create some saved FOAM tree that can be used in between runs. Then one should use the FOAM library for MC event generation, by calling
\beq
\texttt{sidisgen~--generate}~\texttt{<parameter file>}
\label{eqn_eq:sidisgen_gen}
\eeq
to generate variable-weighted events\footnote{Both importance sampling and rejection sampling have been used to reduce the variance of the 
weights. Nevertheless, the scale for rejection sampling can be increased by the user, if uniform event weights are preferred in performing some experimental analyses.}.
Afterwards, the resulting events are provided in a ROOT file, and it can be converted into other formats if required. Any produced event 
ROOT file will store the parameters used to produce it internally, which can be double-checked with \texttt{sidisgen} $--$\texttt{inspect} command. \\

\begin{lstlisting}[float,floatplacement=h!,language=python,caption={\label{lst:sidisgen-parameters} An example parameter file for interfacing with the \texttt{sidisgen} generator.}]
mc.num_events        10000
# structure function parameterization
phys.sf_set          prokudin
# which method of radiative corrections to use
phys.rc_method       approx
# soft threshold dividing non-radiative and radiative cross sections
phys.soft_threshold  0.01
# kinematic threshold for SIDIS process
phys.mass_threshold  1.073249081
# initial conditions
setup.beam           e
setup.target         p
setup.hadron         pi+
setup.beam_energy    10.6
setup.beam_pol       0
setup.target_pol     0 0 0
# cuts on kinematic variables
cut.x                0.1  0.6
cut.Q_sq             2.0  3.5
cut.z                0.3  0.7
cut.W_sq             5.29 1e10
\end{lstlisting}

\hskip -0.65truecm
The events will be written to a ROOT file with the following entries:
\begin{description}
    \item[{\bf\em params}:] A subdirectory storing a record of the parameters used to produce the event file (for later reference).
    \item[{\bf\em events}:] A \texttt{TTree} storing all of the events. It has the following branches:
    \begin{description}
        \item[\texttt{type})] Whether the event is non-radiative \texttt{(1)}, radiative \texttt{(2)}, or exclusive \texttt{(3)}.
        \item[\texttt{weight})] The variable weight of the event.
        \item[\texttt{x}, \texttt{y}, \texttt{z}, \texttt{ph\_t\_sq}, \texttt{phi}, \texttt{phi\_h})] The SIDIS phase-space variables of the event.
        \item[\texttt{R}, \texttt{tau}, \texttt{phi\_k})] Additional phase-space variables for radiative/exclusive events.
        \item[\texttt{p}, \texttt{k1}, \texttt{k2}, \texttt{q}, \texttt{ph}, \texttt{k})] Optional particle four-momenta, which will be included 
        only if the \texttt{file.write\_momenta} is set in the parameter file.
        \item[\texttt{mc\_coords[9]}, \texttt{jac})] Optional internal MC coordinates for the event, which will be included only if the 
        \texttt{file.write\_mc\_coords} is set in the parameter file.
    \end{description}
    \item[{\bf\em stats}:] A subdirectory storing statistics about the generated events. It has the following entries:
    \begin{description}
        \item[\texttt{num\_events[4]})] A number of events in an array that is indexed by the event type (\texttt{[0]} for all events, \texttt{[1]} 
        for non-radiative events, \texttt{[2]} for radiative events, \texttt{[3]} for exclusive events). This number will be greater than the number 
        of entries in \texttt{events} because some events are generated and then later rejected, either by cuts or by rejection sampling.
        \item[\texttt{num\_events\_acc[4]})] A number of accepted events by event type, which is equal to the number of entries in \texttt{events}.
        \item[\texttt{norm[4]})] A normalization needed to calculate the cross section by event type (see note on normalization below).
        \item[\texttt{prime[4]})] A low quality estimate of cross section by event type. There are no guarantees that this is accurate; if a good 
        cross-section estimate is needed, the normalization process should be followed.
    \end{description}
\end{description}

\noindent {\bf\em Normalization{\rm :}}
The produced events have variable weights, although if the various variance-reduction methods are successful, most the weights should be close to 1. 
A normalization is needed to convert the event weights into a cross section. The normalization can be different for events of different types, such as 
a non-radiative normalization $\mathcal{N}_\textrm{nrad}$ and a radiative normalization $\mathcal{N}_\textrm{rad}$. Then the cross section is calculated 
from the event weights $w_{i}$ as
\beq
\sigma^{in}_{\mathrm{SIDIS}} \approx \frac{1}{\mathcal{N}_\mathrm{nrad}} \sum_{i} w_{i,\mathrm{nrad}} + 
\frac{1}{\mathcal{N}_\mathrm{rad}} \sum_{i} w_{i,\mathrm{rad}} + \ldots .
\label{eqn_eq:SIDIS_normalization}
\eeq
The normalizations are calculated by \texttt{sidisgen} and included in the output ROOT file within the array \texttt{stats/norm[4]}. We have, 
$\mathcal{N}_\mathrm{nrad}=\texttt{norm[1]}$, $\mathcal{N}_\mathrm{rad}=\texttt{norm[2]}$, and $\mathcal{N}_\mathrm{excl}=\texttt{norm[3]}$. However, 
\texttt{sidisgen} will attempt to generate events, so that all of the normalizations are nearly equal. The ``average" normalization is provided as 
$\mathcal{N}=\texttt{norm[0]}$, and should be sufficient for most cases. Then the cross section is calculated without paying attention to event types:
\beq
\sigma^{in}_\mathrm{SIDIS} \approx \frac{1}{\mathcal{N}} \sum_{i} w_{i} .
\label{eqn_eq:SIDIS_normalization2}
\eeq

\noindent {\bf\em Inspecting output ROOT file{\rm :}}
The \texttt{sidisgen --inspect} command will give a brief description of an output ROOT file that is produced using \texttt{sidisgen}. This description 
includes the parameters used to produce the output file, and some statistics of the events contained in the output file. \\

\noindent {\bf\em Merging ROOT files{\rm :}}
The \texttt{sidisgen --merge-soft} command will merge multiple output files together into a single ROOT file. The merged file still refers to the other 
output files through filename references, such that the original output files must not be renamed or deleted. The merged file can be used with any analysis 
script that expects a \texttt{sidisgen} output file. Alternatively, the \texttt{sidisgen --merge-hard} command will perform a hard copy, such that the 
original output files can then be renamed or deleted. However, this is not recommended as some information from the original ROOT files will be lost during 
the merging process. It will take much longer to complete than a soft merge. \\

\noindent \underline{\bf{Quick start (library):}} Listing~\ref{lst:libsidis-interface} shows how to compute the Born cross section. Several other 
demonstrations of different features of the \texttt{libsidis} library can be found in the \texttt{example} folder, which is located at 

\url{https://github.com/duanebyer/sidis/tree/master/example} \\

\begin{lstlisting}[floatplacement=h!,language=C++,caption={\label{lst:libsidis-interface} An example program using \texttt{libsidis} to compute the Born cross section.}]
#include <iostream>
#include <sidis/sidis.hpp>
#include <sidis/sf_set/prokudin.hpp>

sidis::Real const PI = sidis::PI;
sidis::Real const M_TH = sidis::MASS_P + sidis::MASS_PI_0;

int main() {
    sidis::part::Particles particles(
        sidis::part::Nucleus::P,   // Target nucleus.
        sidis::part::Lepton::E,    // Beam lepton.
        sidis::part::Hadron::PI_P, // Leading hadron.
        M_TH                       // Threshold mass of undetected part.
    );
    sidis::Real S = 2. * 10.6 * particles.M; // Kinematic variable `S = 2 p k1`.
    sidis::kin::PhaseSpace phase_space {
        0.2,      // Bjorken x.
        0.9,      // Bjorken y.
        0.3,      // Bjorken z.
        2.,       // Transverse momentum of hadron, squared.
        0.5 * PI, // Azimuthal angle of hadron.
        0.,       // Azimuthal angle of transverse target polarization.
    };
    sidis::kin::Kinematics kin(particles, S, phase_space);
    sidis::Real beam_pol = 0.;
    sidis::math::Vec3 target_pol(0., 0., 0.);
    // Compute structure functions with WW-type approximation.
    sidis::sf::set::ProkudinSfSet sf;
    sidis::Real born_xs = sidis::xs::born(kin, sf, beam_pol, target_pol);
    std::cout << "Born unpolarized cross-section is " << born_xs << std::endl;
    return 0;
}
\end{lstlisting}





\end{document}